\documentclass[twocolumn]{aastex61}
\pdfoutput=1
\shorttitle{IMBHs in Globular Clusters}
\shortauthors{Tremou et al.}

\begin{document}

\title{The MAVERIC Survey: Still No Evidence for Accreting Intermediate-mass Black Holes in Globular Clusters}

\correspondingauthor{E.Tremou}
\email{tremou@msu.edu}
\author[0000-0002-4039-6703]{Evangelia Tremou}
\affil{Center for Data Intensive and Time Domain Astronomy,
Department of Physics and Astronomy,
Michigan State University,
   East Lansing, MI 48824, USA}
\author[0000-0002-1468-9668]{Jay Strader}
\affiliation{Center for Data Intensive and Time Domain Astronomy,
Department of Physics and Astronomy,
 Michigan State University,
    East Lansing, MI 48824, USA}
\author[0000-0002-8400-3705]{Laura Chomiuk}
\affiliation{Center for Data Intensive and Time Domain Astronomy,
Department of Physics and Astronomy,
 Michigan State University,
    East Lansing, MI 48824, USA}
\author{Laura Shishkovsky}
\affiliation{Center for Data Intensive and Time Domain Astronomy,
Department of Physics and Astronomy,
 Michigan State University,
    East Lansing, MI 48824, USA}
\author[0000-0003-0976-4755]{Thomas J. Maccarone}
\affiliation{Department of Physics \& Astronomy,
Texas Tech University,
Box 41051, Lubbock, TX 79409-1051, USA}
\author[0000-0003-3124-2814]{James C.A. Miller-Jones}
\affiliation{International Centre for Radio Astronomy Research,
 Curtin University,
GPO Box U1987, Perth, WA 6845, Australia}

\author[0000-0003-4553-4607]{Vlad Tudor}
\affiliation{International Centre for Radio Astronomy Research,
 Curtin University,
GPO Box U1987, Perth, WA 6845, Australia}

\author[0000-0003-3944-6109]{Craig O. Heinke}
\affiliation{Department of Physics, University of Alberta, CCIS 4-181, Edmonton, AB T6G 2E1, Canada}

\author[0000-0001-6682-916X]{Gregory R. Sivakoff}
\affiliation{Department of Physics, University of Alberta, CCIS 4-181, Edmonton, AB T6G 2E1, Canada}

\author[0000-0003-0248-5470]{Anil C. Seth}
\affiliation{Department of Physics and Astronomy, University of Utah, Salt Lake City, UT, USA}

\author[0000-0002-4537-3367]{Eva Noyola}
\affiliation{McDonald Observatory, University of Texas at Austin, Austin, TX 78712, USA}



\begin{abstract}
We present the results of an ultra-deep, comprehensive radio continuum survey for the accretion signatures of intermediate-mass black holes in globular clusters. The sample, imaged with the Karl G.~Jansky Very Large Array and the Australia Telescope Compact Array,
comprises 50 Galactic globular clusters.  No compelling evidence for an intermediate-mass black hole is found in any cluster in our sample. In order to achieve the highest sensitivity to low-level emission, we also present the results of an overall stack of our sample, as well as various subsamples, also finding non-detections. These results strengthen the idea that intermediate-mass black holes with masses $\gtrsim 1000 M_{\odot}$ are rare or absent in globular clusters.

\end{abstract}




\section{Introduction} \label{sec:intro}

Intermediate-mass black holes (IMBHs) have been proposed as a population of black holes with masses between those of supermassive black holes that reside at the centers of galaxies (typically $\gtrsim 10^{6} M_{\odot}$) and the stellar-mass black holes created through the deaths of massive stars ($\lesssim 100 M_{\odot}$). IMBHs ($10^{2}-10^{5} M_{\odot}$; e.g., \citealt{2006noyola, 2011feng}) draw ongoing interest as 
promising seeds for the growth of supermassive black holes \citep{2005Volonteri}. Many possible origins for IMBHs have been suggested, including the collapse of metal-free Population III stars
\citep{2001madau}  or direct collapse from gas in low-mass halos \citep{1994loeb, 1995eisenstein, 2003bromm, 2006lodato}. Massive young star clusters have also been suggested as important IMBH formation sites, perhaps via runaway stellar mergers \citep{2002portegies, 2004gurkan, 2004portegies, 2009vanbe}. \citet{2010vespe} has suggested that a central BH can grow through accretion of gas lost in these star clusters. Another potential channel for forming IMBHs is via sequential mergers of stellar-mass BHs in globular clusters (GCs; \citealt{2002millerhamilton}). These possibilities make GCs primary targets for the search for IMBHs.

Efforts to investigate the presence of IMBHs in GCs, and subsequently constrain their masses, utilize two main approaches: dynamical or accretion signatures.
Stars residing near the cluster center, inside the putative sphere of influence of any IMBH, have been studied using kinematic measurements such as radial velocities and proper motions \citep{1976newell, 2005geb, 2006mcLaughlin, 2010noyola, 2010anderson, 2014kamann}. Compared to dynamical measurements of supermassive black holes in galaxy centers, these IMBH studies are much more challenging owing to the relatively low black hole masses: there are few stars within the IMBH sphere of influence, and, especially for low-mass IMBHs, the signature of a central point mass can be confused with a concentrated population of stellar remnants \citep{2014denbrok}. A more recent approach is to constrain the potential
using precise timing of one or more radio pulsars \citep{2017perera,2017kiziltan,2017freire}. Overall, despite claims of dynamical evidence for IMBHs in many individual Galactic GCs (e.g., \citealt{2009ibata, 2010noyola, 2011lutz, 2013feldmeier, 2013lutz}), there is no consensus about the presence of an IMBH in any particular object.

An alternative approach, again drawing a parallel to studies of supermassive black holes, is to search for accretion evidence for IMBHs. The source HLX-1, located in the halo of galaxy ESO 243-49 at distance of 95 Mpc, offers the best current such case. X-ray, optical, and radio data suggest the likely presence of a $\sim 10^4$--$10^5 M_{\odot}$ IMBH accreting at close to the Eddington rate \citep{2012webb}, possibly located at the center of a tidally stripped young nuclear star cluster \citep{2014farrell, 2015musaeva, 2017webb, 2017soria}. 

By contrast, any so-far undiscovered IMBHs in  Galactic clusters would necessarily be in quiescence, accreting at an extremely low rate, akin to Sgr A$^{\ast}$. They would therefore be best observed as radio continuum sources due to synchrotron emission from relativistic jets \citep{2004maccarone, 2005maccarone,2008maccarone}, though in a few nearby clusters with very deep X-ray data, these observations can offer comparable or better constraints on accretion than radio data (e.g., Haggard et al.~2013). 

The radio methodology has been applied to deep radio imaging of the Galactic GCs M15, M19, and M22 by \citet{2012strader} to set $3\sigma$ upper limits of $< 1000 M_{\odot}$ on the masses of IMBHs in these clusters, and extended by \cite{2015wrobel, 2016wrobel} to extragalactic GCs in M81 and NGC~1023 (where non-detections were also found, at higher mass limits than for Galactic clusters) using a stacking analysis.

In this paper, we use deep radio observations from the Karl G.\ Jansky Very Large Array (VLA) and the Australia Telescope Compact Array (ATCA) to search for accretion signatures of IMBHs in 50 Galactic GCs, using data obtained as part of the MAVERIC (Milky-way ATCA and VLA Exploration of Radio-sources in Clusters) survey.  Our paper is organized as follows: in Section \ref{data}, we present the sample of GCs and the radio observations. Section \ref{analysis} describes our methodology for determining IMBH mass constraints.  In Section \ref{res}, we present the results for each cluster and the stack of all 50 Milky Way GCs. In Section 5 we {discuss} and analyze these results. An Appendix considers in detail the radio data for NGC 6624, which has a number of bright known radio sources near its center and a recent IMBH claim \citep{2017perera}.

\section{Radio Data and Reduction}\label{data}

\subsection{Target selection}

The MAVERIC sample was chosen primarily by distance and mass, and was intended to be complete for GCs with masses $> 10^5 M_{\odot}$ and distances $< 9$ kpc, keeping in mind that many GCs near the bulge have uncertain distances. Massive clusters ($\gtrsim 5 \times 10^5$ M$_{\odot}$) at larger distances were added to search for IMBHs. Objects at declinations $\delta > -35^{\circ}$ were primarily observed with the VLA and more southerly sources with ATCA, though this division is not precise and a few clusters were observed with both arrays. 

Here, we use the most recent updated distance measurements to sample GCs; distances and references are listed in Tables \ref{tab:vla} and \ref{tab:atca}.

More details on the sample selection, observing setup, and source catalogs will follow in separate survey papers (Shishkovsky et al., in preparation; Tudor et al., in preparation). 

\subsection{VLA}
VLA observations took place during May--Aug 2011 (NRAO/VLA Program IDs 10C-109 and 11A-269),  Sep 2012 (12B-073) Feb--Jun 2014 (13B-014), and May--Aug 2015 (15A-100 and 15A-225). Data for 18 GCs were obtained in the most extended A configuration, with 7 more southerly clusters observed in the BnA configuration. The remainder of the GCs were observed during ``move time" in or out of A configuration.
With these extended configurations, we obtained angular resolutions $\lesssim 1\arcsec$ for nearly all VLA globular clusters, which facilitates the comparison between radio images and the optical centers. 
The median synthesized beam in the 5 GHz VLA image (and in the averaged 6 GHz images; see below) is $0.7\arcsec \times 0.4\arcsec$.

VLA observations were made with C band receivers (4--8 GHz). Data taken from 2011-2014 used 8-bit mode, with two independent 1024-MHz wide basebands centered at 5.0 GHz and 6.75 GHz. In 2015, we used 3-bit mode, with two independent 2048-MHz wide basebands centered at 5.0 GHz and 7.0 GHz. The amount of usable continuum for the 3-bit observations was less than double that in the 8-bit observations owing to significant radio frequency interference (RFI) in the band.  In both modes, the bandwidth was divided into 128-MHz wide spectral windows, and each spectral window was sampled by 64 channels. All observations were obtained in full polarization mode.

We were approved to obtain 10 hr of observations per VLA cluster, which nominally would result in 7--8 hr on source, depending on the length of the individual observing blocks. The median on-source observing time was consistent with this goal, at 7.4 hr. Not all the requested blocks were successfully observed, and three clusters had final on-source times less than 5 hr (M54, Liller 1, and NGC 6522), with correspondingly higher rms noise levels. Observations typically alternated between 10 min on source and short observations of a phase calibrator (selected to be within 10$^{\circ}$ degrees of the science target). A bandpass and flux calibrator was observed at the start or end of each block.

Standard procedures were followed for flagging, calibration and imaging of the data with the Common Astronomy Software Application package (CASA; \citealt{2007mcmullin}) and the Astronomical Image Processing System (AIPS; \citealt{2003greisen}). Imaging of the data was carried out using a Briggs weighting scheme (robust=1). To mitigate artifacts from large fractional bandwidth, the $uv$ data were divided into two frequency chunks (basebands) and imaged separately. In CASA, frequency dependent clean components (with two Taylor terms; \verb|nterms=2|) were also used in imaging to mitigate large-bandwidth effects \citep{2012rau}. For the GCs with bright sources that generated considerable artifacts, we applied self-calibration to optimize the image quality.

Of the 29 GCs with VLA data, 17 were taken primarily or only in 3-bit mode, which had 3--3.4 GHz of bandwidth remaining after RFI flagging. The median rms noise in each of the baseband images centered at 5.0 and 7.0 GHz was 2.0 $\mu$Jy bm$^{-1}$. For nine GCs observed earlier in the project, the data were taken in 8-bit mode, which typically results in 1.8 GHz bandwidth after flagging. Given the lower bandwidth, the baseband rms noise values were slightly higher, 2.5 $\mu$Jy bm$^{-1}$ (5.0 GHz) and 2.1 $\mu$Jy bm$^{-1}$ (6.75 GHz). The 3 GCs with comparable exposure times in 3-bit and 8-bit mode had noise levels intermediate between the 3-bit and 8-bit median values, as expected. 

Given the expected flat spectrum of the radio continuum emission from low-luminosity accreting IMBHs, we also combined the individual basebands into a single image to maximize sensitivity. To do this, the higher frequency image was convolved to the resolution of the lower frequency image using the AIPS tasks {\tt CONVL}. The high and low-frequency images were then averaged using the AIPS task {\tt COMB}, weighting by the variance. The median rms noise at 6.0 GHz is 1.5 $\mu$Jy bm$^{-1}$ (3-bit data) and 1.7 $\mu$Jy bm$^{-1}$ (8-bit data). The distribution of the observed image noise in the frequency averaged images is presented for all GCs in Figure \ref{rmsimage}.

\subsection{ATCA}
27 GCs were observed with ATCA (Project Code: C2877) in the extended 6A or 6D configurations in runs from 2013 to 2015. Archival data for 47 Tuc and $\omega$ Cen from \cite{2011lukong} are also included in our analysis, imaged together with our newer data for these clusters. All observations were carried out using the Compact Array Broadband Backend (CABB; \citealt{2011wilson}), allowing simultaneous observations in two bands centered at 5.5 and 9.0 GHz, each with 2 GHz of bandwidth. Observations were typically made over 2--3 separate blocks on different days. In each epoch, the target cluster and phase calibrator were alternately observed for 10--20 min and 1.5--2 min integration times respectively, depending on atmospheric stability and calibrator brightness.

\begin{figure}[t!]
\includegraphics[width=0.5\textwidth]{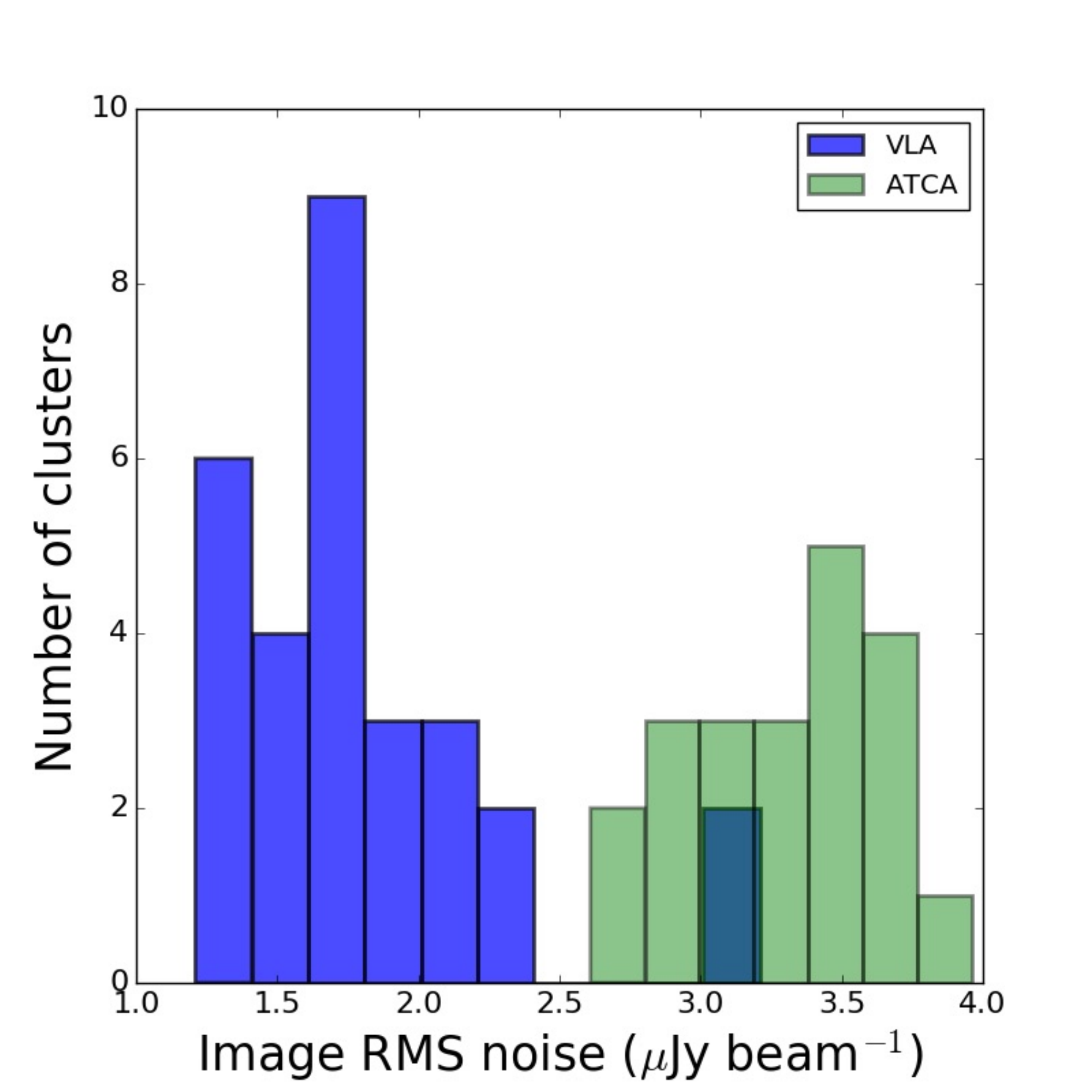}
\caption{{The distribution of the observed rms noise in our frequency-averaged images for 50 Galactic GCs. VLA data are blue and ATCA data green.}\label{rmsimage}}
\end{figure}

Flagging and calibration were performed in MIRIAD \citep{1995sault}, and imaging in CASA. As with the VLA, the two frequency bands were imaged separately, and with a Briggs robust value of 1. The median synthesized beam in the 5.5 GHz images (and in the frequency-averaged images) is $3.2\arcsec \times 1.6\arcsec$. The median on-source integration time was 17.0 hr, yielding median rms noise levels of 4.1 and 4.6 $\mu$Jy bm$^{-1}$ in the 5.5 and 9.0 GHz bands, respectively. As with the VLA images, we averaged images in the two frequency bands together, and the median rms was 3.3 $\mu$Jy bm$^{-1}$ in the resulting 7.25 GHz images. This is about twice the median rms of the VLA images, but still very sensitive, allowing tight constraints on the presence of central radio sources. The distribution of the observed rms of the ATCA images is shown in Figure \ref{rmsimage}. 

5 GCs in the sample were observed by both VLA and ATCA; for all but NGC 6522 (which had only 2.5 hr on-source with the VLA), the beam size is smaller and the rms is substantially lower in the VLA images, and hence we use the VLA data for subsequent analysis.

\subsection{Stacking Analysis} \label{stackanalysis}
We also stacked all the clusters to give the highest sensitivity to low-level emission under the assumption that IMBHs exist in the centers of globular clusters. For the stack, we only used clusters that do not have any central radio emission (from unresolved pulsars or bright X-ray binaries), in order to make a weighted-mean image of these fields. In total, we used 24 clusters from VLA and 14 from ATCA observations, which are noted in boldface in Tables \ref{tab:vla} and \ref{tab:atca}. {12 clusters that were excluded from the deep stack are discussed extensively in sections \ref{exclclu} and \ref{masstack}.}

To do this, the frequency-averaged images were convolved to a common resolution (2.2$\arcsec$ for VLA and 6.0$\arcsec$ for ATCA data) using the AIPS tasks {\tt CONVL}. Images were aligned so that the cluster centers align at the image center (see Tables \ref{tab:vla} and \ref{tab:atca} for GC centers). We then used the AIPS task {\tt STACK} to produce the weighted mean image (see \cite{2015lindroos} for details of the stacking technique).  Separate stacks were made for VLA and ATCA clusters (see subsection \ref{masstack}). 

\section{Data analysis}\label{analysis}

\subsection{Linking IMBH Mass and Radio Luminosity} \label{formalism}
The ``fundamental plane'' of BHs describes the relation between X-ray luminosity ($L_{X}$), radio luminosity {($L_{R}=4\, \pi\, d^{2}\, \nu\, S_{\nu} $)} and the mass of the BH ($M_{BH}$) for objects ranging from stellar-mass BHs to supermassive BHs \citep{2003merloni, 2004falcke, 2012plotkin}. It implies that accreting BHs of all masses show a correlation between X-ray and radio luminosity, but that more massive BHs have a higher ratio of $L_R/L_X$ at fixed $L_X$.

Since no IMBHs have been dynamically confirmed, their consistency with the fundamental plane has not been tested: their behavior is an interpolation between the extreme low and high BH masses. For this paper, we assume accreting IMBHs would behave in a manner consistent with the fundamental plane, and use our radio observations to constrain the masses of IMBHs at the centers of Galactic GCs.

We adopt the form of the fundamental plane with black hole mass as the dependent variable \citep{2012miller, 2012plotkin}:
\begin{eqnarray}\label{fp}
 {\rm log}\, M_{BH} & = & (1.638 \pm 0.070){\rm log}\, L_{R}\nonumber \\ 
 & &  - (1.136 \pm 0.077){\rm log}\, L_{X}\nonumber \\
  & & - (6.863 \pm 0.790)
\end{eqnarray}

where the black hole mass $M_{BH}$ is given in units of $M_{\odot}$, and $L_X$ and $L_R$ are given in erg~s$^{-1}$. This form of the fundamental plane is derived using BHs in the ``low/hard"
accretion state ($\lesssim$ a few percent of the Eddington rate), in which the X-ray emission is thought to originate in a corona or possibly near the base of the jet, while the radio emission
is partially self-absorbed synchrotron radiation from the jet. If IMBHs are present in GCs and accreting, they must be in the low state, or else they would be easily observed as bright X-ray sources. The respective luminosities in this plane are formally defined between 0.5--10 keV ($L_X$) and at 5 GHz ($L_R$). In all cases we assume flat radio spectra {consistent with that observed for low-luminosity AGNs, e.g: \citealt{2000nagar} (see also \cite{1979blandford,1988hjellming,2005gallo})} to translate the observed flux density ($L_{R}$) limits into 5 GHz luminosities.

Consistent with previous work, we assume that the putative IMBH accretes from ambient intracluster gas in a manner similar to low-luminosity central supermassive BHs in galaxies {\citep{2003maccarone,2005pellegrini, 2012strader,2013haggard}}. In globular clusters, gas should be present due to the high density of red giants and their associated winds. Direct evidence of this gas has been seen in 47~Tuc and M15, revealed by the radial distribution of the dispersion measure of pulsars, which traces the free electron density inside the GC \citep{2001freire}. These observations and the lack of \ion{H}{1} in GCs implies that the gas must be mostly ionized \citep{2006vanloon}; here we assume that intra-cluster gas is evenly distributed and it is fully ionized and isothermal at a temperature $T = 10^4$ K. We assume a gas number density of $n = 0.2$ cm$^{-3}$, corresponding to an electron density $n_{e} = 0.1$ cm$^{-3}$, consistent with the aforementioned pulsar data and expectations for freely-expanding red giant winds \citep{2001pfahl, pooley2006}.

The Bondi accretion rate for an isothermal gas onto a black hole is given by: $\dot{M}_{\rm Bondi} = e^{3/2}\, \pi\, G^2\, M^2\, \rho\, c_s^{-3}$, where $e^{3/2}$=$2(5-3\gamma)^{(5-3 \gamma)/2(\gamma-1)}$, from $\gamma=1$ for an isothermal gas, $G$ is the gravitational constant, $M$ is the BH mass, and $\rho = n \, \mu \, m_H$ is the gas density \citep{1952bondi}. The sound speed $c_s$ in an isothermal gas is given by $c_s = (k_B T/ \mu m_H)^{1/2}$ for mean molecular mass $\mu$. Here we take $\mu = 0.59$, appropriate for ionized gas with the usual composition of a GC {\citep{1985fall}}.  Substituting in typical values, the Bondi rate in units of g s$^{-1}$ is: 
\begin{equation}
\dot{M}_{\rm Bondi} = 1.18 \times 10^{17} \left({{M}\over{2000 M_{\odot}}}\right)^2  \left({{n}\over{0.2 \, \textrm{cm}^{-3}}}\right) \left({{T}\over{10^4 \, \textrm{K}}}\right)^{-3/2}
\end{equation}
To yield the Bondi rate in units of $M_{\odot}$ yr$^{-1}$, the pre-factor on the above equation is instead $1.88 \times 10^{-9}$.

The results of \citet{2005pellegrini} show that low-luminosity AGN are not as luminous
as would be expected if they were accreting at the Bondi rate in a radiatively efficient manner (see also recent theoretical work on sub-Bondi accretion by \citet{2017inayoshi}). The quantity directly constrained by X-ray measurements is not the accretion rate itself, but a combination of the radiative efficiency and the accretion rate. The X-ray luminosity is given by the standard equation $L_X = \epsilon \dot{M} c^2$, for radiative efficiency $\epsilon$ and speed of light $c$. As in our previous work \citep{2008maccarone,  2012strader}, we assume that the accretion flow is radiatively inefficient, as for example in an advection-dominated accretion flow \citep{1995narayan}, such that the radiative efficiency $\epsilon = \eta \, (\dot{M}/\dot{M}_{\rm edd}$) for some normalization $\eta$ and $\dot{M}_{\rm edd}$ the Eddington accretion rate.

If we parameterize the accretion rate $\dot{M}$ as a fraction $f_b$ of the Bondi rate (so that $\dot{M} = f_b \, \dot{M}_{\rm Bondi}$), then the observationally-determined quantity is $\eta \, f_b^2$. Using the $\dot{M}_{\rm Bondi}$ and $L_X$ measurements of 
\citet{2005pellegrini} for low-luminosity AGN, and adjusting for their use of $\gamma = 4/3$ rather than our isothermal $\gamma = 1$, we find that the quantity $\eta \, f_b^2$ has a median of about 0.005, but an extremely large spread of about 1.5 dex in the log. Considering its large uncertainty, this value is equivalent to the $\eta \, f_b^2 = 0.0045$ implicitly used in \citet{2012strader}, and for consistency we use this latter value. We discuss the uncertainty in this assumption and its implications for the mass limits below.

Measuring the upper limits on the radio luminosity, $L_R$, for a GC IMBH, we use the fundamental plane assuming the BH follows the $L_R$-$L_X$ relationship and the accretion efficiency in order to determine the predicted $L_X$ and subsequently to constrain the $M_{BH}$. We emphasize that our methodology uses predicted $L_X$ rather than observed $L_X$ in the context of the fundamental plane to obtain IMBH mass constraints. This is because most GCs in our sample do not have published X-ray limits on a central IMBH. For one cluster ($\omega$ Cen) the published X-ray limit is more stringent than the limit inferred from our formalism (see \S \ref{omegacen}). Given the scatter in the fundamental plane, additional observed X-ray limits would also provide valuable constraints on the presence of IMBHs in our GC sample, and our team is currently in the process of conducting a uniform X-ray analysis of the sample.

For the rest of this paper, we use $3\sigma$ upper limits as a constraint on the radio luminosity to constrain the IMBH mass. The plane between $M_{BH}$ (in $M_{\odot}$), distance $d$ (in kpc), and observed 5 GHz flux density or upper limit $S$ (in $\mu$Jy) implied by the above formalism is: log $M_{BH}$ = 0.743 log $d$ + 0.372 log $S$ + 2.134. 

\vspace{-3mm}
\subsubsection{Uncertainties in this Formalism}

As discussed in \citet{2012strader}, the main uncertainty in the mass limits predicted by this formalism is the combination of the radiative efficiency and the actual accretion rate, here parameterized by $\eta \, f_b^2$. The scatter in the fundamental plane itself is nearly negligible. The observational scatter of about 1.5 dex in $\eta \, f_b^2$ translates into a 0.39 dex scatter in log $M_{BH}$. Another source of uncertainty is the gas density $\rho$, but this is much harder to quantify, given the scarcity of observational data. An uncertainty of 0.3 dex in log $\rho$, suggested by the few Galactic globular cluster data points as well as the theoretical considerations discussed above, barely increases the overall uncertainty in log $M_{BH}$ (to 0.43 dex). But we emphasize that the distribution of $\rho$ is not well-constrained.

A systematic uncertainty is our assumption of an isothermal gas, rather than an adiabatic gas or one with an intermediate index. The isothermal assumption is conservative: it produces higher limits on $M_{BH}$ than other choices. If we instead used the opposite extreme, an adiabatic gas with $\gamma = 5/3$, log $M_{BH}$ would be lower by 0.52 dex.
This is of the same order as the variation due to $\eta \, f_b^2$, though the quantities are not necessarily independent.

The above discussion illustrates that the exact predictions depend on the assumptions made, and that these limits should not be compared to other radiative limits on X-ray or radio emission without an appropriate consideration of the assumptions.

\subsection{IMBH location} \label{sec:location}

We adopt the photometric center of each GC as the best current estimate of its center of mass. These are listed in Tables \ref{tab:vla} and \ref{tab:atca}. The photometric centers come primarily, though not exclusively, from fitting models to \emph{Hubble Space Telescope (HST)} star count observations. For example, \cite{2010goldsbury} determined the photometric centers of 65 Milky Way globular clusters by performing an analysis of star counts in \emph{HST}/ACS survey images, while \cite{2013miocchi} supplemented \emph{HST} observations with ground-based data. 

Dynamical friction leads an IMBH to spiral to the center of mass of its host cluster on a short timescale \citep{2007matsubayashi}. From this location, encounters with stars or stellar remnants can perturb the IMBH, especially if the BH mass is relatively low. Using the principle of Brownian motion, \citet{2002chatterjee} analyze the dynamics in the context of a BH at the center of a dense stellar cluster. Assuming a Plummer potential for the stellar system surrounding the central BH, the predicted variance of mean-squared one-dimensional deviations will be:
{$<x^{2}>=(2/9)(M_{*}/M_{BH}\, r_{c}^{2})$}, where $M_{*}$ is the average mass of a star in the cluster core ($\sim 1 M_{\odot}$) and $r_{c}$ is the cluster core radius \citep[see also][]{2012strader}. {The core of a GC can be depleted of lower-mass stars due to mass segregation; therefore, we adopt a conservative value for the mass of the observable stars in the core of GCs \citep{2002fregeau}.} The actual motion of an IMBH would depend, of course, on the detailed mass profile of the inner regions, but this basic estimate gives a guide to how much the IMBH might wander as a function of cluster parameters. To calculate the Brownian radii for our GCs, we use the $3\sigma$ mass limits as given in Tables \ref{tab:vla} and \ref{tab:atca}. For GCs with a cluster centroid uncertainty larger than the Brownian radii, we use the former for our analysis. 

We note that when the current paper was close to submission, 
\cite{2018vita} published a more sophisticated model, based on N-body simulations, for predicting the random motion of IMBHs in GC centers. Nonetheless, their results are generally consistent with simple the Brownian motion model we use.

\subsection{Notes on clusters with radio sources near their centers}\label{exclclu}

Our basic result is that we find no compelling evidence for accreting IMBHs in any of the 50 GCs in our sample. However, some clusters \emph{do} have radio sources near their photometric
centers, which are associated with X-ray binaries or pulsars. Here we discuss these sources and why we favor non-IMBH explanations for the emission in all cases. Furthermore, previous studies regarding the IMBH existence in individual clusters are discussed extensively in section \ref{liter}.

\subsubsection{Liller 1}

This massive, dense cluster shows several steep spectrum sources near the center (but outside the Brownian radius of a putative IMBH). Given the high interaction rate inferred
for Liller 1 \citep{2015saracino} and the strong steep-spectrum emission as identified by \cite{2000fruchter}, these sources are likely to be pulsars. Their presence does not affect the IMBH limits for the cluster, but we exclude it from the VLA stack.

\subsubsection{M15}
Our non-detection of an IMBH in M15 has previously been published in \citet{2012strader}. As discussed there, there are two previously-known radio bright X-ray binaries and a pulsar near the cluster center (within $4\arcsec$; \citealt{1989Wolszczan, 1991Johnston, 2011Miller-Jones}). However, there is no radio emission at the center of M15 within the Brownian radius expected for an IMBH. See \citet{2012strader} for more discussion.

\subsubsection{M62}

M62 has six pulsars known in its core \citep{2001damico, 2003possenti, 2012lynch}, two within $2.5\arcsec$ of the cluster center. One of these, J1701$-$3006F, is a $4.5\sigma$ detection in our 4.7 GHz image, but is not detected at 7.4 GHz. There is no significant detection at either frequency within the Brownian radius of an IMBH.

\subsubsection{NGC 6388}
NGC 6388 was previously observed with ATCA to search for an IMBH, and no central radio emission was detected to a level $< 42\ \mu$Jy at 5.5 GHz \citep{2011bozzo}. 

In our ATCA 5.5 GHz image, there is a source located near the cluster center with flux density $20.2 \pm 3.6\ \mu$Jy and a J2000 location of (R.A., Dec.) = (17:36:17.276, $-44$:44:09.03), just $1.1\arcsec$ from the cluster photometric center (Figure \ref{im:atca}). The source is not significantly detected in the 9.0 GHz image, with a flux density at its location of $4.6 \pm 3.6\ \mu$Jy. If we conservatively assume a uniform prior on $\alpha$ between --3 and 3 and use the 3$\sigma$ upper limit of the 9.0 GHz flux density ($< 10.8 \mu$Jy), then the most likely value of $\alpha$ is --2.1, with a $3\sigma$ upper limit to the spectral index of --0.2. Therefore, the most probable interpretation is that this central source is a pulsar in the cluster, but a flat-spectrum source cannot be definitively ruled out.

The presence of a pulsar near the cluster center would be far from surprising, as a large pulsar population is expected in NGC 6388 based on the \emph{Fermi}/LAT detection of GeV $\gamma$-rays \citep{2010abdo}, its high stellar encounter rate \citep{2013bah}, and its large X-ray source population \citep{2011bozzo,2012maxwell}. In addition, the centroid uncertainty for a IMBH of mass $> 1000$ M$_{\odot}$ in NGC 6388 is $\lesssim 0.3\arcsec$. We therefore conclude that the central point source is not an IMBH, but could be a pulsar. We proceed by using the 9.0 GHz image to constrain the mass of an IMBH; the limit listed in Table \ref{tab:atca} corresponds to the sensitivity of the 9.0 GHz image alone.

\subsubsection{NGC 6624}
The center of the cluster NGC 6624 has two neutron star sources that emit in the radio: the bright pulsar PSR 1820-30A and the ultracompact low-mass X-ray binary 4U 1820-30 (Figure \ref{n6624}). Radio timing observations of the pulsar and X-ray timing of the X-ray binary have been used in a number of recent papers to argue for the presence of an IMBH \citep{2014peuten,2017perera}, though \citet{2018gieles} argues that standard dynamical models of this globular cluster can explain these observations without a massive black hole.

Consistent with previous work, we observe a bright radio source near the center of NGC 6624 (Figure \ref{im:atca}). Using these ATCA data and new $HST$ observations, we find that all of the radio emission observed is consistent with being from the X-ray binary 4U 1820-30. We also note that this is in agreement with \citet{2004migliari} finding that 4U 1820-30 dominates above 2 GHz, and the pulsar below that. As the level of detail necessary to reach this conclusion is out of the main line of our paper, we place most of it in an Appendix, and here focus only on the radio emission from an IMBH.

To search for residual emission from a possible IMBH, we subtracted the X-ray binary 4U 1820-30 from each of the 5.5 and 9.0 GHz images, assuming it was a point source. We fit a Gaussian component in the image - plane using the AIPS task {\tt JMFIT} and then we subtracted it, without any assumption on the spectral index value. No residuals are apparent in either image after the subtraction, suggesting this assumption is reasonable. We re-measured the rms noise in a 24\arcsec\ box around the cluster center after the subtraction, using this value for the IMBH limits. We find an rms noise of 4.2 $\mu$Jy bm$^{-1}$ at 5.5 GHz, 3.9 $\mu$Jy bm$^{-1}$ at 9.0 GHz, and 3.3 $\mu$Jy bm$^{-1}$ in the averaged image at 7.25 GHz. We use this latter limit to determine the IMBH mass limit using our standard formalism.

\subsubsection{NGC 6652}
NGC 6652 has a high central density, suggesting the efficient production of dynamically-formed compact object binaries \citep{2012stacey} and several relatively bright X-ray sources have been detected by Chandra in this cluster. Only one radio pulsar has yet been detected in this cluster \citep{2015decesar}, but NGC 6652 is detected with \emph{Fermi}/LAT at GeV energies, suggesting a substantial total population of millisecond pulsars \citep{2010abdo}.

We detect a steep spectrum source in the ATCA image (present at 5.5 GHz and absent at 9.0 GHz) that is offset by about $1.5\arcsec$  from the cluster center, so we can conclude that it is not associated with an IMBH. Just as for the central source in NGC 6388, the inferred steep spectrum implies that this source is likely a pulsar.  This radio source is also $1.5\arcsec$  from the nearest X-ray source in the cluster (source D of \citealt{2012stacey}), ruling out an association. To place limits on the presence of an IMBH in NGC 6652, we only use the 9.0 GHz image, which does not show emission from this source.

\subsubsection{Terzan 1}
There is a radio source near the center of Terzan~1, with position J2000 (R.A., Dec.) = (17:35:47.204, $-30$:28:54.89). This source may be associated with the quiescent X-ray binary CX2 \citep{2006Cackett}.  However, we measure flux densities of $90.5 \pm 4.1$ $\mu$Jy at 5.5 GHz and $34.5 \pm 4.1$ $\mu$Jy at 9.0 GHz, implying a steep spectral index of $\alpha = -2.0\pm0.3$, more consistent with a pulsar than an X-ray binary. We will revisit this source in future work; in any case, the steep spectrum is inconsistent with the expectations for low-level accretion onto an IMBH.

To search for any residual emission that might be present from an IMBH, we subtracted this source from the 5.5 and 9.0 GHz images under the assumption that it is a point source, applying the same technique as in NGC 6624. No residuals are apparent in the 5.5 GHz image. In the 9.0 GHz image, a second source is visible at the $\sim 3\sigma$ level (in fact, this source is approximately one beam away from the brighter source in the original 9.0 GHz image, and is clearly detectable there as well). This source is $3\arcsec$ from the cluster center, far outside the Brownian radius of a $> 1000 M_{\odot}$ IMBH, and hence cannot be attributed to an IMBH.

We re-estimate the rms noise from these residual images in a region 24$^{\prime\prime}$ wide centered on the cluster center. We find rms sensitivities of $4.8\ \mu$Jy beam$^{-1}$ at 5.5 GHz, $4.1\ \mu$Jy beam$^{-1}$ at 9.0 GHz, and $3.9\ \mu$Jy beam$^{-1}$ in the combined frequency image, and use these revised values for our analysis.

\subsubsection{Terzan 5}
Terzan~5 hosts a large population of pulsars, with 38 known in the cluster core (\citealt{2005ransom, 2017prager, 2018cadelano}, P.\ Freire's ``Pulsars in Globular Clusters'' page\footnote{http://www.naic.edu/$\sim$pfreire/GCpsr.html}), and our images show many point sources. However, assuming a 1$^{\prime\prime}$ uncertainty on the position of the cluster center and an IMBH Brownian radius of 0.12$^{\prime\prime}$, we find no VLA radio sources located right at the cluster center. ATCA images are also consistent with no flux at the cluster center, but are less constraining because of the lower image resolution and poorer sensitivity.

\subsubsection{Terzan 6}
Terzan~6 hosts the known transient neutron star X-ray binary GRS 1747-312 near its center (at a projected distance of $0.8\arcsec$; \citealt{1991predehl, 1994pavlinsky, 2003zand}).
 This source was in outburst and radio bright during our observation (its X-ray/radio correlation will be reported elsewhere), measured at $21.4\pm 5.1\ \mu$Jy in our ATCA 5.5 GHz image and $17.3\pm5.3\ \mu$Jy in the 9.0 GHz image. We subtracted this point source and there is no residual emission near the cluster center. We remeasured the rms sensitivities of these residual images in a 24$^{\prime\prime}$ region, and found $5.3\ \mu$Jy beam$^{-1}$ at 5.5 GHz, $5.3\ \mu$Jy beam$^{-1}$ at 9.0 GHz, and $4.2\ \mu$Jy beam$^{-1}$ in the frequency-averaged image, which we use for our IMBH limits.

\section{Results}\label{res}

\subsection{IMBH mass limits per cluster}
We followed the method described in Section \ref{sec:location} to search for IMBHs in cluster centers. The clusters discussed in \S \ref{exclclu} were considered individually, as detailed above, due to confusing sources near their centers. 

For the rest of the targets, no radio emission was detected above 3$\sigma$ that matches the photometric center within the cluster centroid uncertainty or Brownian radius. Figures \ref{im:vla} and \ref{im:atca} show the radio continuum images of these clusters, zoomed in on their centers.

\begin{figure*}[t]
\includegraphics[width=1.0\textwidth]{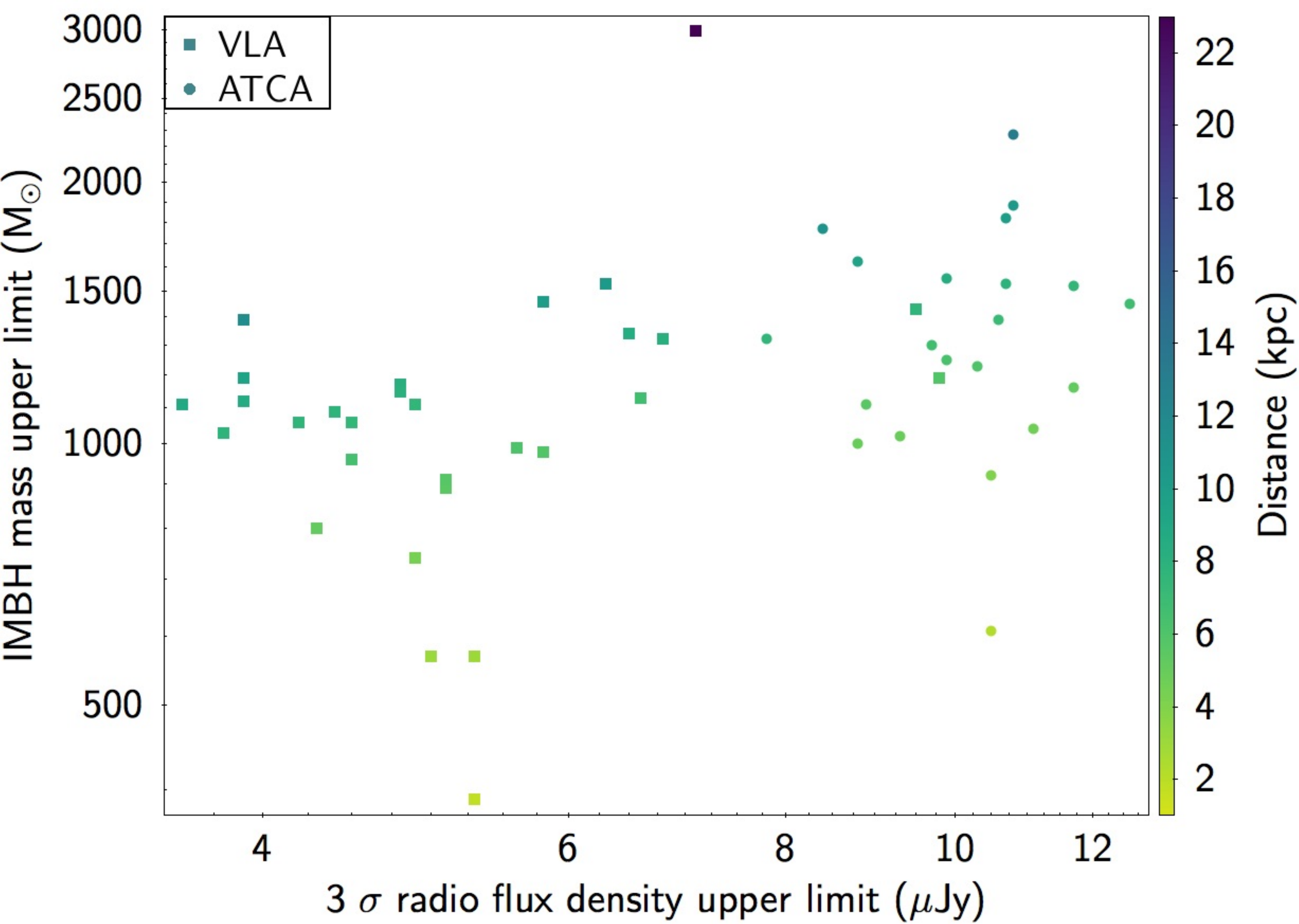}
\caption{{IMBH mass upper limits as a function of radio flux density upper limits for VLA (squares) and ATCA (circles). The colors of symbols denote distance, as shown in the color-bar at right. Both axes are plotted in logarithmic scales. M4 shows the lowest IMBH mass upper limit while the next-lowest IMBH upper limits from the VLA are for NGC 6544 and M22. From ATCA data, NGC 6397 has the lowest IMBH mass upper limit.}
\label{uplim}}
\end{figure*}

The 3$\sigma$ flux density upper limit was translated to an upper limit on luminosity assuming distances in Tables \ref{tab:vla} and \ref{tab:atca} and flat radio spectra. The median radio luminosity $3\sigma$ upper limit is $L_{R} \lesssim 1.9\times10^{27}$ erg s$^{-1}$. We then use the formalism described in \S \ref{formalism} to convert this luminosity upper limit to an IMBH mass upper limit. In Figure \ref{uplim}, we plot the IMBH mass upper limits as a function of radio flux density, with cluster distance included as the color scale. The most distant cluster is M54 at $\sim 24$ kpc, which was excluded from the cluster stack owing to its much larger distance (\S \ref{masstack}).

Our median IMBH mass limits are $<1110 M_{\odot}$ (VLA clusters) and $<1320 M_{\odot}$ (ATCA clusters). Limits on individual clusters are listed in Tables \ref{tab:vla} and \ref{tab:atca}, with the extreme limits of $< 390 M_{\odot}$ and $< 2990 M_{\odot}$ for the nearest and most distant clusters with VLA data: M4 and M54, respectively.

In these Tables we have also listed the predicted $3\sigma$ X-ray luminosity limits corresponding to the IMBH mass limits and $3\sigma$ radio limits in the context of our formalism. The median predicted X-ray limits for the VLA and ATCA samples are $< 3.4\times 10^{30}$ and $< 5.8\times 10^{30}$ erg s$^{-1}$, respectively. For most clusters the radio limits on the presence of an IMBH are deeper than published X-ray limits, with the exception of $\omega$ Cen, which we discuss below.

These Tables also list the 3$\sigma$ IMBH mass limits expressed as a percentage of the total cluster mass. The GCs masses are mostly from the recent work of \citet{2018baumgardt}, excepting four GCs not in that paper, for which we use the $M_V$ listed in \cite{1996harris} and assume $M/L_V = 2$. For the VLA sample, the median 3$\sigma$ IMBH mass limit is $< 0.36$\% of the cluster mass; for the ATCA sample, it is $<0.52$\%. We discuss these limits in the context of theoretical predictions in \S \ref{sec:disconcl}.

\subsection{Deep Limits from Image Stacking}\label{masstack}

The analysis above reflects mass limits on accreting IMBHs in individual GCs. If we instead assume that the IMBH occupation fraction is high, then we can set deeper limits on the presence of such IMBHs by stacking the cluster images.

\begin{figure}[t!]
\epsscale{1.0}
\plotone{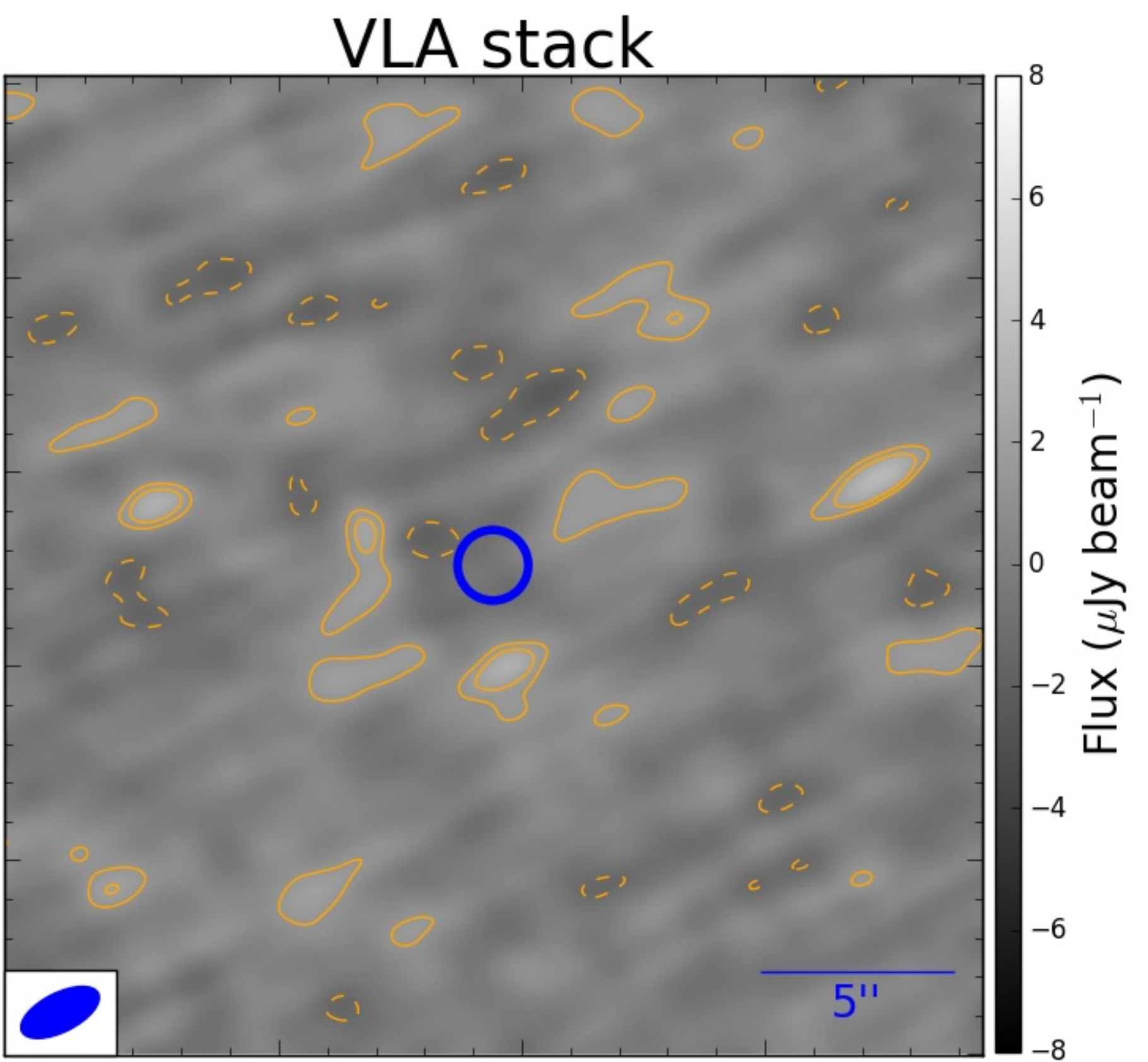}
\caption{{Weighted-mean stack of 24 clusters observed with VLA showing the central 25.2 $\arcsec$ $\times$ 25.2 $\arcsec$ area. The stacked image has an rms noise of 0.65 $\mu$Jy beam$^{-1}$. The contours represent flux densities of --$2\sigma$ (dotted), $2\sigma$, and $3\sigma$. The blue circle shows the mean IMBH wander radii (0.91 $\arcsec$), and the radio beam is shown in the bottom left corner.}\label{vla_st}}
\end{figure}

\begin{figure}[t!]
\epsscale{1.0}
\plotone{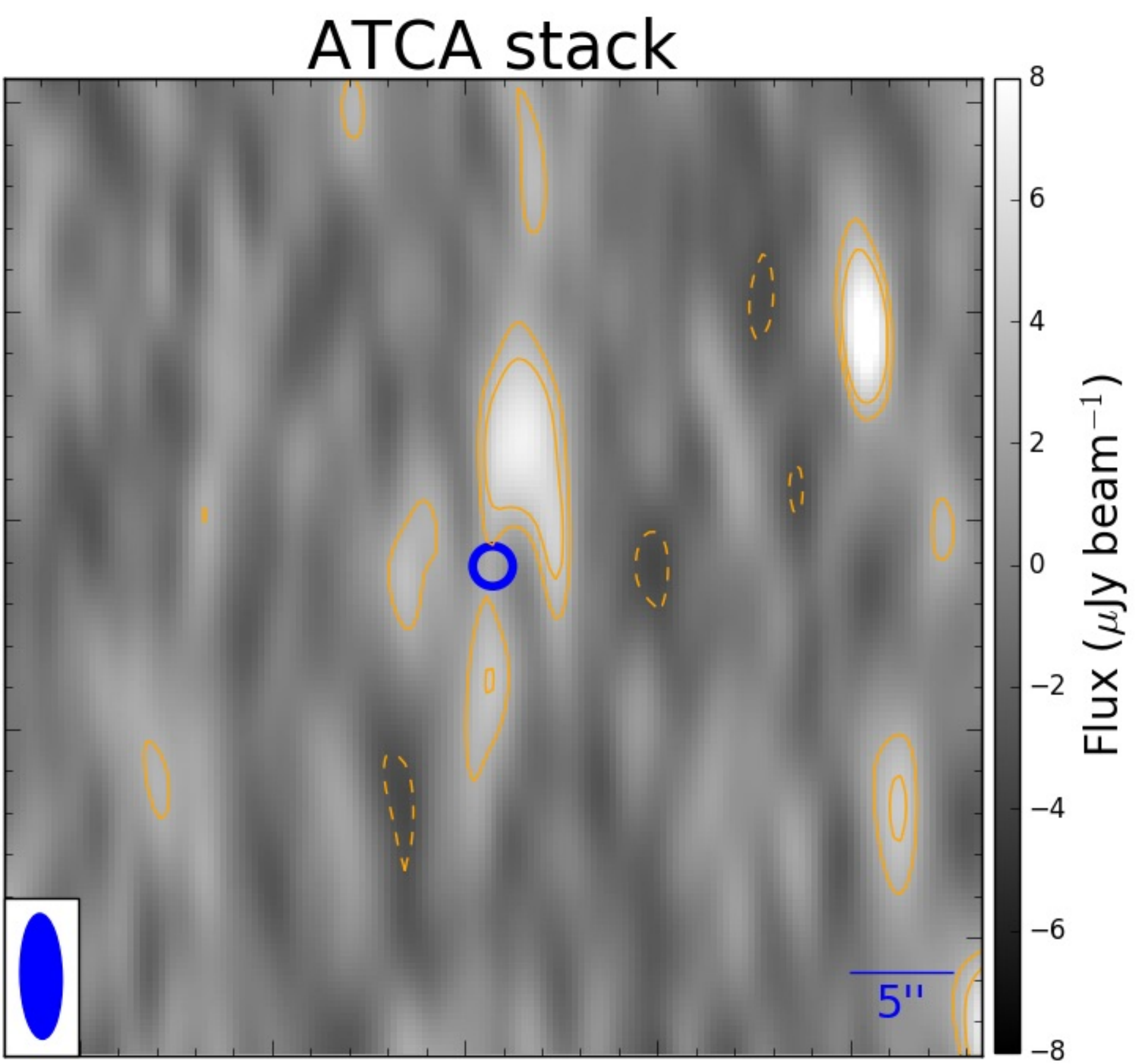}
\caption{{ Weighted-mean stack of 14 clusters observed with ATCA showing the central 47.8$\arcsec$ $\times$ 47.8$\arcsec$ area. The stacked image has an rms noise of 1.42 $\mu$Jy beam$^{-1}$. The contours represent flux densities of --$2\sigma$ (dotted), $2\sigma$, and $3\sigma$. The blue circle shows the mean IMBH wander radii (0.96 $\arcsec$), and the synthesized beam is shown in the bottom left corner.}\label{atca_st}}
\end{figure}

Section \ref{stackanalysis} describes our technique for co-adding the cluster images, resulting in the deep images presented in Figures \ref{atca_st} and \ref{vla_st}. The distant globular cluster M54 and GCs with stellar radio sources coinciding with their photometric centers (\S \ref{exclclu}) were excluded from these stacks. We also excluded NGC~6139, because there is a 3.2$\sigma$ flux peak $0.6\arcsec$ from the cluster center at 9.0 GHz. There is no emission at this location at 5.5 GHz and the 9.0 GHz emission appears to be an artifact associated with a bright source $3.4\arcmin$ from the center. Djorg 2 does not have a central source, but also shows artifacts associated with a bright source $2.5\arcmin$ from the center, and is likewise excluded. Finally, NGC 4372 does not contain a significant central source, but shows diffuse ``fuzz'' in the frequency-averaged image. NGC 4372 does not have a particularly large interaction rate \citep{2013bah}, so it is not clear that a population of pulsars is expected; we defer a detailed comparison of interaction rates to radio source populations to a future work, and for now exclude NGC 4372 from the ATCA stack.

For the VLA stack, we included 24 clusters, which have a median (mean) distance of 7.7 (6.9) kpc. The rms sensitivity of the co-added image is 0.65 $\mu$Jy beam$^{-1}$, and there is no significant source detected at the center of the stacked image. Using our formalism, this corresponds to a 3$\sigma$ VLA stack limit of $< 800 M_{\odot}$ ($< 730 M_{\odot}$). For this limit, the implied $L_X/L_{\rm edd} \sim 10^{-11}$.

For the ATCA stacked image, we averaged 14 clusters, with a median (mean) distance of 6.8 (6.5) kpc. The image rms sensitivity is 1.42 $\mu$Jy beam$^{-1}$, and no central source is
detected. The corresponding $3\sigma$ ATCA stack limit is $< 970 M_{\odot}$.

{Since many authors have argued that the densest (``core collapse'') clusters are unlikely to contain IMBHs (e.g., \citealt{2005baumgardt,2006trenti,2011noyola}), we also created VLA and ATCA stacks
excluding those GCs typically identified as core-collapsed. Many of these were already excluded for individual reasons as listed above. As expected, these new stacked images had slightly higher rms values than the full VLA and ATCA stacks (about 1.0 and 1.5 $\mu$Jy beam$^{-1}$, respectively), and also do not show significant central sources. Hence the exclusion of these GCs does not affect any of our conclusions.}

\section{Discussion and Conclusions}

Our main result is that there is no accretion evidence for IMBHs with masses $\gtrsim 1000 M_{\odot}$ in any Milky Way globular cluster. We first discuss general uncertainties in our analysis and then specific cases of GCs where IMBHs have been claimed in previous work.

\subsection{Uncertainties on Mass Limits}
The logical basis for our constraints can be divided into three parts: if an IMBH is present, (a) is gas also present; (b) is this gas accreted by the IMBH; (c) does this accreting gas produce the assumed radio signature?

That \emph{some} gas is present in the core of globular clusters is certain---the winds of red giants supply an ongoing flux of material, and ionized gas has been observed in 47 Tuc, with some evidence in M15 \citep{2001freire}. As 47 Tuc is rich in millisecond pulsars, this obviates the suggestion that such energy sources will reduce the plasma density to negligible amounts \citep{1991sp}, though many mechanisms might well be responsible for clearing out much of the gas lost from stars (see, e.g., the discussion in \citealt{2013nai}). Additional measurements of the ionized gas density in clusters and more sophisticated models of the intracluster medium are desirable.

Since no $\sim 1000 M_{\odot}$ IMBHs are known, all discussions of their properties necessarily involve indirect inferences. ``Low-mass'' central black holes with $\sim 10^5$--$10^6 M_{\odot}$ are present and evidently accreting gas in the nuclear star clusters of nearby galaxies \citep{2017nguyen, 2017nyland}, with even lower mass black holes detected at higher accretion rates in more distant galaxies \citep{2015baldassare}. At least in terms of their mass, these systems probably represent the nearest analogues to IMBHs in GCs, and nothing prevents accretion onto the central black hole. We have also assumed that the radiative efficiency and accretion rate, parameterized in terms of the 
Bondi rate, have typical values comparable to nearby low-luminosity accreting central black holes. These observed systems have a large dispersion in the accretion rate and/or efficiency. Some may not have appreciable accretion at all, though others could be accreting at much higher rates or with higher efficiency than assumed. If the dispersion in these quantities is high, then the non-detection of IMBHs in 50 GCs strongly suggests that IMBHs are rare, if they exist at all. We emphasize that the actual accretion rates and radiative efficiencies of IMBHs are the chief uncertainty in this analysis.

Depending on the assumptions for radiative efficiency, our typical VLA limits correspond to very low accretion rates of a few $\times 10^{-11} M_{\odot}$ yr $^{-1}$, which is only a few percent of the mass loss rate of one red giant \citep{2009dupree}. We note that \citet{2016macleod} argue that some fraction of GC IMBHs should have non-degenerate companions that could supply a higher rate of gas to the IMBH than accretion from ambient material. 

The use of the fundamental plane to convert radio limits to masses is an interpolation between stellar-mass and supermassive black holes, rather than an extrapolation. 
This fact is heartening, but the accretion behavior of hypothetical IMBHs is still unknown. \cite{2015cseh} found that HLX-1 was more radio-bright than predicted by the fundamental plane, assuming that the mass inferred from X-ray spectral fitting was correct.
The large (0.4 dex; \citealt{2012plotkin}) scatter in the fundamental plane also limits precise statements about accretion implications for specific systems. 

\subsection{Revisiting IMBHs from the Literature}\label{liter}

We begin with $\omega$ Cen and M54, as these massive GCs are often argued to be stripped galaxy nuclei (the case for M54 is especially strong; \citealt{2000layden}) and hence might be the most likely to host IMBHs.

\subsubsection{$\omega$ Cen} \label{omegacen}

$\omega$ Cen has many contrasting claims of dynamical evidence for an IMBH \citep{2008noyola,2010noyola,2010van,2017baumgardt}, with some papers finding dynamical evidence for an IMBH with mass $\gtrsim 10^{4} M_{\odot}$. We do not revisit this work here, but solely focus on the accretion constraints. Our $3\sigma$ ATCA upper limit of $< 8.9$ $\mu$Jy on a central radio source implies a $3\sigma$ IMBH mass upper limit of $< 1000 M_{\odot}$ using our formalism. The corresponding 0.5--10 keV X-ray luminosity limit is $< 2.5 \times 10^{30}$ erg s$^{-1}$ {(Table \ref{tab:atca})}, which can be compared to the observed (95\%) upper limit of $< 1.7 \times 10^{30}$ erg s$^{-1}$ \citep{2013haggard}. Because of the large amount of \emph{Chandra} data on this cluster, this is one case where the X-ray limit is as (or more) constraining than the radio limit. As discussed in \cite{2013haggard} and \cite{2012strader}, if an IMBH of mass $\gtrsim 10^4 M_{\odot}$ is present in $\omega$ Cen, then it must be accreting at a relative rate below any other central black hole known in the universe, with $L_{\rm bol}/L_{\rm edd} \lesssim 2 \times 10^{-11}$. There is no accretion evidence for an IMBH in $\omega$ Cen.

\subsubsection{M54}

In M54, the $3\sigma$ radio upper limit of $< 7.2$ $\mu$Jy gives a mass limit $< 3000 M_{\odot}$, far below the $9400 M_{\odot}$ suggested dynamically \citep{2009ibata}. The new radio limit is about a factor of 7 stronger than the one presented in \cite{2011wrobel}, due entirely to the improved sensitivity of the post-upgrade VLA. 
M54 is the nucleus of the Sagittarius dwarf galaxy (e.g. \citealt{2005monaco}) with a V-band stellar luminosity of ~$10^{8} L{_\odot}$ and a halo mass $> 6\times 10^{10} M{_\odot}$ \citep{2017gibbons}.  While the occupation fraction of BHs is known to be high at the centers of slightly higher mass galaxies \citep{2017nguyen}, little is known about the BH occupation fraction at these low masses.
Owing to its identity as the  closest confirmed galaxy nucleus beyond Sgr A$^{*}$, in our view M54 presents a strong case for even deeper radio observations.

\subsubsection{New Pulsar Evidence: 47 Tuc and NGC 6624}

For many years, 47 Tuc was a rare case where most papers agreed that there was no substantial evidence for an IMBH; most prominently, \cite{2006mcLaughlin} set a $1\sigma$ dynamical upper limit of $< 1000$--1500 M$_{\odot}$ on an IMBH. This was challenged by \cite{2017kiziltan}, who used timing of millisecond pulsars in the core of 47 Tuc to argue for the presence of a $2300 M_{\odot}$ IMBH. \cite{2017freire} have disputed this interpretation of the observations (partially on the basis of the assumed cluster distance) and argue that no IMBH is necessary to explain the pulsar timing data. Our new ATCA $3\sigma$ radio upper limit of $< 11.2$ $\mu$Jy corresponds to a mass limit of $< 1040 M_{\odot}$, suggesting that an IMBH of the mass published by \cite{2017kiziltan} is not present, or that it is accreting at a rate or efficiency lower than assumed in our formalism. {47 Tuc is another cluster, like $\omega$ Cen, where there are extremely deep \emph{Chandra} data \citep{2001grindlay}, which limit a central X-ray source to $< 10^{31}$ erg s$^{-1}$ (0.5--10 keV). The corresponding 
3$\sigma$ limit from our formalism is $L_{X}< 2.8\times 10^{30}$ erg s$^{-1}$, and a deeper \emph{Chandra} constraint should be possible in the future through the analysis of archival data.}

A unique recent claim of dynamical evidence for an IMBH in the GC NGC 6624 comes from \citet{2017perera}, who argue, on the basis of precise, long-term timing, that the pulsar PSR 1820-30A is in a wide, eccentric orbit around an IMBH. The observational interpretation of this cluster is complicated and we discuss it in detail in the Appendix. Here we only mention our ATCA radio limit on a central IMBH: a $3\sigma$ value of $< 9.8$ $\mu$Jy, giving a $3\sigma$ IMBH mass of $< 1550$ M$_{\odot}$, compared to an dynamical IMBH mass of 7500--10000 M$_{\odot}$ in \citet{2017perera}. With our formalism, the radio flux density of a 7500
M$_{\odot}$ IMBH would be predicted to be about 700 $\mu$Jy, which is about a factor of 70 larger than our ATCA limit.

\subsubsection{Other Clusters}

NGC 6388 is a case similar to $\omega$ Cen where there is disagreement in the literature about dynamical evidence for the presence of a $(2-3)\times 10^{4} M_{\odot}$ IMBH \citep{2011lutz,2013lanzoni,2015lutz}. Previous ATCA observations gave $3\sigma$ upper limits of $< 81$ $\mu$Jy \cite[at 9 GHz;][] {2010cseh} and $< 42$ $\mu$Jy \cite[at 5.5 GHz;][] {2011bozzo}. As discussed above, we do observe emission near (but not coincident with) the center of the cluster at 5.5 GHz, which we attribute to unresolved pulsars. Such emission is not unexpected given the strong \emph{Fermi} GeV flux from the cluster. At 9.0 GHz, no emission is detected, with a $3\sigma$ upper limit of $< 8.5$ $\mu$Jy, about a factor of 5 lower than the \cite{2011bozzo} limit. This corresponds to a $3\sigma$ IMBH mass limit of $< 1770 M_{\odot}$, about a factor of 16 lower than the dynamical mass inferred from \citet{2015lutz}.

M15 is a cluster where there were early dynamical hints for an IMBH \citep{1970newell,2002gers,2003gers} for which the interpretation was immediately disputed \citep{1977illingworth, 2003baumgardt}, and subsequent works have generally agreed that no IMBH is required \cite[e.g.][] {2014denbrok}. Our M15 radio data are the same as presented in \cite{2012strader}, which found no accretion evidence for an IMBH.

There are other clusters for which single studies have suggested dynamical evidence for IMBHs. \citet{2013lutz} present $2\sigma$ dynamical evidence for a $\sim 2000 M_{\odot}$ IMBH in M62, for which our formal VLA $3\sigma$ limit is $<1130 M_{\odot}$. \cite{2016kamann} suggest that NGC 6397 could host a 600 $M_{\odot}$ IMBH. This does not conflict with our ATCA limit for this cluster ($< 630 M_{\odot}$). For both of these clusters, \citet{2017baumgardt} argues that the surface brightness profiles and kinematic data do not require the presence of an IMBH.

\subsection{Conclusions and Implications} \label{sec:disconcl}
We have presented the deepest radio observations to date for a sample of 50 Milky Way globular clusters, with a goal of searching for evidence of central accreting IMBHs. While a few clusters do have radio sources near or in their centers, we do not find any credible IMBH candidates. A stacking analysis of GCs observed with VLA or ATCA also reveals no emission that can be attributed to IMBHs.

We emphasize that for any particular GC, it is possible to conceive of mechanisms that would reduce or eliminate accretion of ambient gas, rendering the IMBH undetectable in current radio or X-ray observations. Yet it is difficult to argue that such conditions should apply in many or most GCs. The most straightforward conclusion to draw from our work is that IMBHs with masses $\gtrsim 10^3 M_{\odot}$ are either not present or at least not common in GCs. 

The recent detections of merging binary black holes through gravitational waves \citep{2016abb} may help explain the observed lack of IMBHs in GCs. If such binary stellar-mass black holes are formed dynamically in GCs \citep{2016rod}, then it could indicate that single and binary black holes are preferentially ejected from GCs, rather than merging with a more massive seed black hole to form an IMBH \citep{2002millerhamilton,2004gult,2006gult,2008baker,2009moody}. Of course, there are many ways to grow IMBHs in GCs that do not depend on mergers of smaller BHs (e.g., \cite{2004portegies}).

We also cannot constrain IMBHs that might have been ejected from GCs \citep{2008holley}, nor the presence of less massive IMBHs (those of a few hundred $M_{\odot}$) through accretion signatures, especially since such objects may wander far from the cluster center{\citep{2013lutz1}}. Theoretical predictions for IMBH masses range widely, from 0.1\% to 1\% or more of the cluster mass (e.g., \citealt{2002millerhamilton,2006oleary,2006portegies,2015giersz,2017woods}). Our median VLA and ATCA limits are in the middle of this range (0.36\% and 0.54\%), though the extreme values range as low as 0.03\% and as high as 2.3\%. If IMBHs typically make up only 0.1\% of the mass of a GC, then (in the context of our formalism for radio emission) they would be difficult to detect outside the most massive GCs, unless the accretion is more radiatively efficient than we assume. We note that IMBHs well below 1000 $M_{\odot}$ are also not easily detected via standard dynamical techniques \citep{2017baumgardt}. Future gravitational wave observatories, including eLISA, offer more hope for detecting such IMBHs \citep{2016hast}---if they exist.

\appendix
\section{NGC 6624}

\subsection{The position of 4U 1820-30}
A continuing discussion about radio continuum imaging of NGC 6624 is whether central radio emission can be attributed to the low-mass X-ray binary 4U 1820-30, the bright pulsar PSR 1820-30A, or to a frequency-dependent combination of the two. Owing to the steep spectral slope of the pulsar, it may significantly contribute at low frequencies, but at the higher frequencies of our ATCA observations, the contribution of the pulsar is expected to be minimal \citep{2004migliari}. One issue with this argument in the past is that the location 
of the radio emission was not entirely consistent with that of 4U 1820-30, independently determined via \emph{Hubble Space Telescope (HST)} imaging of the UV-bright optical counterpart to the X-ray binary.

We first revisit the position of the optical counterpart to 4U 1820-30 using new $HST$ data taken with WFC3 (Program GO-13297; P.I.~Piotto). 4U 1820-30 is the brightest source in the core of the cluster in $F275W$; however, many of the \emph{Gaia} stars in the initial DR1 data release in the field of view of this image are not well-detected. Hence we instead focus on $F336W$, in which 4U 1820-30 is still bright but the number of well-detected \emph{Gaia} stars is larger. We correct the astrometry of the $F336W HST$ image using the \emph{Gaia} stars, achieving a solution with an rms uncertainty of 13--14 mas per coordinate. The J2000 position of this star is (R.A., Dec.) = (18:23:40.4975$\pm0.0010s$, --30:21:40.096$\pm0.017\arcsec$).

To determine the position of the radio source in our ATCA images, we use the 9.0 GHz image, which has the best resolution, finding a J2000 position of (R.A., Dec.) = (18:23:40.4978$\pm0.0005s$,
--30:21:40.081$\pm0.024\arcsec$). The difference between our ATCA 9.0 GHz position and the $HST\ F336W$ position is $0.016\arcsec$---the two positions agree even within their small uncertainties. Hence we conclude that the radio emission at 9.0 GHz is entirely due to 4U 1820-30 and that the position of this binary is well-determined. 4U 1820-30 is located $0.43\arcsec$ from the cluster center.

We next compare this position to that found in previous work. It is just outside of the larger error circle of the VLA source found by \citet{2004migliari} at 4.9 and 8.4 GHz---consistent at the 1.3$\sigma$ level. 
However, it is entirely inconsistent with the previous $HST$ position by \citet{1995sosinking}, differing by $0.58\arcsec$. This has important implications for the interpretation of the properties of 4U 1820-30: rather than being more distant from the cluster center than PSR 1820-30A, it is at a similar distance.

We note that our new position for 4U 1820-30 differs from the ALMA position found by \citet{2017diaz} by $0.36\arcsec$; this difference is nominally highly significant given the stated positional uncertainties. However, both positions are approximately the same distance from the cluster center (about $0.4\arcsec$ in both cases) so they do not change the interpretation of the X-ray binary properties in the context of an IMBH. Understanding the source of this discrepancy is beyond the scope of the current paper.

\begin{figure}[t!]
\epsscale{0.8}
\plotone{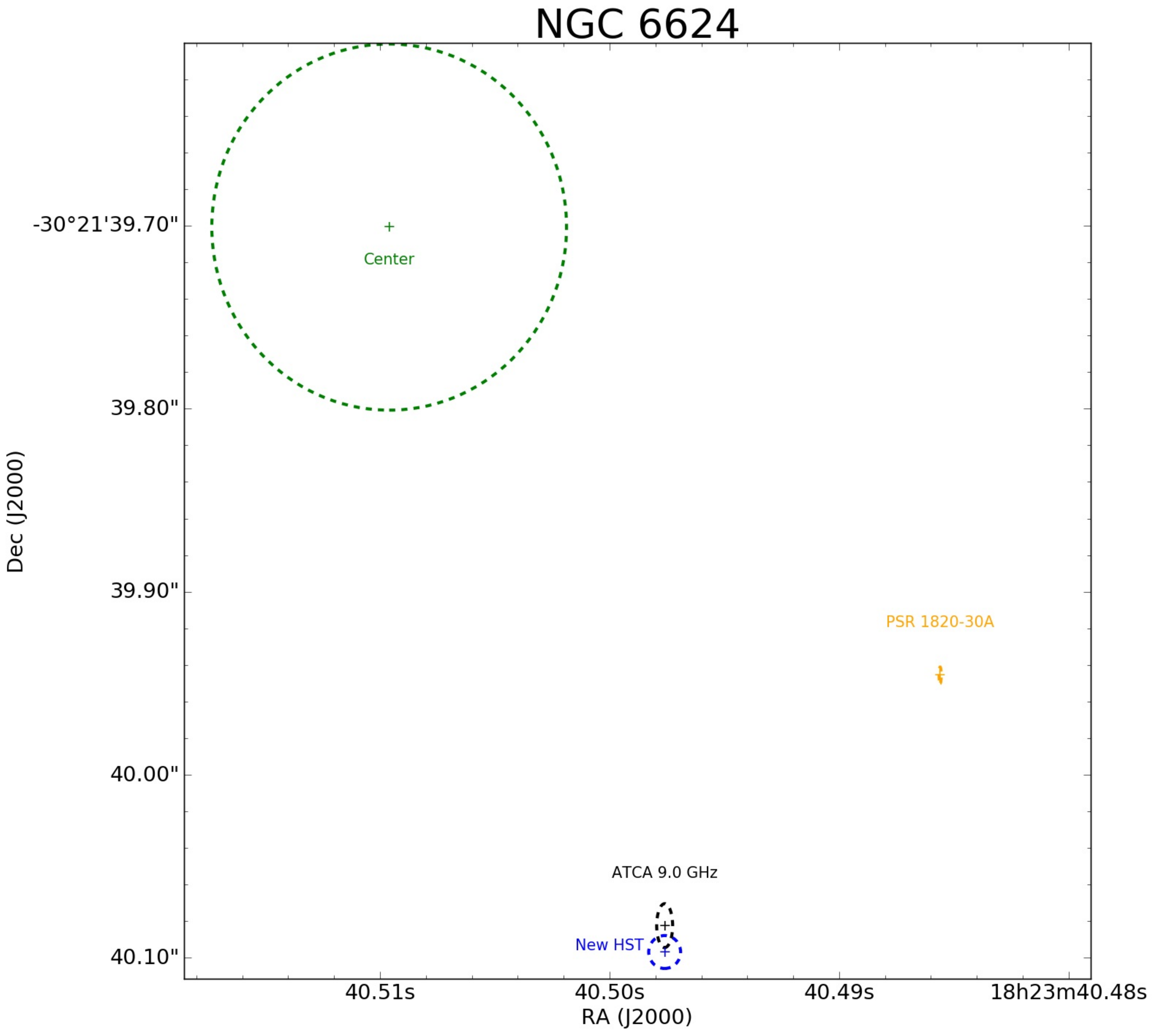}
\caption{{The central region of NGC 6624. The photometric center is marked with a green cross, and the position uncertainty is represented with the green dashed circle ($\sim$0.1$\arcsec$ in radius). At radio wavelengths, the center of NGC 6624 is dominated by the well-known neutron star sources PSR 1820-30A and 4U 1820-30. The position of the optical counterpart to 4U 1820-30 is shown with the blue dashed circle; the radius denotes the uncertainty on the position. The position and uncertainty of the radio source in our ATCA images is shown as a black ellipse. The position of the PSR 1820-30A is marked in orange \citep{2017perera}.}\label{n6624}}
\end{figure}

\subsection{Implications}

\cite{2017perera} study long-term radio timing observations of PSR 1820-30A, located close to the center of NGC 6624. These observations admit two possibilities: that PSR 1820-30A
is in an extremely wide, eccentric orbit around a massive IMBH, or that it is in an less eccentric orbit with a normal star or stellar remnant. One of their arguments in favor of the IMBH interpretation is that the unexpected negative orbital period derivative of 4U 1820-30 is due to an acceleration that requires the addition of an IMBH to the cluster potential.

From reviewing the literature it is clear that a series of errors has propagated since the earliest identification of 4U 1820-30 with $HST$. The first paper to locate the source was \citet{1993king}, who found that 4U 1820-30 was located $0.66\arcsec$ N of the cluster center (from the abstract; the listed positions actually imply only 0.6\arcsec, possibly due to rounding). \citet{1995sosinking} ``corrected'' a $1.8\arcsec$ error in the absolute astrometry of \citet{1993king} and refined the cluster center, but the relative distance of 4U 1820-30 from the center was essentially unchanged at $0.67\arcsec$.

\citet{2014peuten} report that if the updated \citet{2010goldsbury} center is used, then 4U 1820-30 is 0.046 pc from the cluster center ($1.2\arcsec$ at their assumed distance of 7.9 kpc). We are unable to determine the origin of this value. The distance of the \citet{1993king} position from the \citet{2010goldsbury} center is $0.98\arcsec$; if the X-ray binary position from \citet{1995sosinking} is used, the distance from the new center is $0.83\arcsec$.

\cite{2017perera} cite \citet{2004migliari} for the position of 4U 1820-30, who in turn take the position from \citet{1995sosinking}. However, the value published by \citet{2004migliari}  is rounded off, and hence the separation of 4U 1820-30 from the center implied is slightly larger than the correct value ($0.87\arcsec$ instead of $0.83\arcsec$). Notwithstanding this rounding, the positions in their Table 1 and Figure 1 are accurately rendered, and would imply a separation of 0.032 pc (for 7.9 kpc). However, in their Figure 9 that summarizes the dynamical constraints, and in their Erratum Figure 1, the separation plotted for 4U 1820-30 is simply the physical value from \citet{2014peuten}, 0.046 pc. The incorrect separation from \citet{2014peuten} is repeated in \citet{2018gieles}, who argue, contra \cite{2017perera}, that in any case most or all of the period change can be attributed to intrinsic factors.

The propagation of this mistake through the literature caused an over-interpretation of the 4U 1820-30 period derivative when only considering the information available. Our new $HST$ position doubles down: we find that 4U 1820-30 is $0.43\pm0.10\arcsec$ from the \citet{2010goldsbury} center. This is equivalent to $0.0175\pm0.0041$ pc using our assumed distance of 8.4 kpc (using 7.9 kpc instead would not change any of the conclusions). This is nearly identical to the separation between PSR 1820-30A and the center using the updated position of the pulsar from \cite{2017perera}; given the uncertainties in the center, either source might actually be closer. We do not present updated dynamical models of NGC 6624 in this paper, but only note that using the correct position of 4U 1820-30 would lower the inferred IMBH mass---if its period derivative is interpreted as being dominated by the potential of an IMBH, rather than intrinsic factors.

\subsection{The Radio Properties of 4U 1820-30}

Considering the radio properties of 4U 1820-30 itself: the flux density of 4U 1820-30 is $235\pm4$ $\mu$Jy (5.5 GHz) and $207\pm4$ $\mu$Jy (9.0 GHz), yielding a spectral index of $\alpha = -0.26\pm0.06$ for a power-law $S_{\nu} \propto \nu^{\alpha}$. Previous radio continuum observations at similar frequencies have been made with the VLA (mean flux densities of $130\pm40$ and $100\pm20$ $\mu$Jy at 4.9 and 8.4 GHz, respectively; \citealt{2004migliari}) and with ATCA (flux densities of $236\pm27$ and $<200$ $\mu$Jy at 5.5 and 9.0 GHz; \citealt{2017diaz}). Our new flux densities are a factor of $\sim 2$ higher than those measured with the old VLA, possibly related to the well-known superorbital modulation in X-rays \citep{2001chu}. Our 5.5 GHz ATCA measurement is spot on with the 2014 ATCA measurement made simultaneously with ALMA observations \citep{2017diaz}. Our 9.0 GHz measurement is nominally inconsistent with the upper limit of $< 200$ $\mu$Jy reported by \citet{2017diaz}, but only marginally so. In any case, the flux density of 4U 1820-30 and its spectral slope are extremely well-measured in these new observations.

Given that we find an identical flux density at 5.5 GHz to \citet{2017diaz}, it seems reasonable to combine their 302 GHz ALMA flux density of $400\pm20$ $\mu$Jy with our ATCA measurements at 5.5 and 9.0 GHz to determine the radio/mm spectral energy distribution of the binary.
In this case, we find that the 5.5, 9.0 and 302 GHz flux densities are strongly inconsistent with a single power law. This suggests either that the ALMA observations (taken 1 week from the 2014 ATCA data) were taken during a flare unobserved at other wavelengths, or that there is another source of 302 GHz emission. Truly simultaneous radio and mm observations of 4U 1820-30 appear necessary to determine an accurate spectral energy distribution for this binary.

\acknowledgments
We thank an anonymous referee and B.~Stappers and B.~Perera for useful comments. The National Radio Astronomy Observatory is a facility of the National Science Foundation operated under cooperative agreement by Associated Universities, Inc. The Australia Telescope Compact Array  is part of the Australia Telescope National Facility which is funded by the Australian Government for operation as a National Facility managed by CSIRO.

JS acknowledges support from the Packard Foundation. We thank the National Science Foundation through support from grants AST-1308124 and AST-1514763. JCAMJ is the recipient of an Australian Research Council Future Fellowship (FT140101082). COH and GRS are supported by NSERC Discovery Grants, and COH also by an NSERC Discovery Accelerator Supplement. ACS acknowledges financial support from NSF AST-1350389.

\vspace{5mm}
\facilities{VLA, ATCA}

\software{CASA \citep{2007mcmullin}, AIPS \citep{1985wells}, TOPCAT \citep{2005taylor}, APLpy \citep{2012robit}, Astropy \citep{2013astropy}.
          }

\begin{figure*}[htb]
\centering
  \begin{tabular}{@{}ccc@{}}
     \includegraphics[width=.33\textwidth]{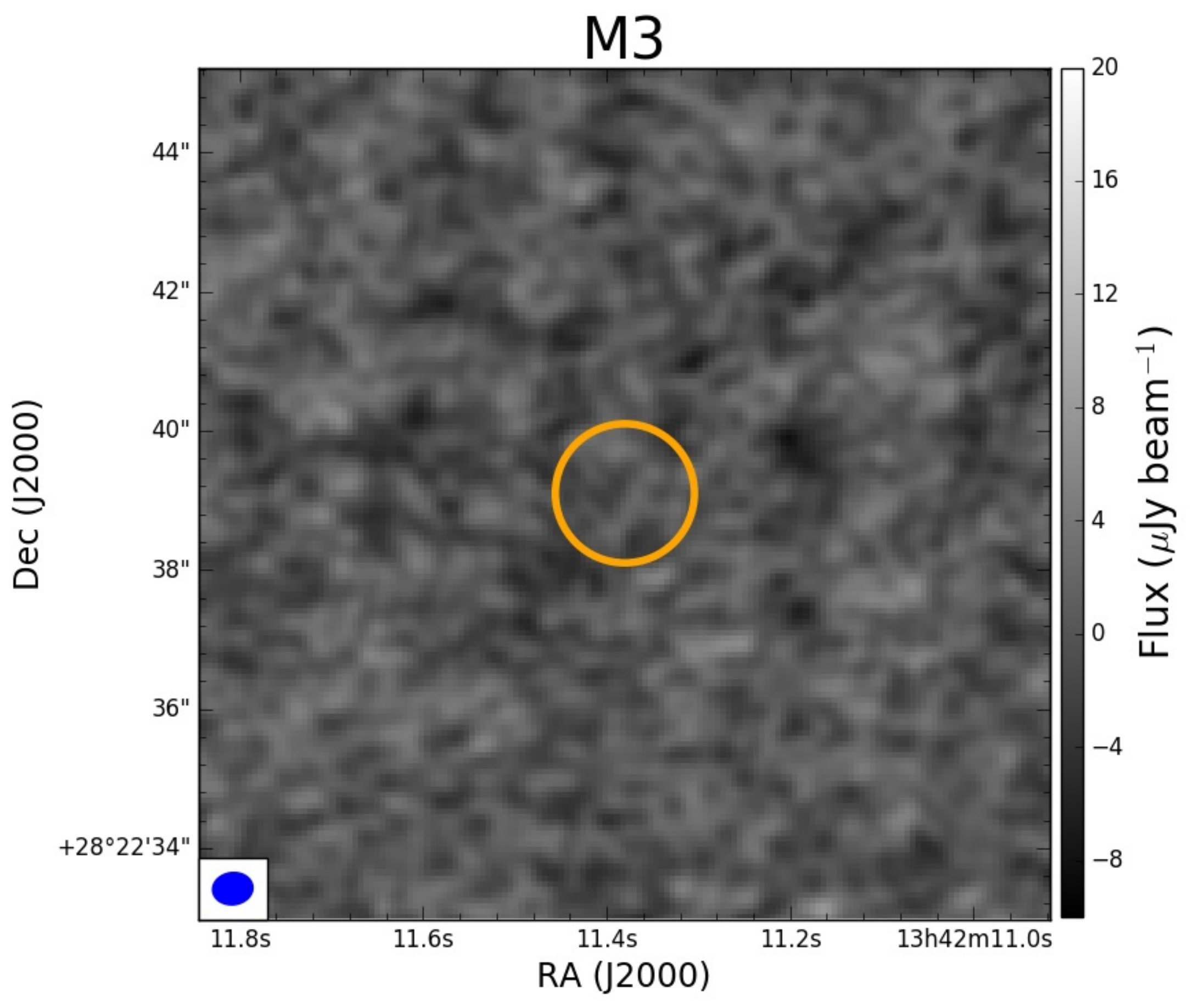} &
  \includegraphics[width=.33\textwidth]{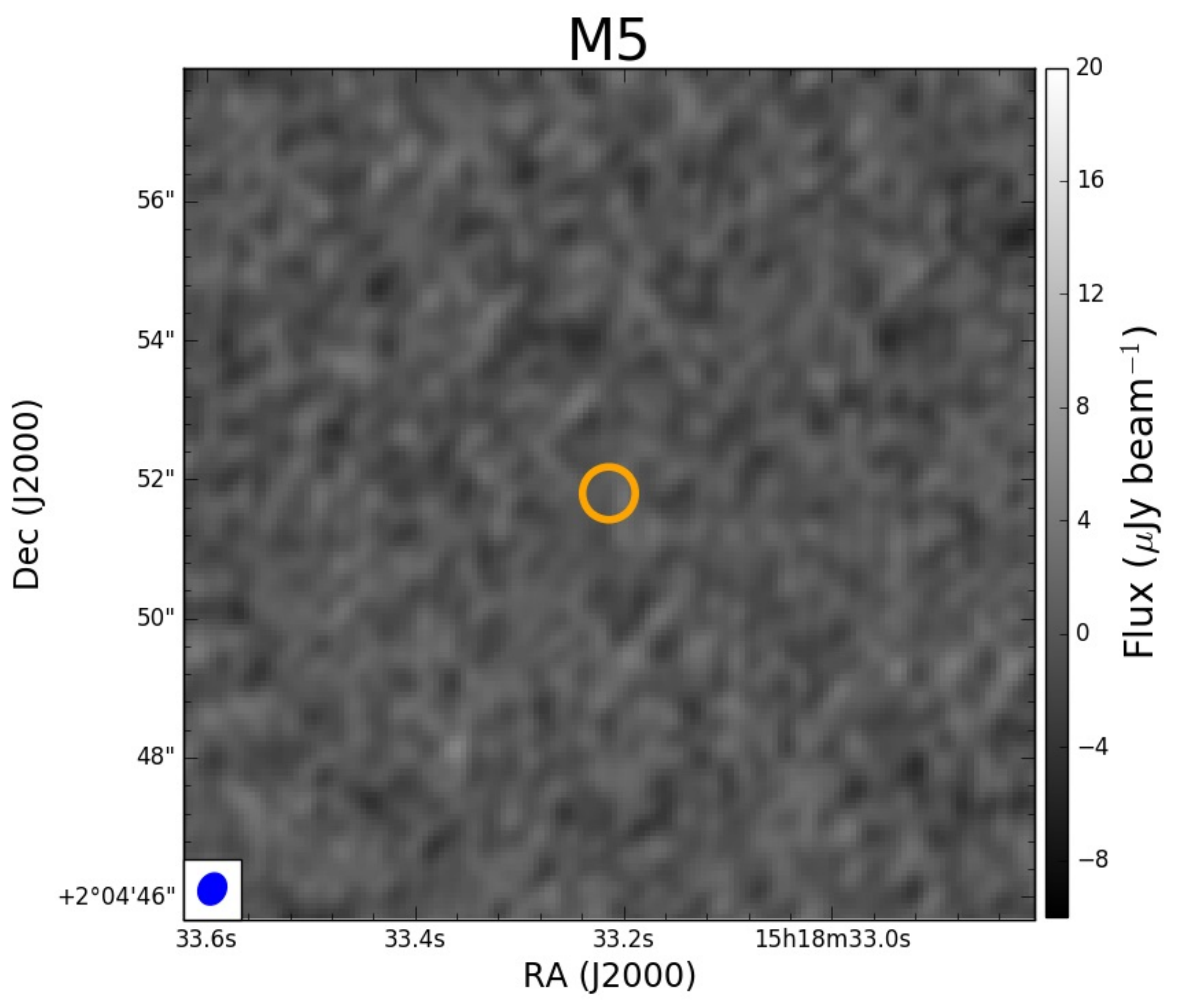} &
  \includegraphics[width=.33\textwidth]{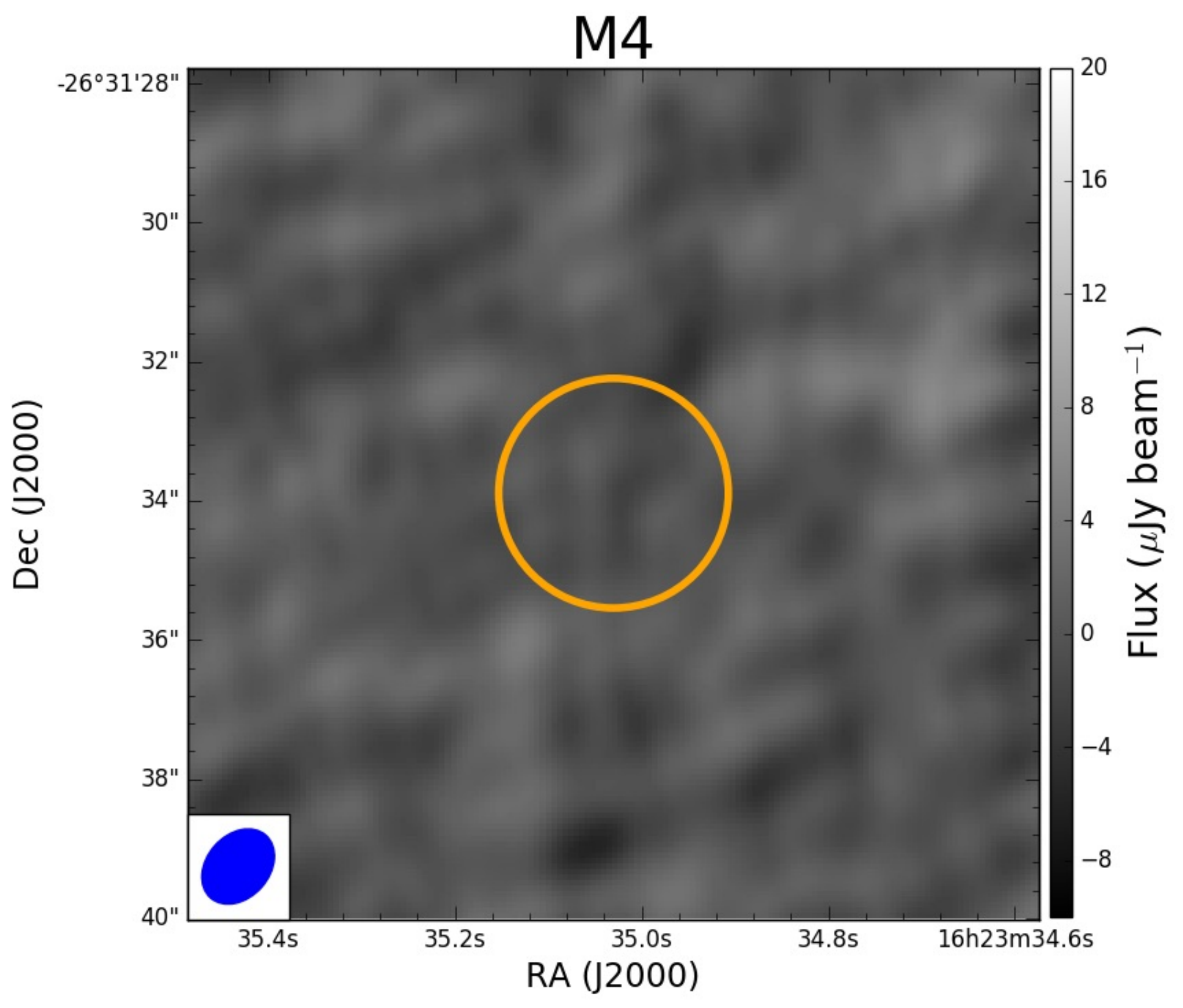} \\
  \includegraphics[width=.33\textwidth]{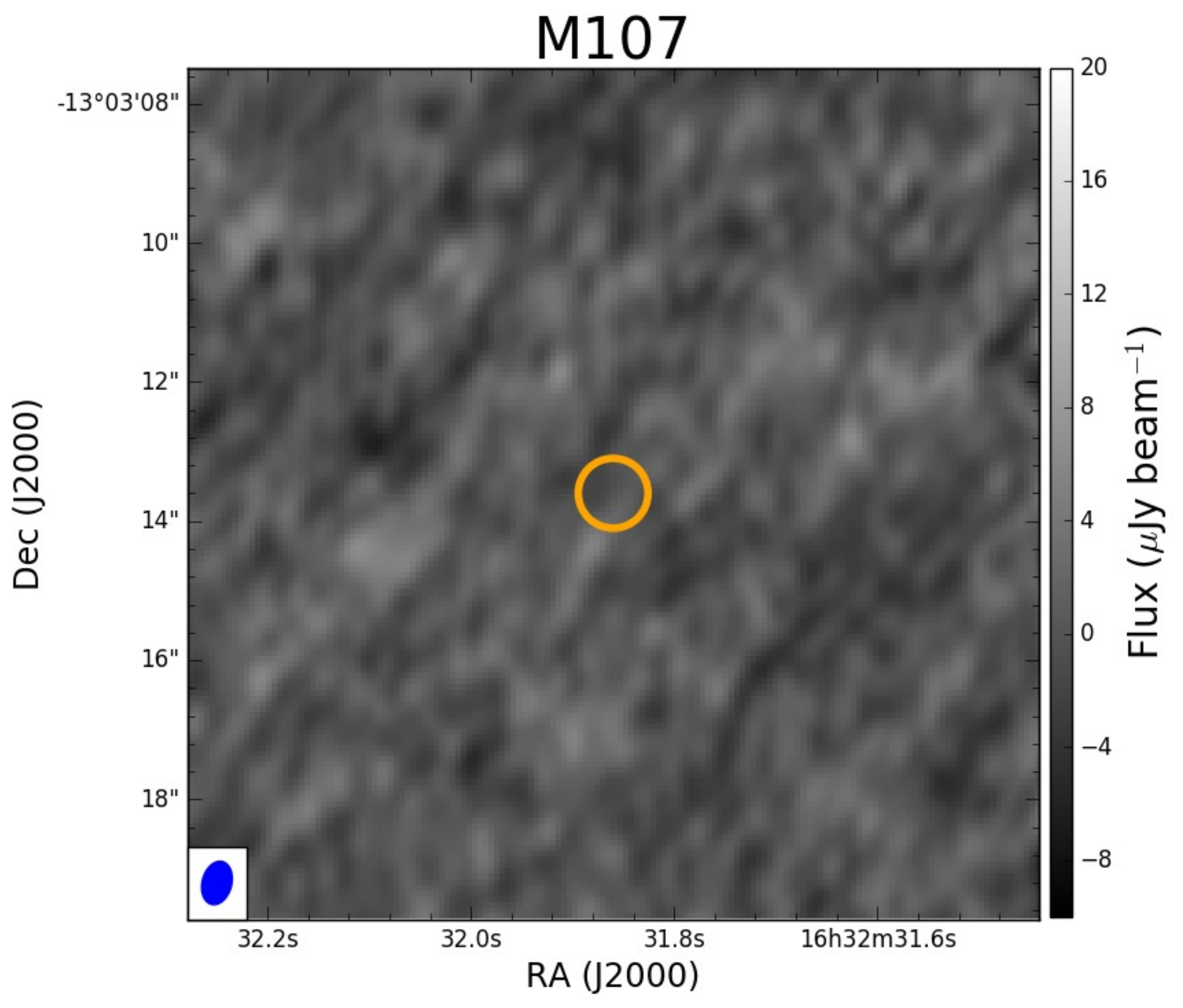}   &
  \includegraphics[width=.33\textwidth]{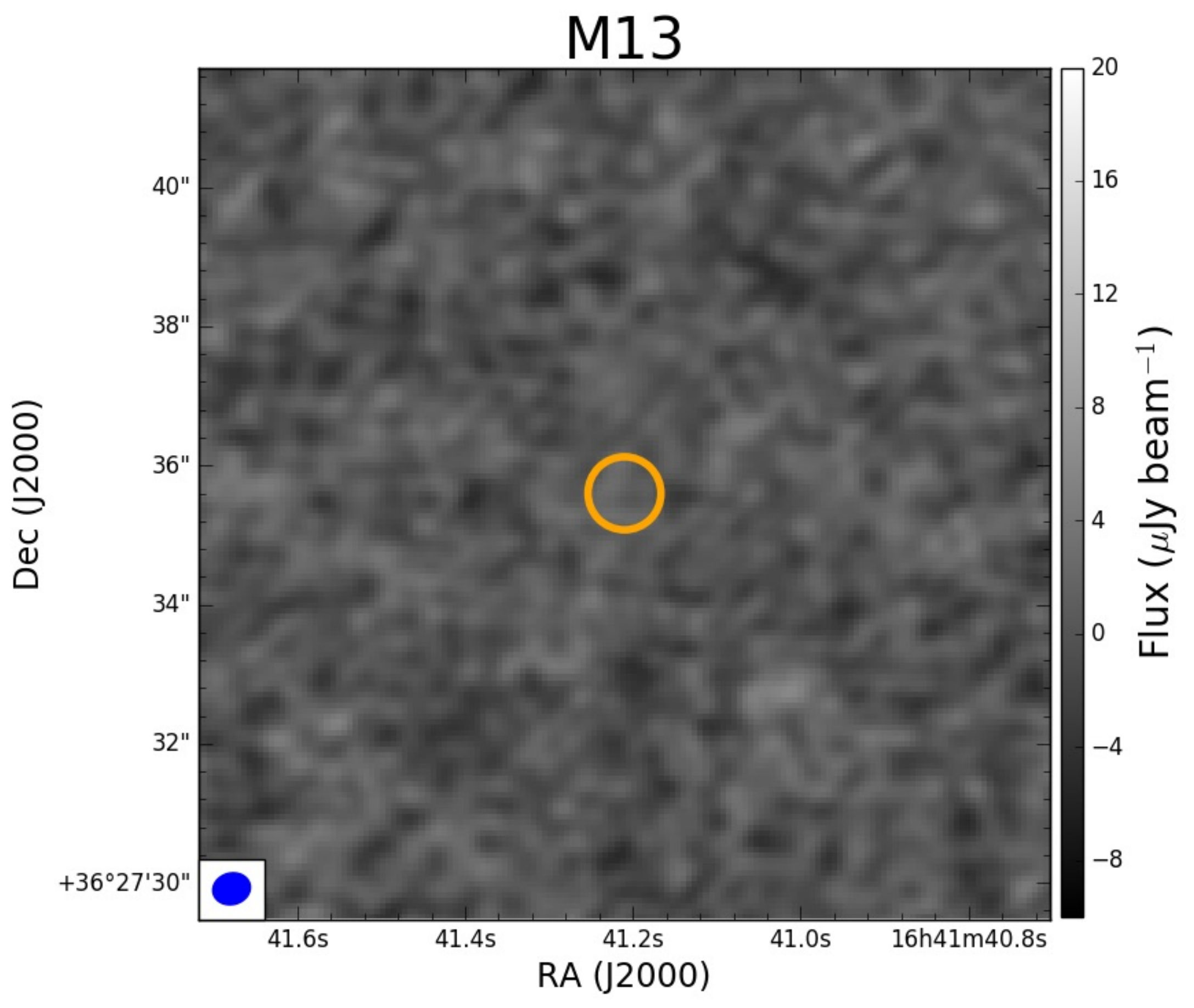} &
  \includegraphics[width=.33\textwidth]{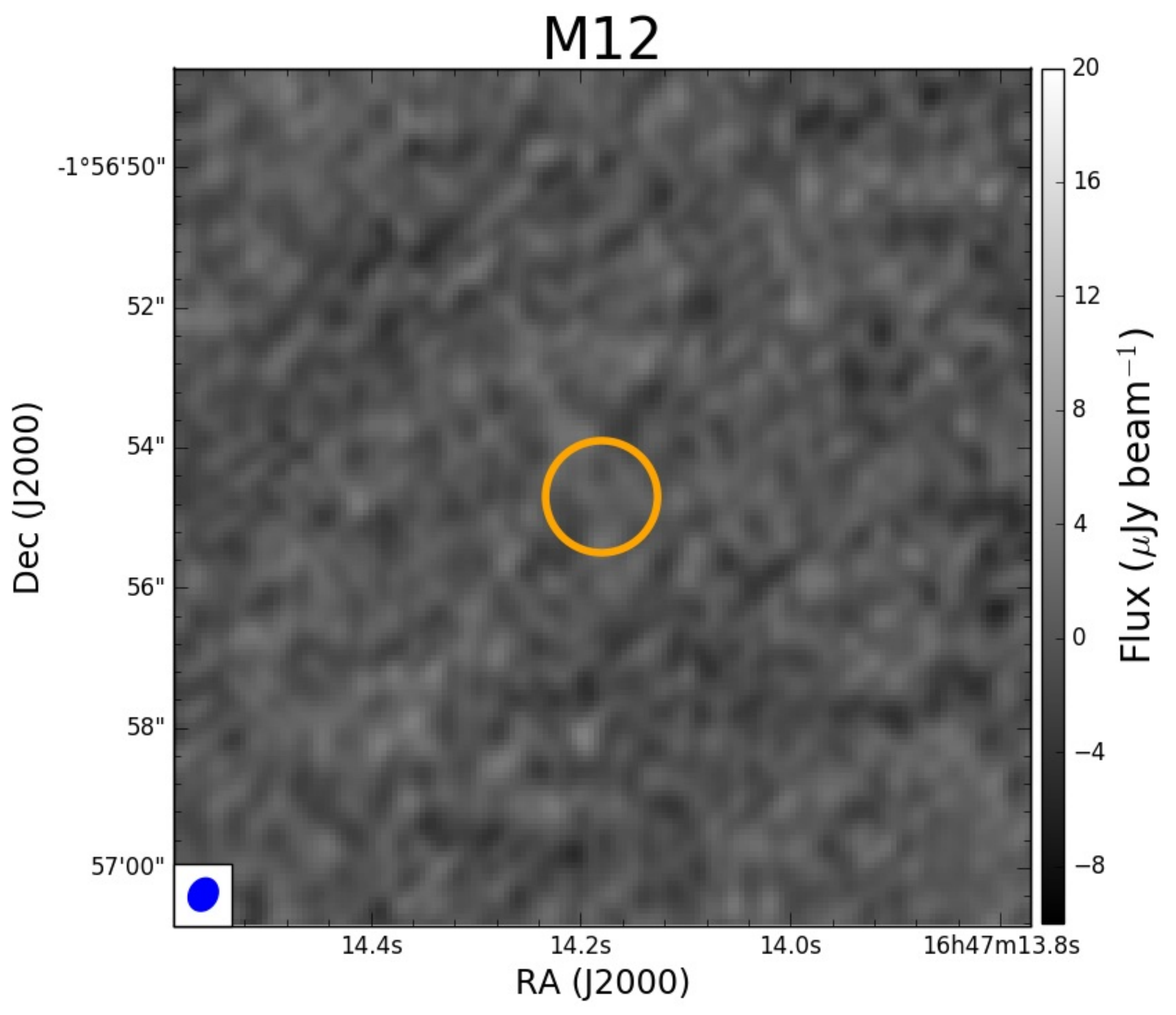} \\
  \includegraphics[width=.33\textwidth]{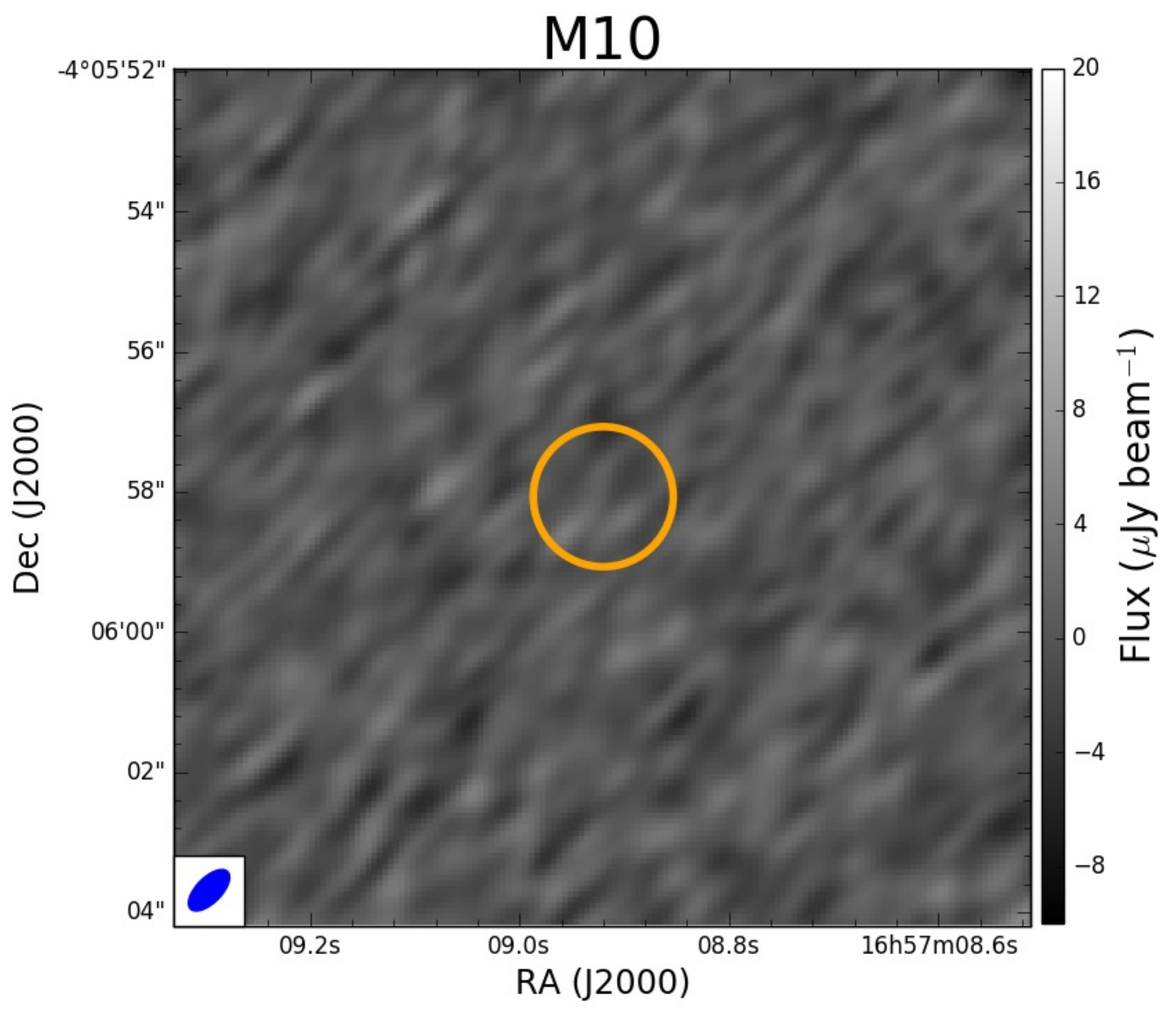} &
  \includegraphics[width=.33\textwidth]{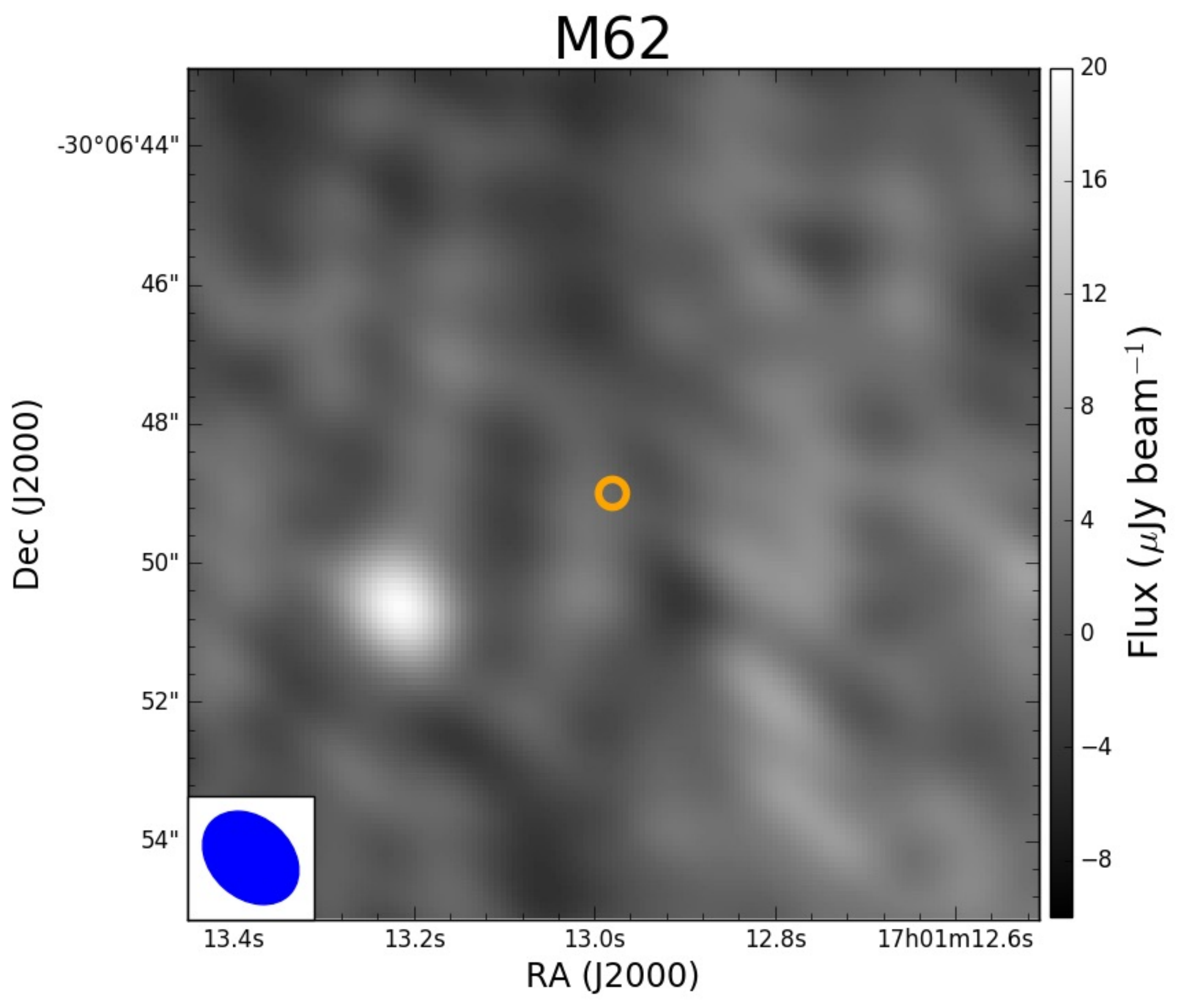}   &
  \includegraphics[width=.33\textwidth]{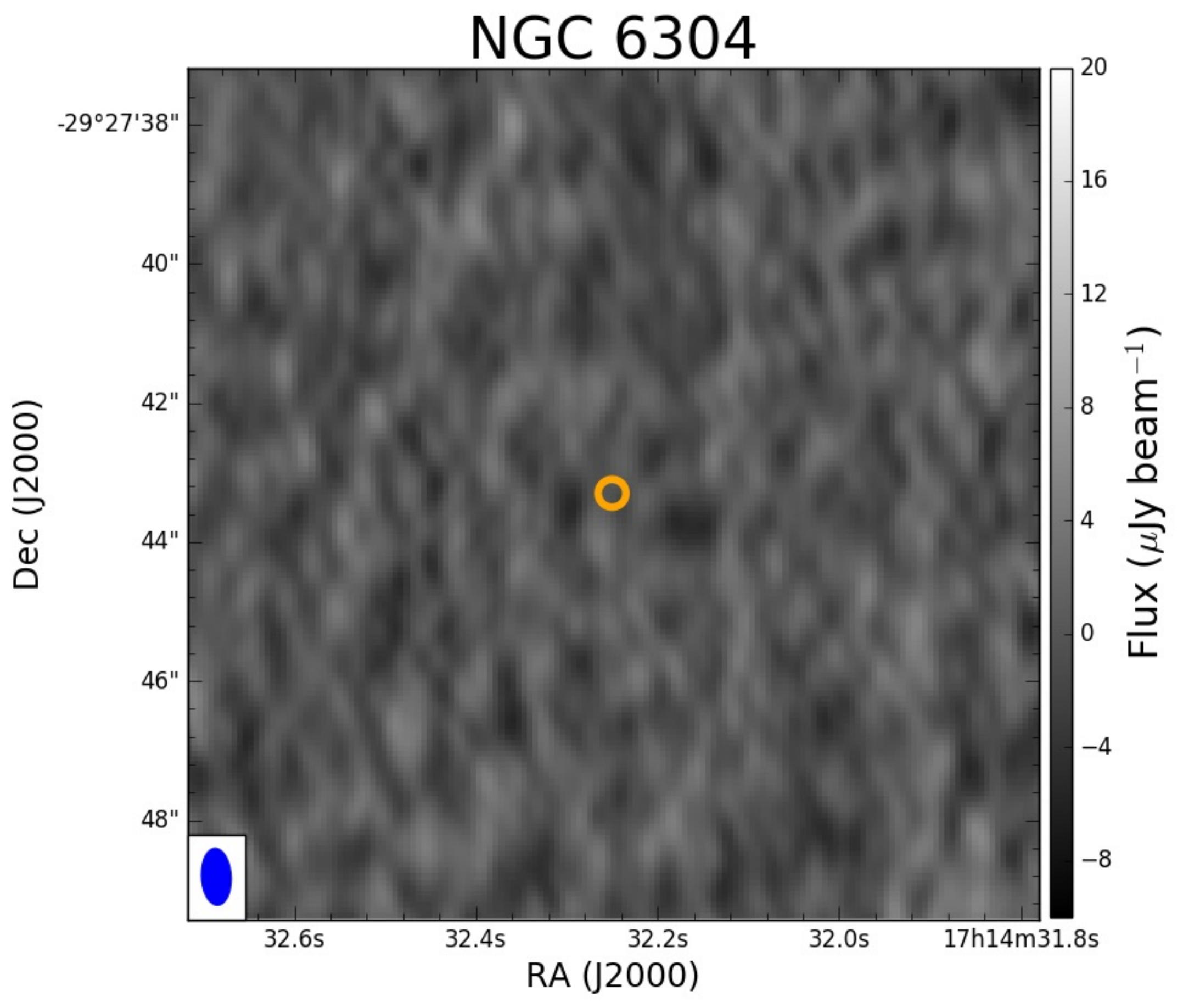} \\
  \includegraphics[width=.33\textwidth]{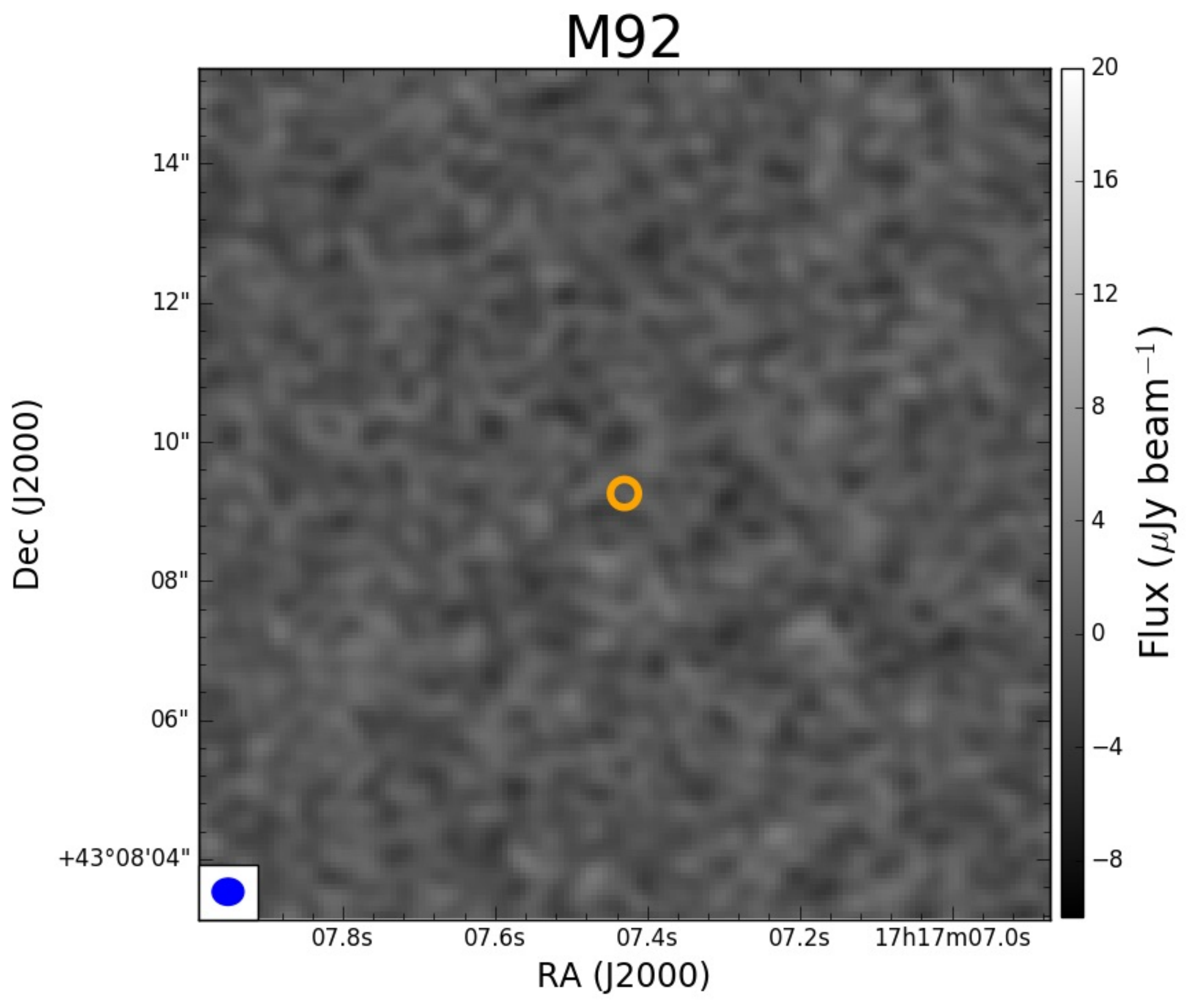} &
  \includegraphics[width=.33\textwidth]{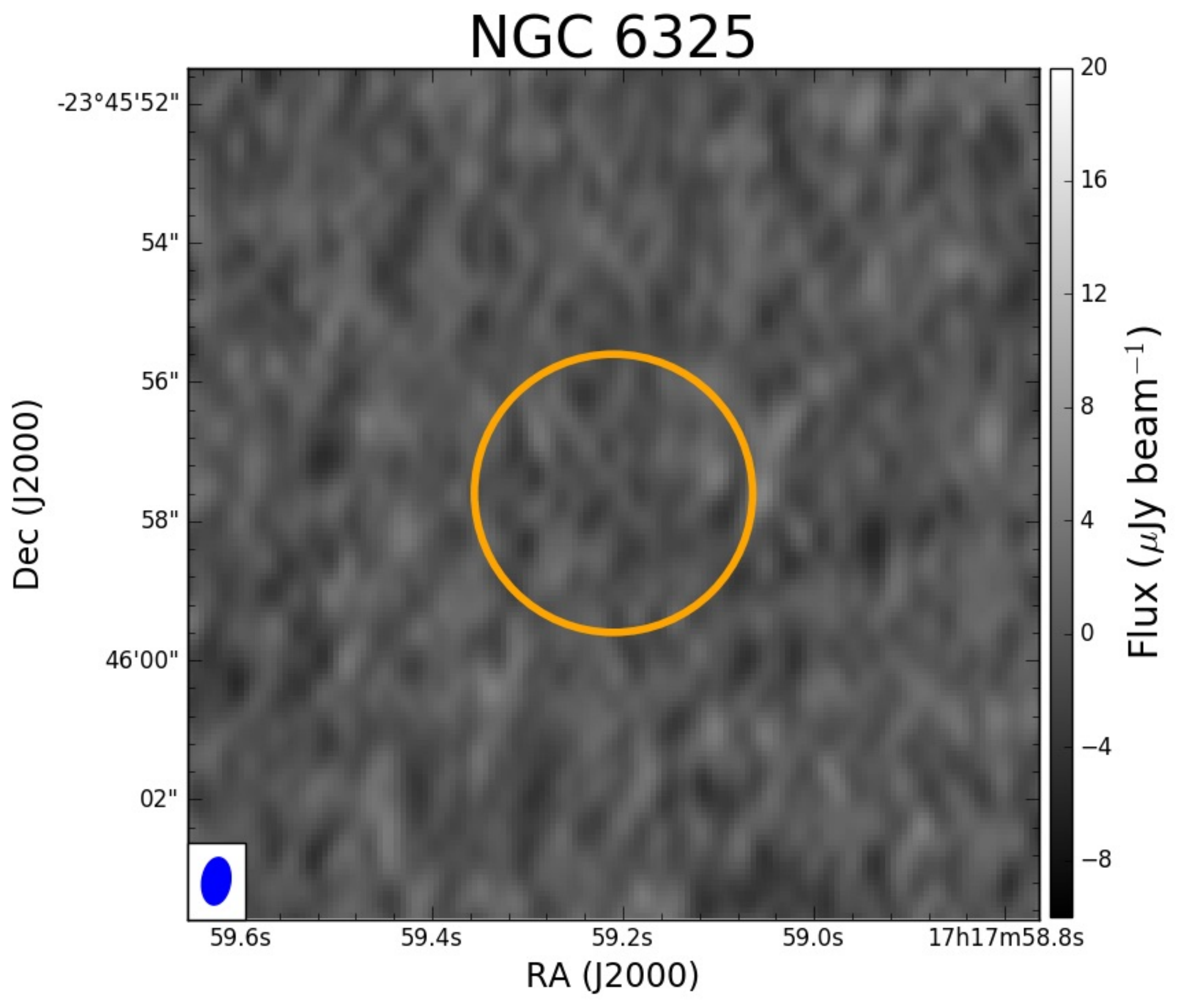} &
  \includegraphics[width=.33\textwidth]{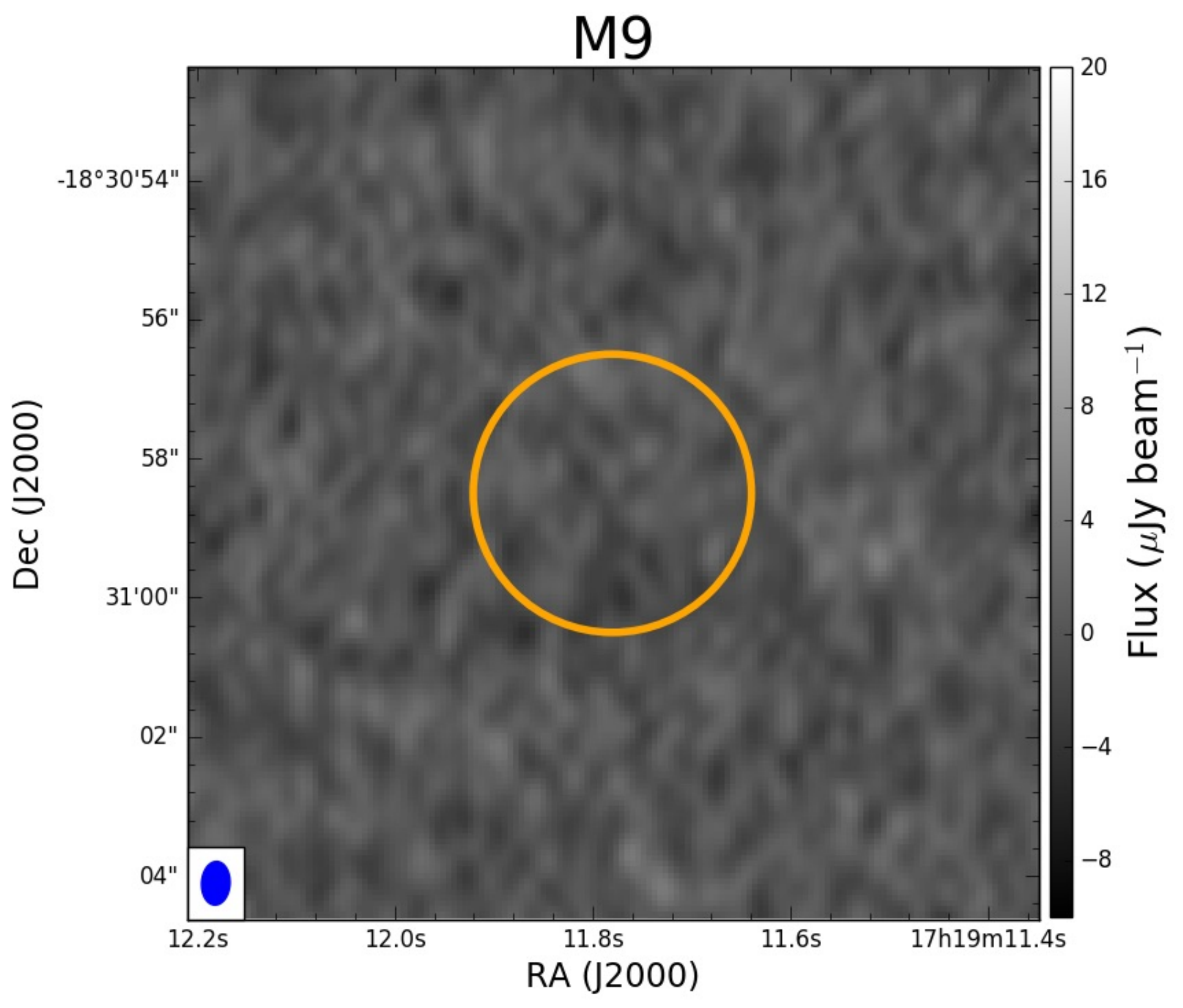} 
     
  \end{tabular}
 \end{figure*} 

 \begin{figure*}[htb]
  \begin{tabular}{@{}ccc@{}}  
   \includegraphics[width=.33\textwidth]{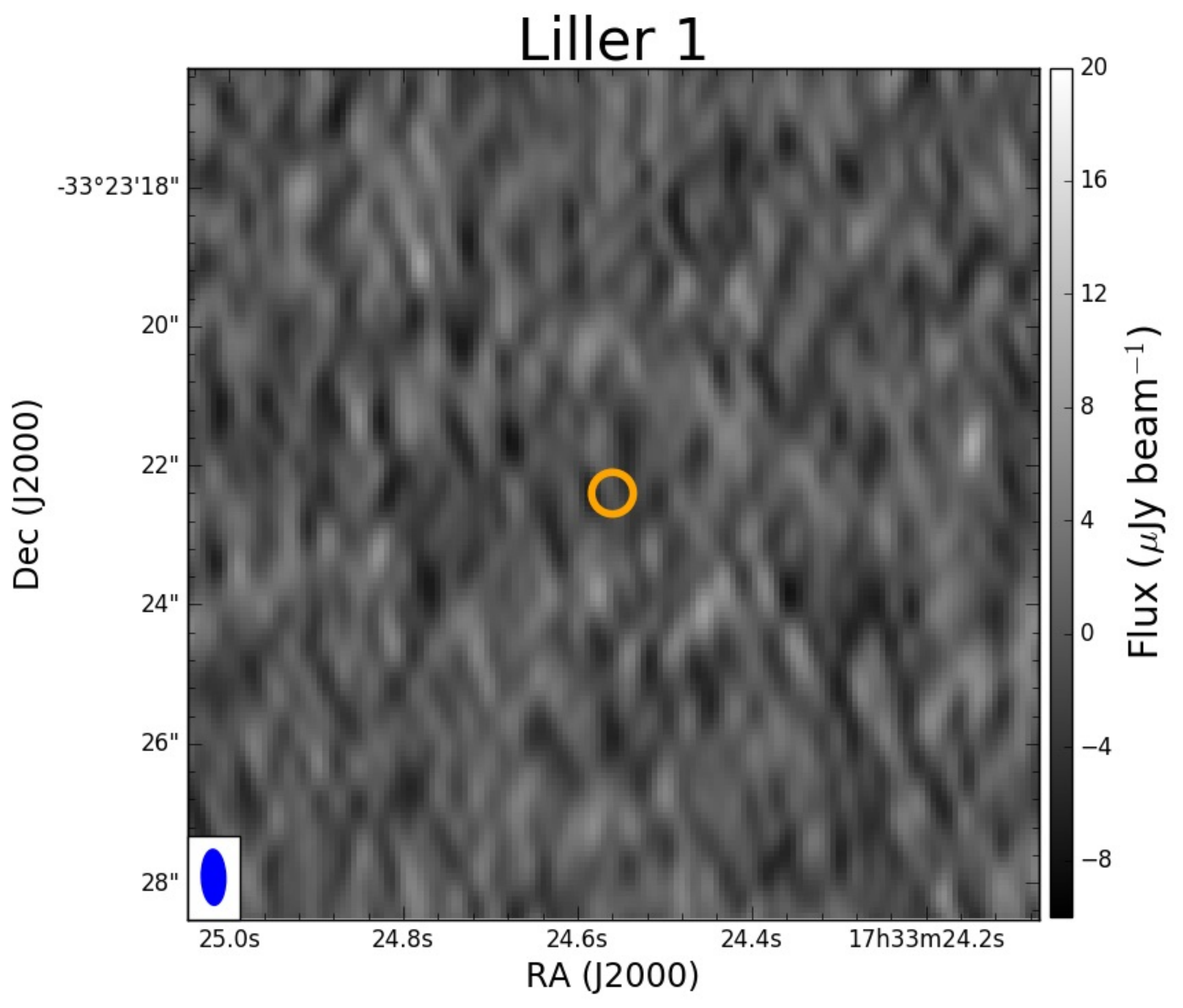} &
  \includegraphics[width=.33\textwidth]{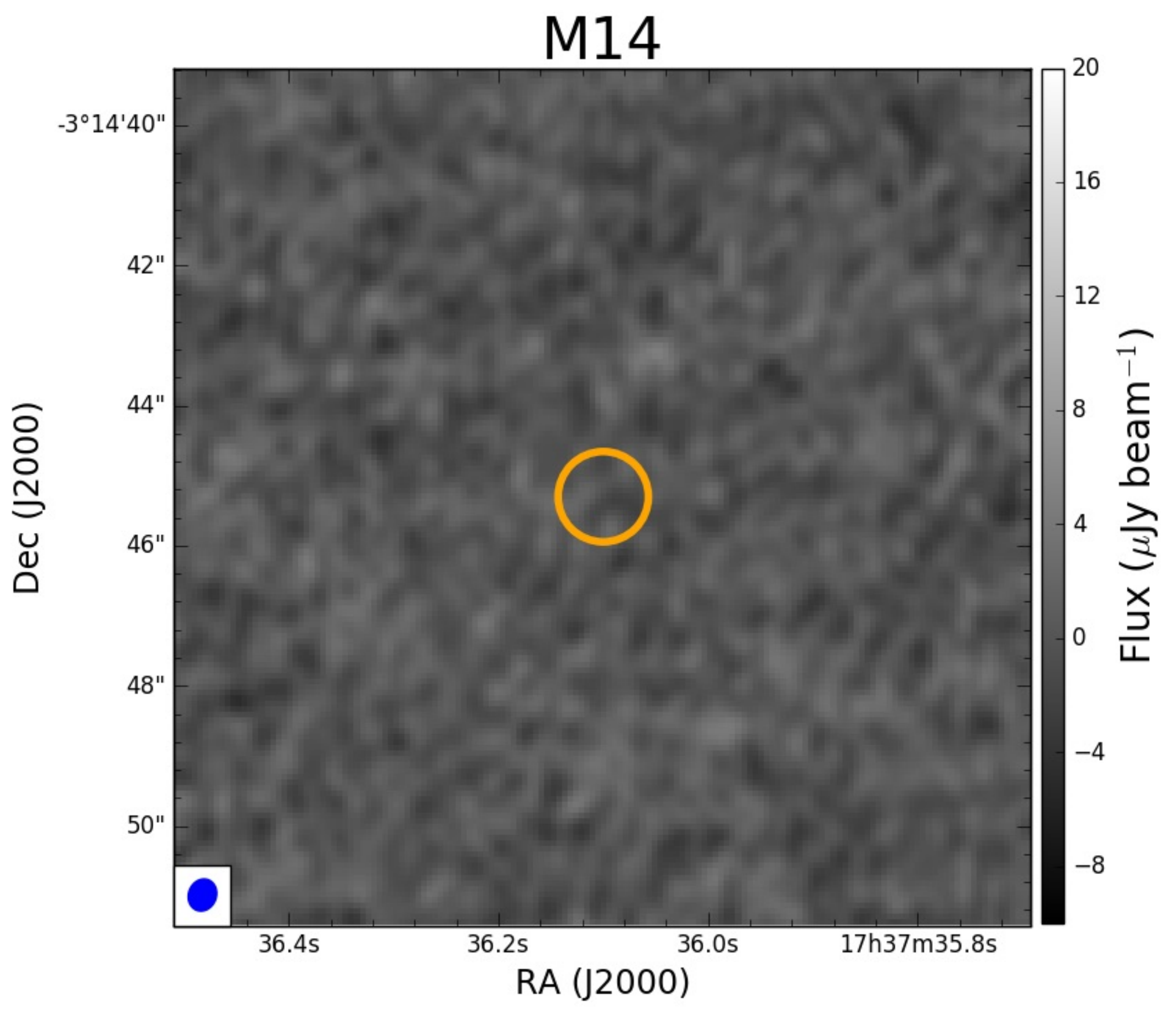} &
  \includegraphics[width=.33\textwidth]{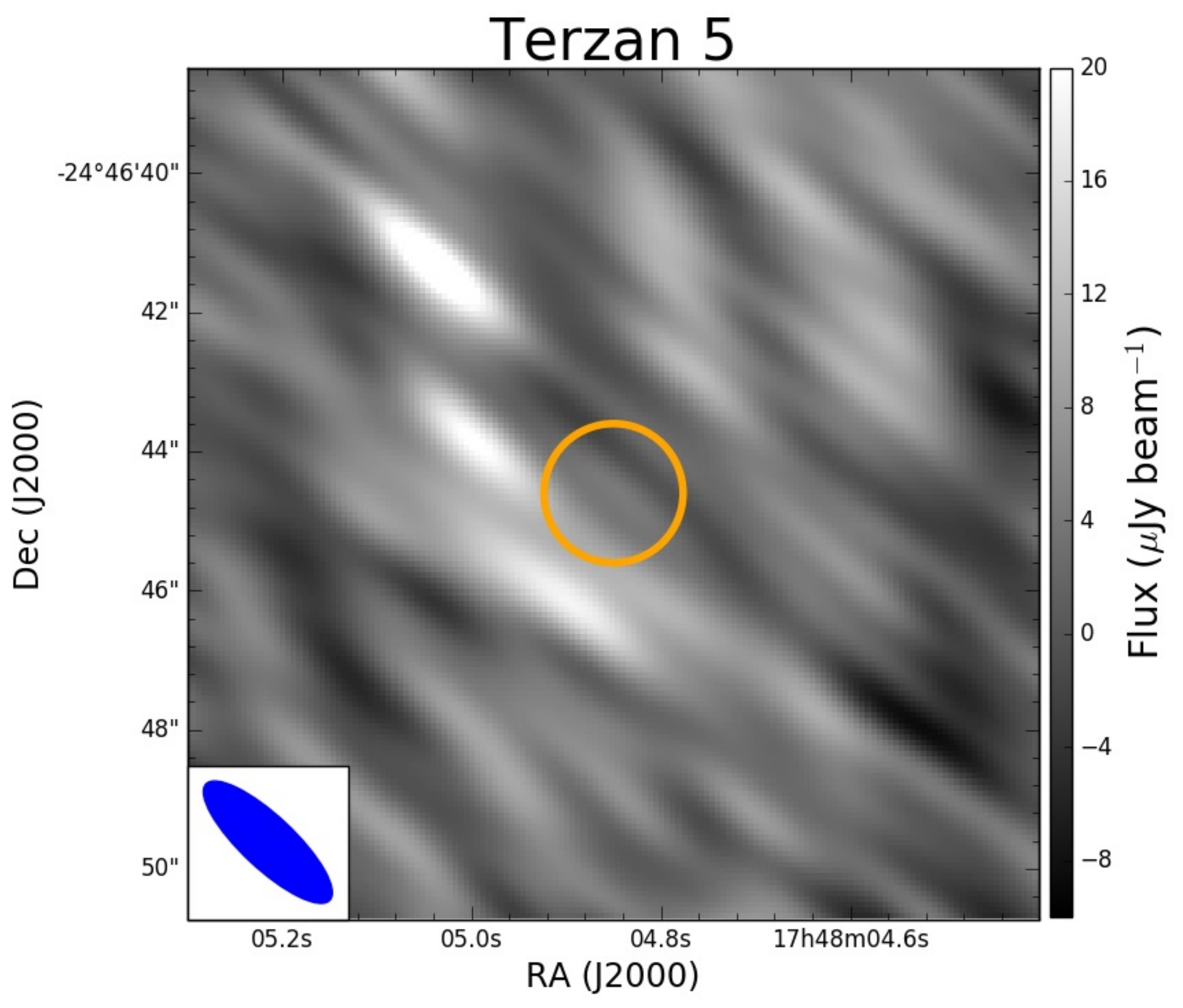} \\
   \includegraphics[width=.33\textwidth]{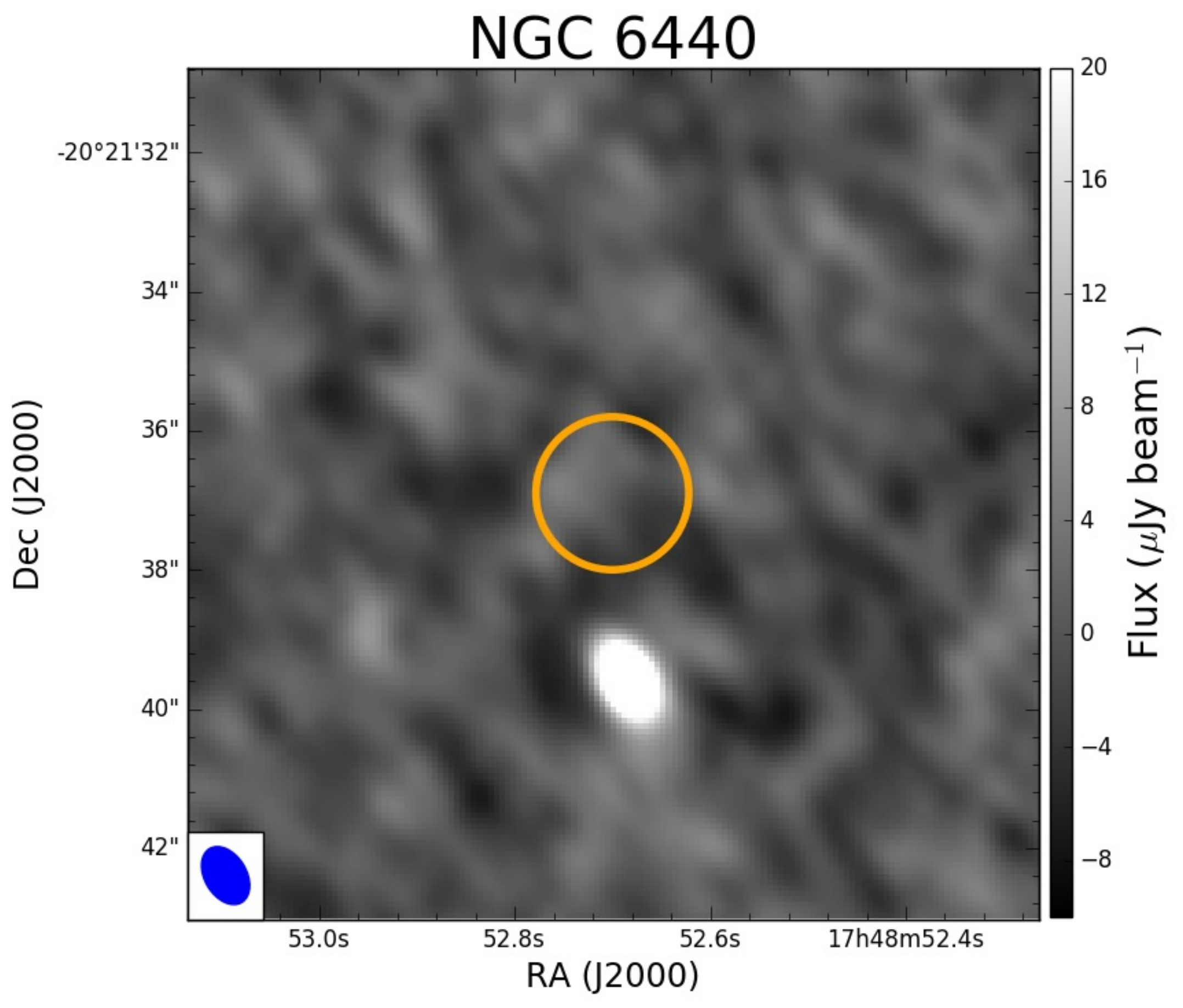} &
 \includegraphics[width=.33\textwidth]{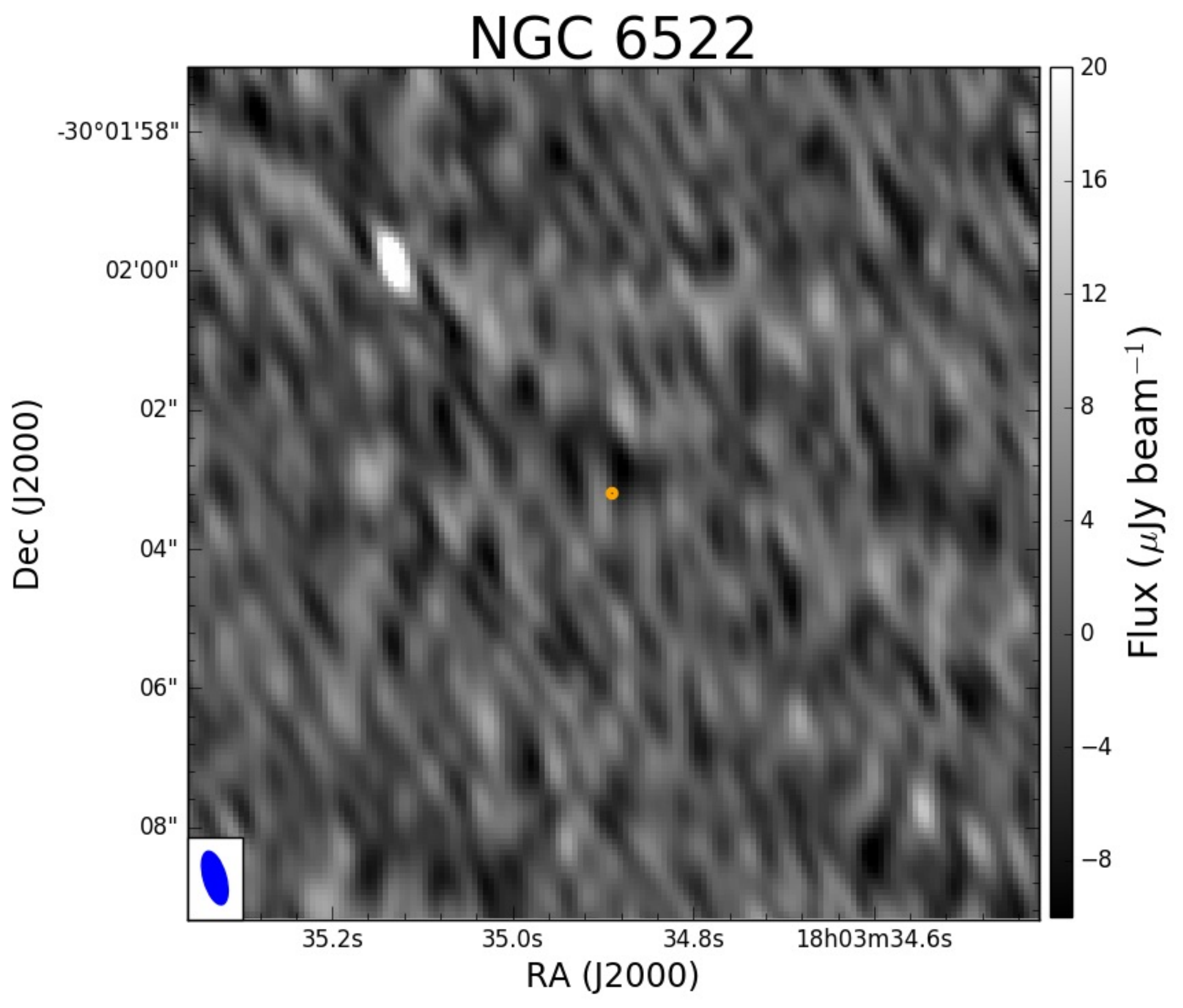} &
 \includegraphics[width=.33\textwidth]{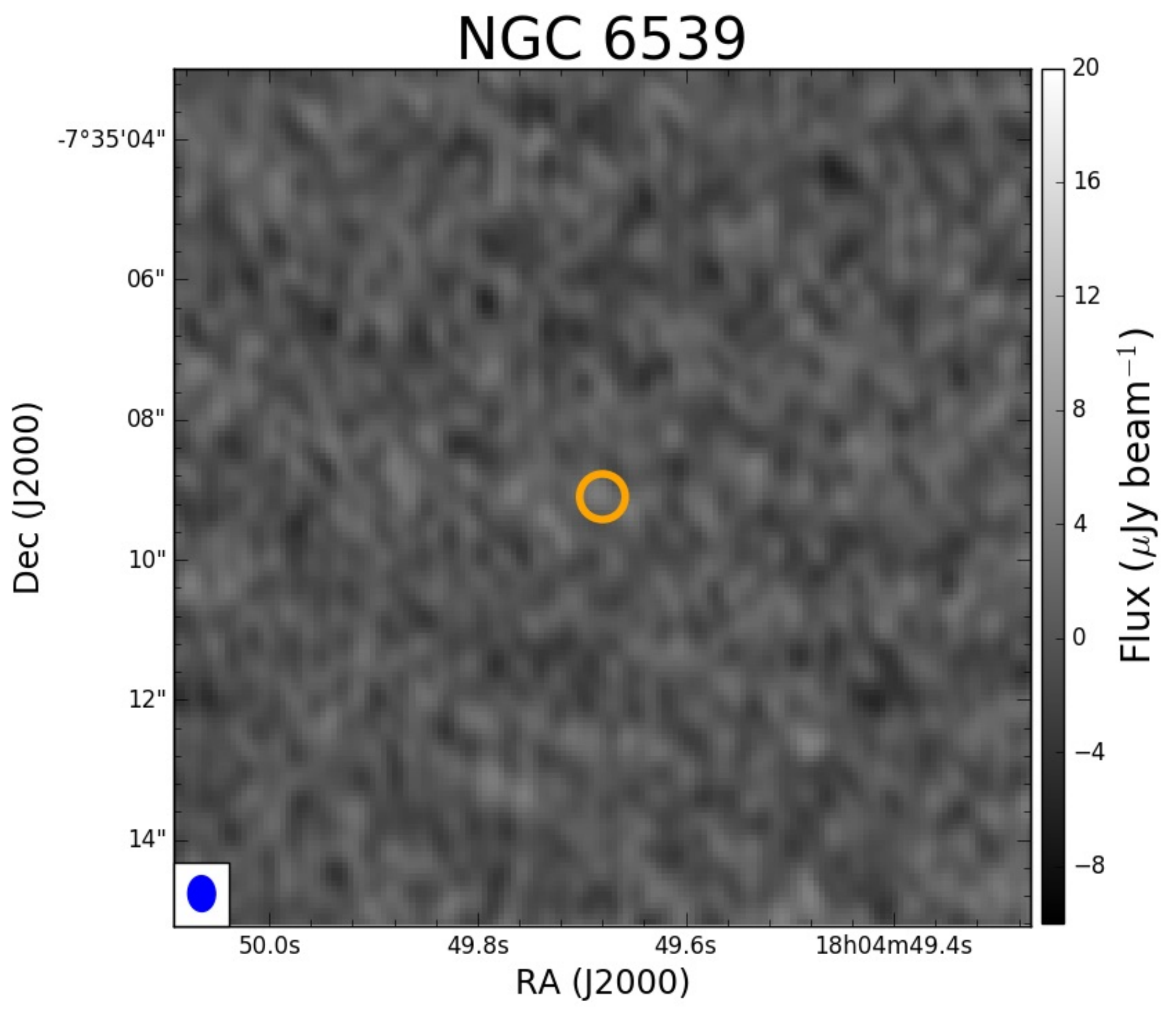} \\
 \includegraphics[width=.33\textwidth]{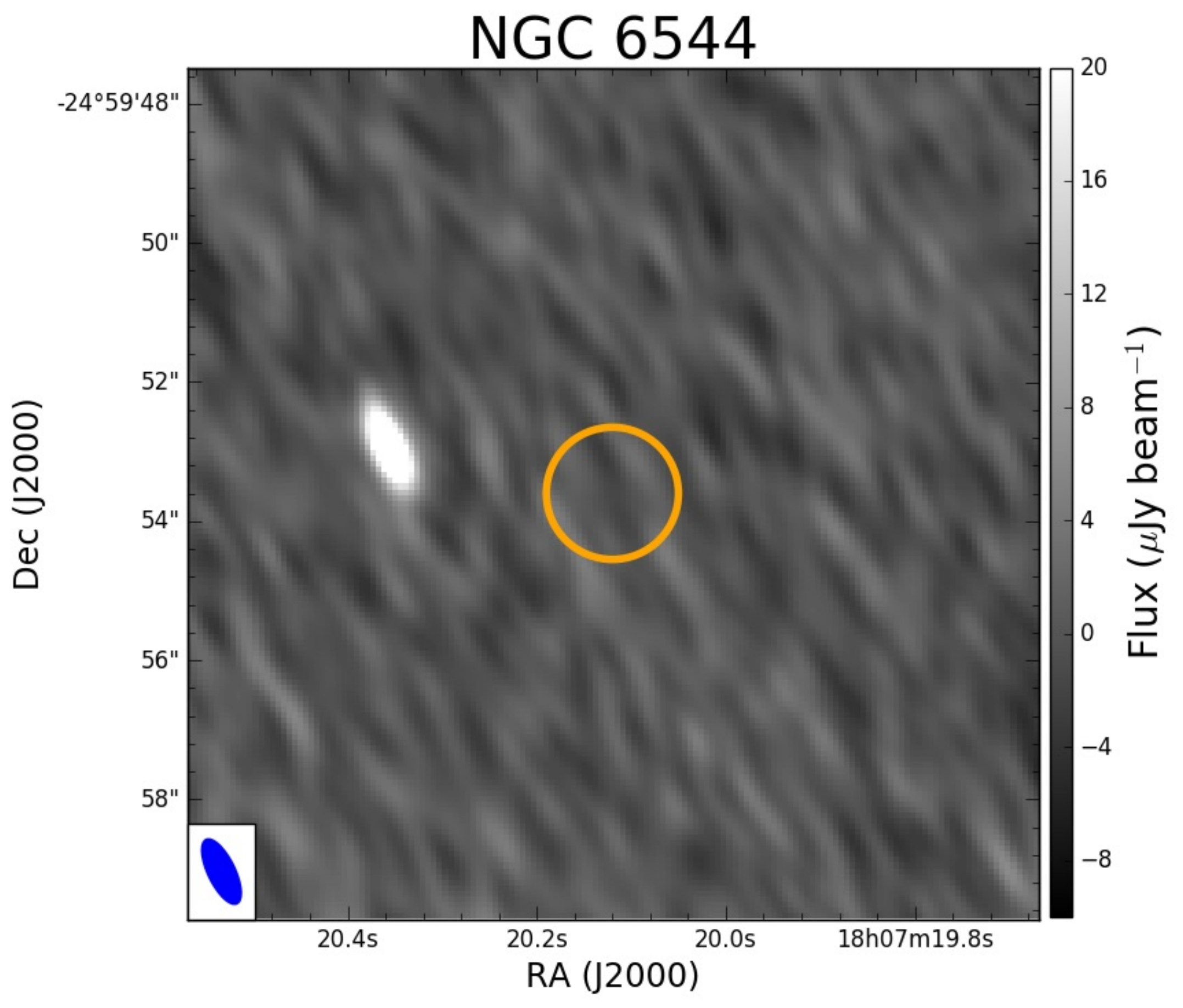}   &
  \includegraphics[width=.33\textwidth]{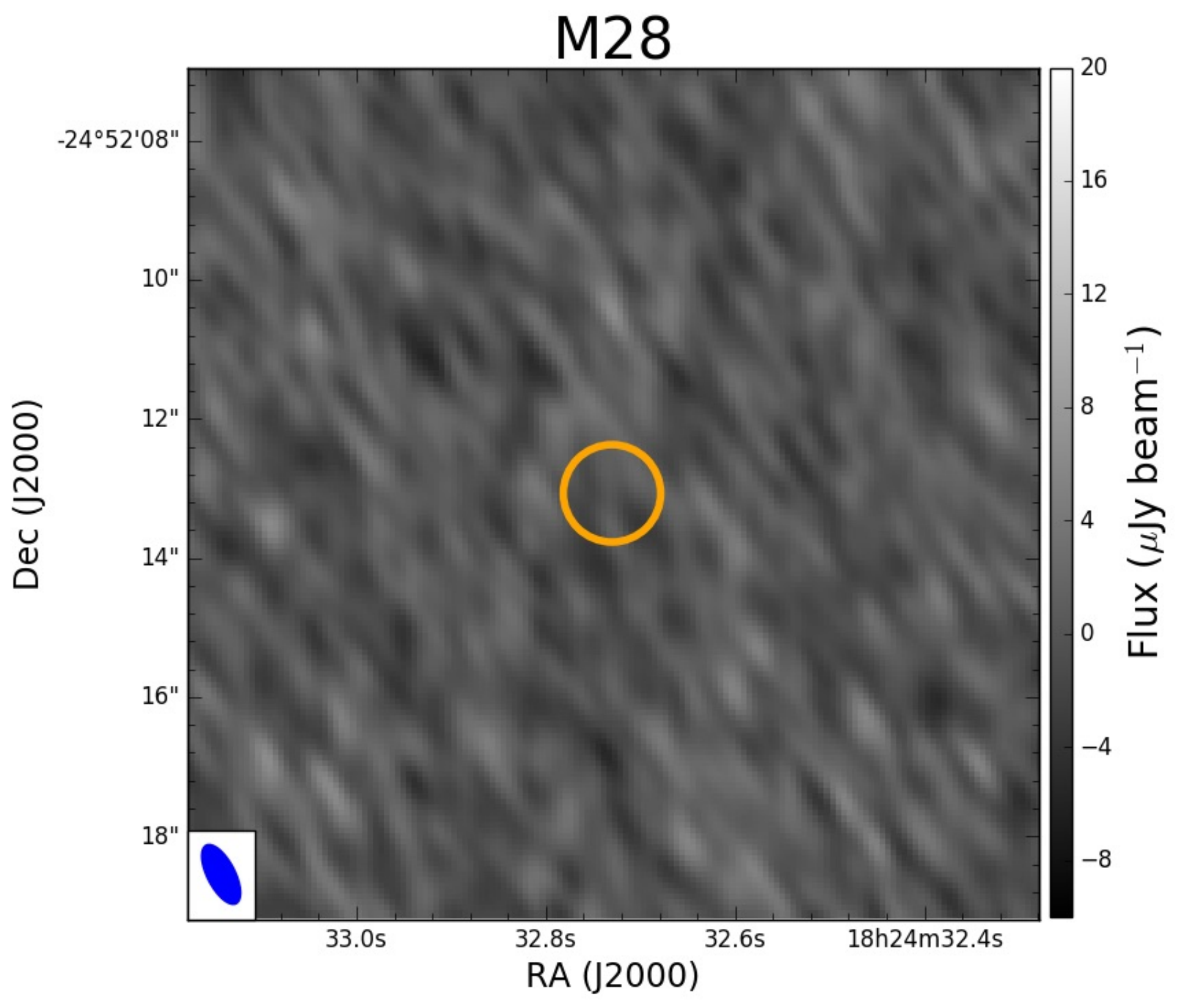}   &
 \includegraphics[width=.33\textwidth]{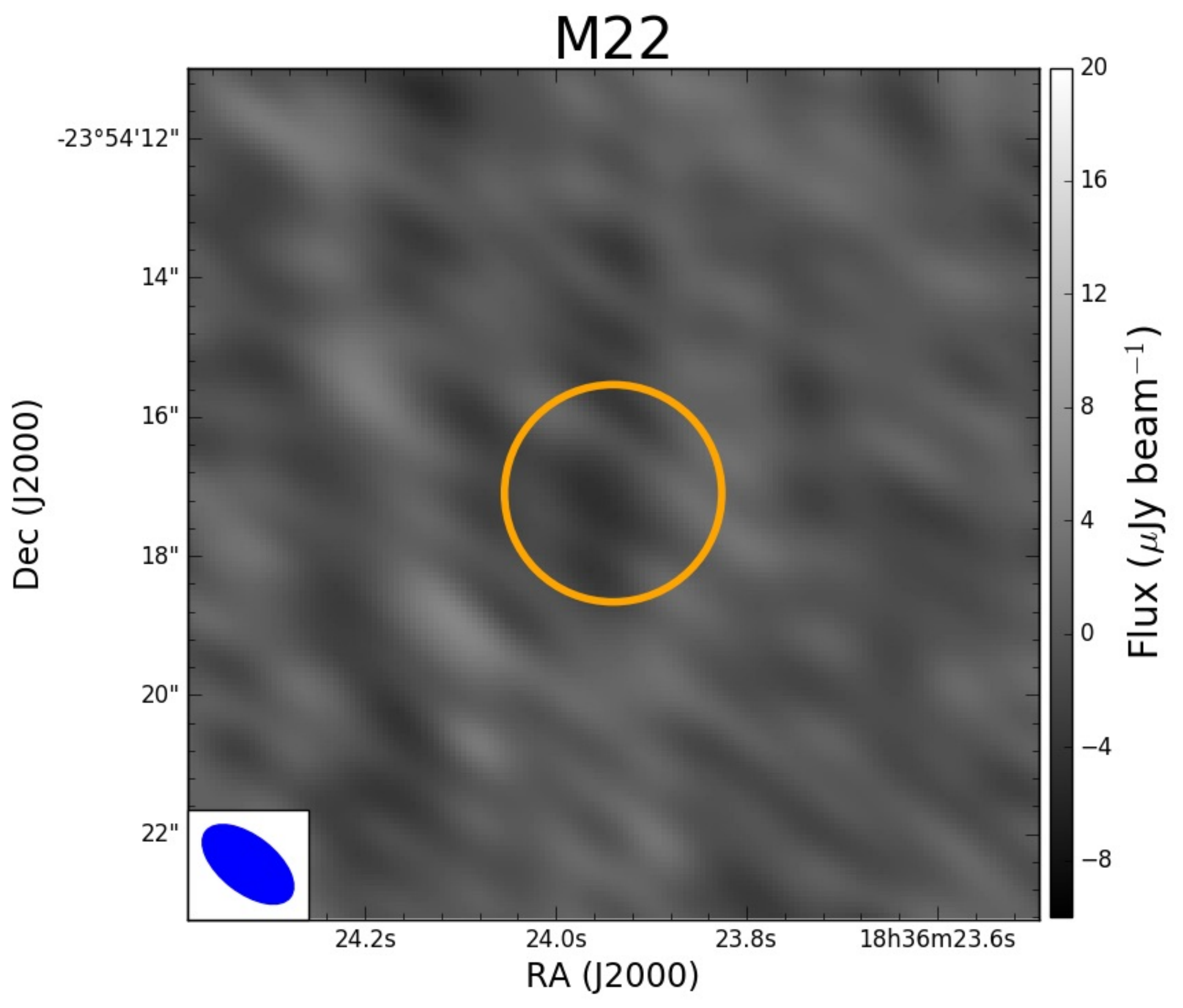}\\
 \includegraphics[width=.33\textwidth]{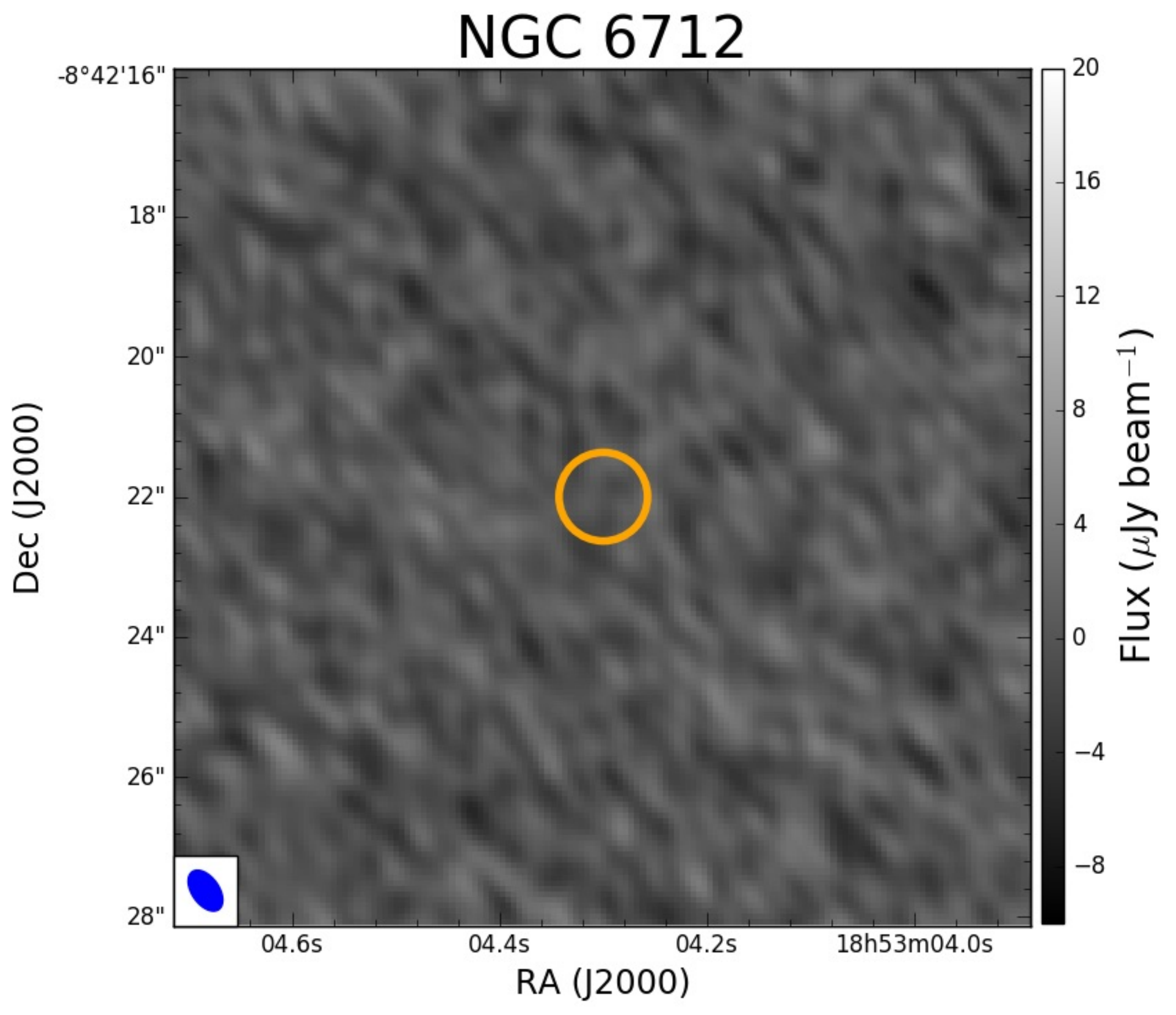} &
 \includegraphics[width=.33\textwidth]{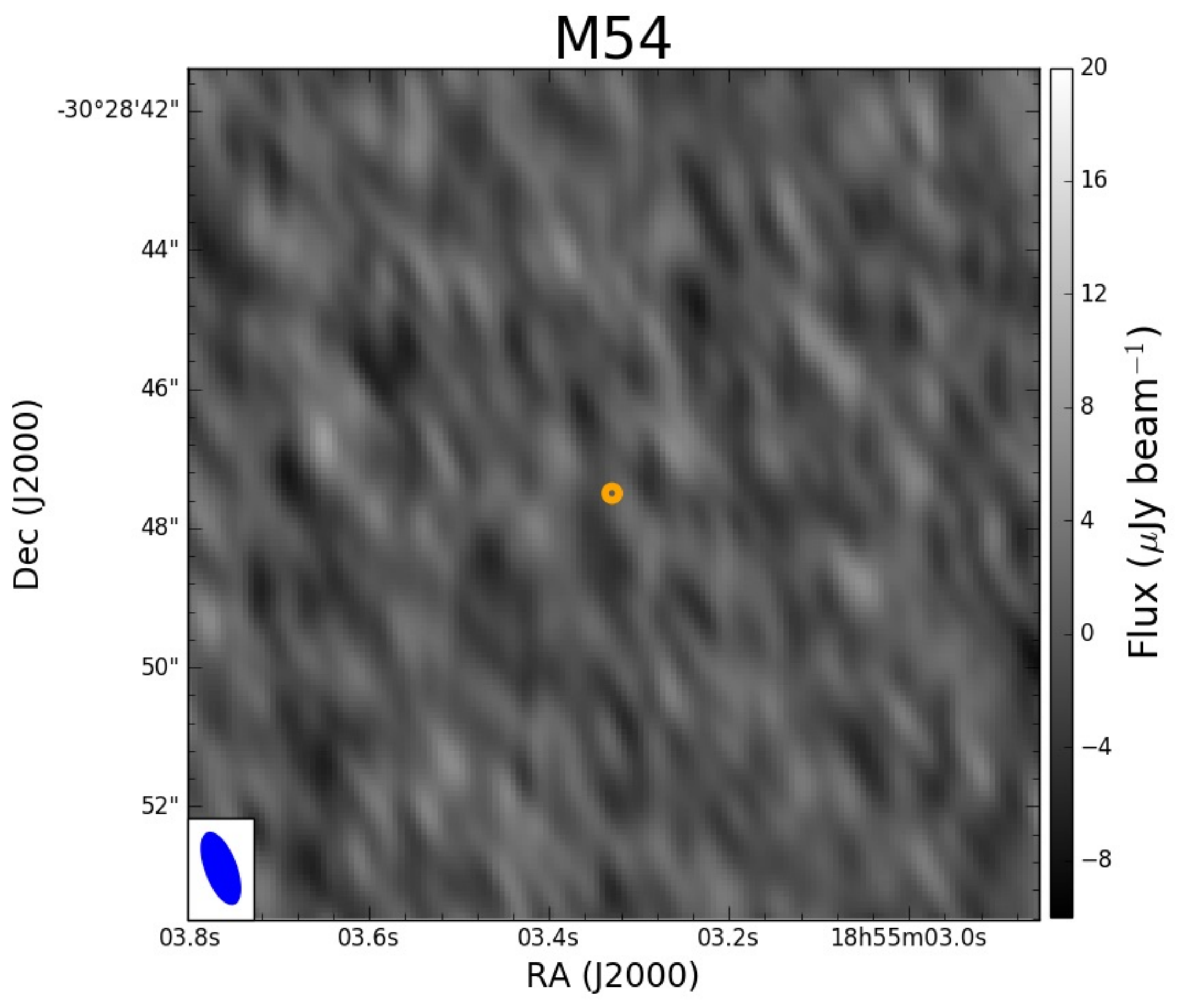} &
 \includegraphics[width=.33\textwidth]{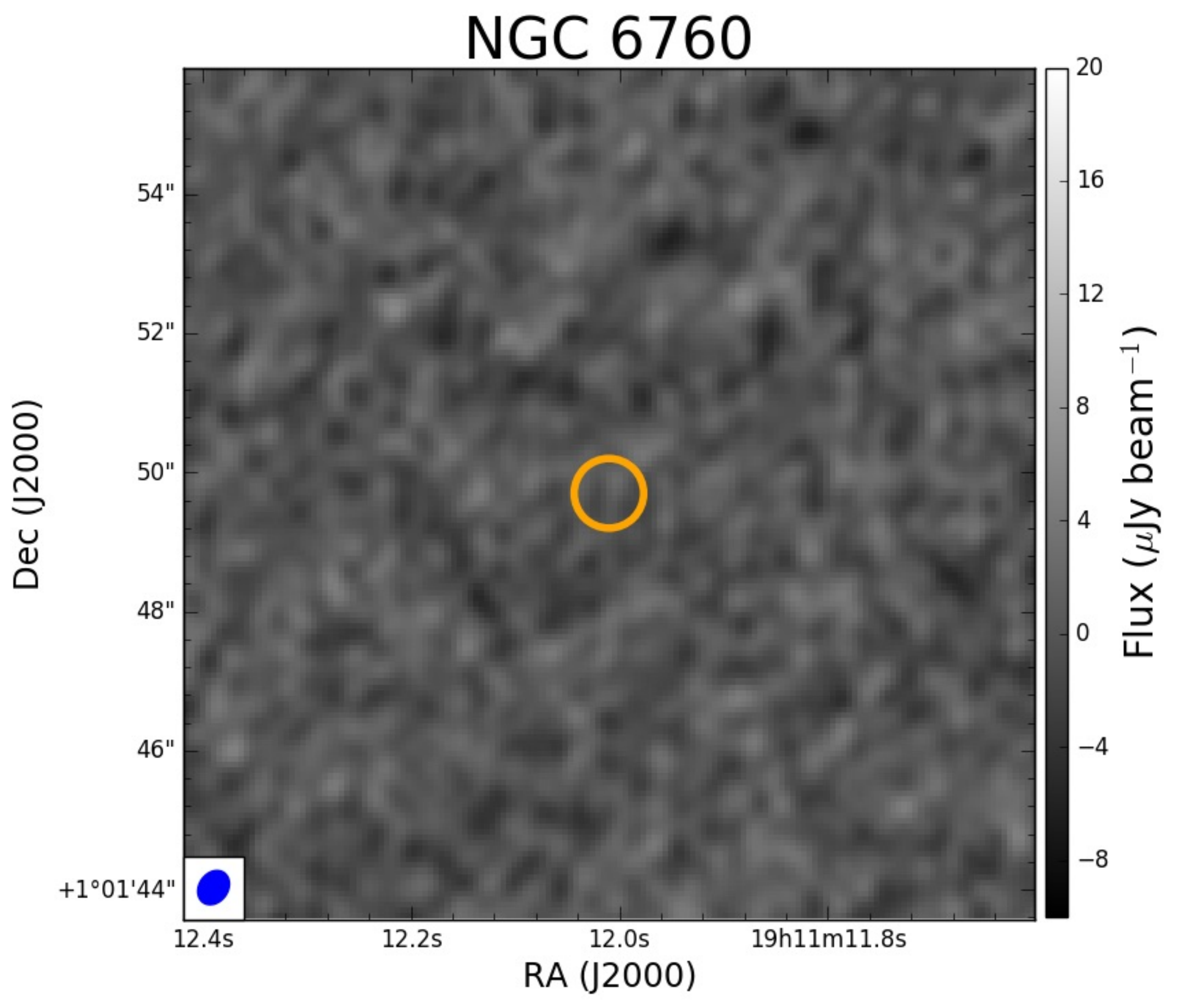}   
   \end{tabular}
 \end{figure*} 
 
\begin{figure}[htb]
  \begin{tabular}{@{}ccc@{}} 

 \includegraphics[width=.33\textwidth]{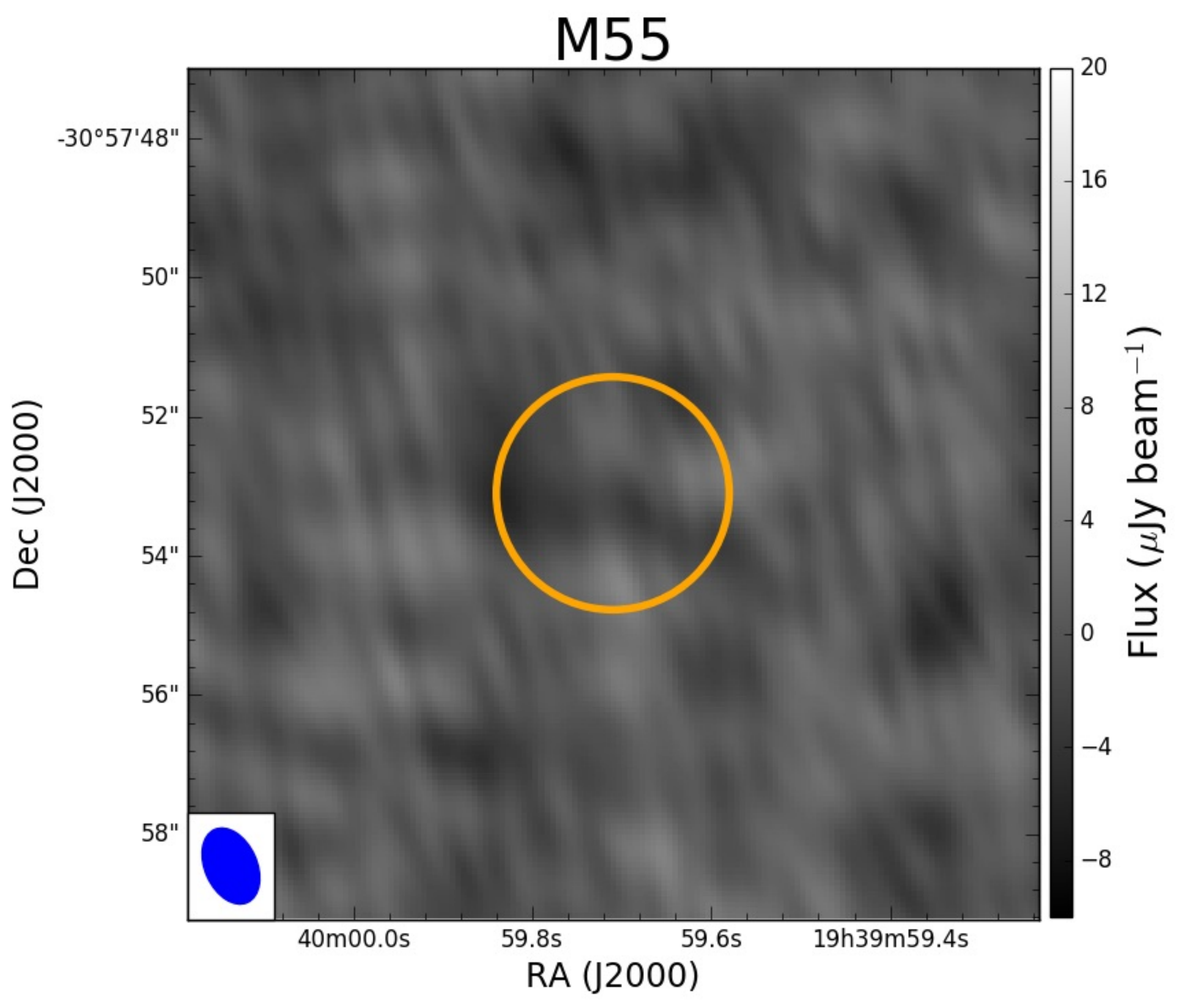} &
 \includegraphics[width=.33\textwidth]{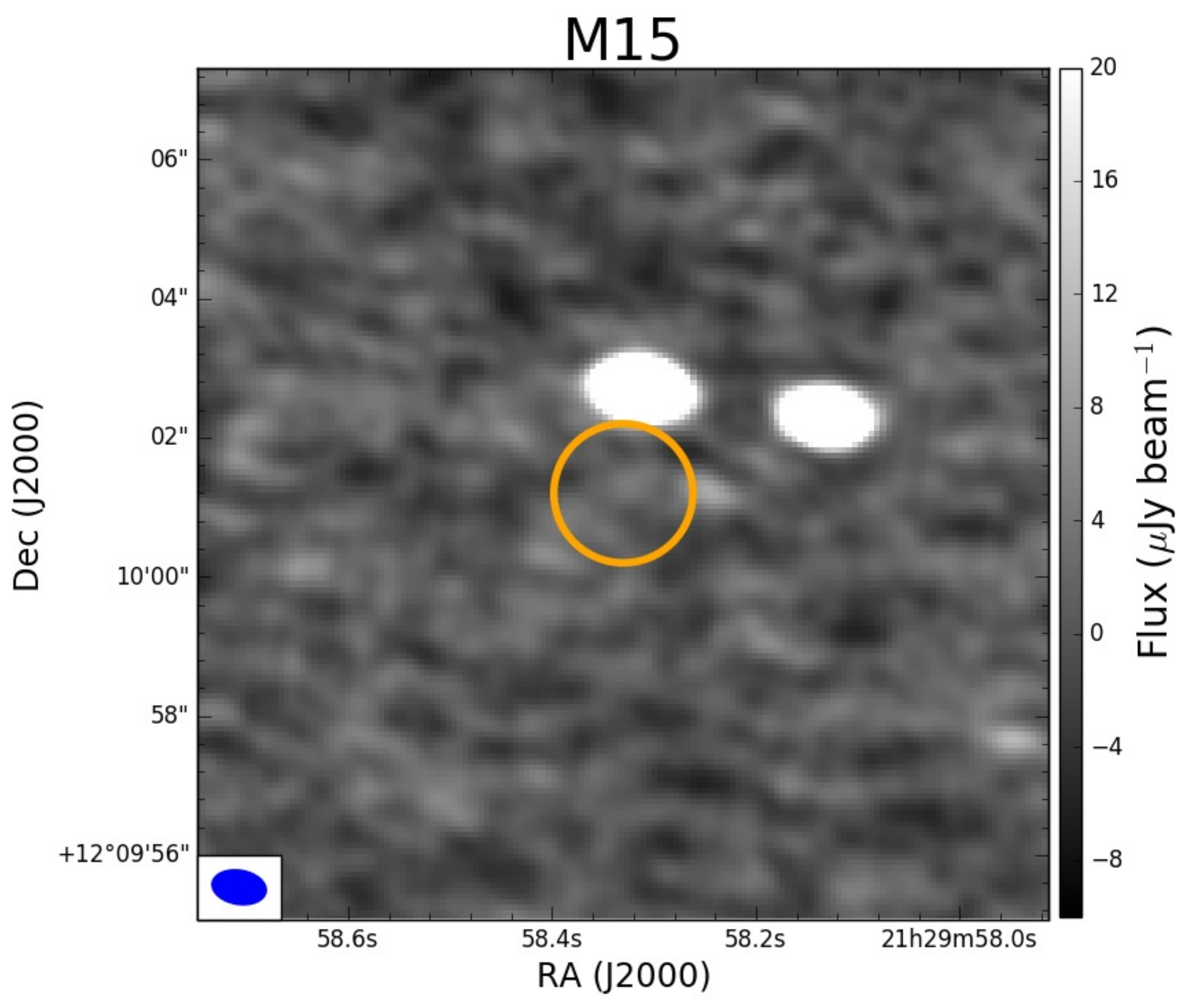} &
 \includegraphics[width=.33\textwidth]{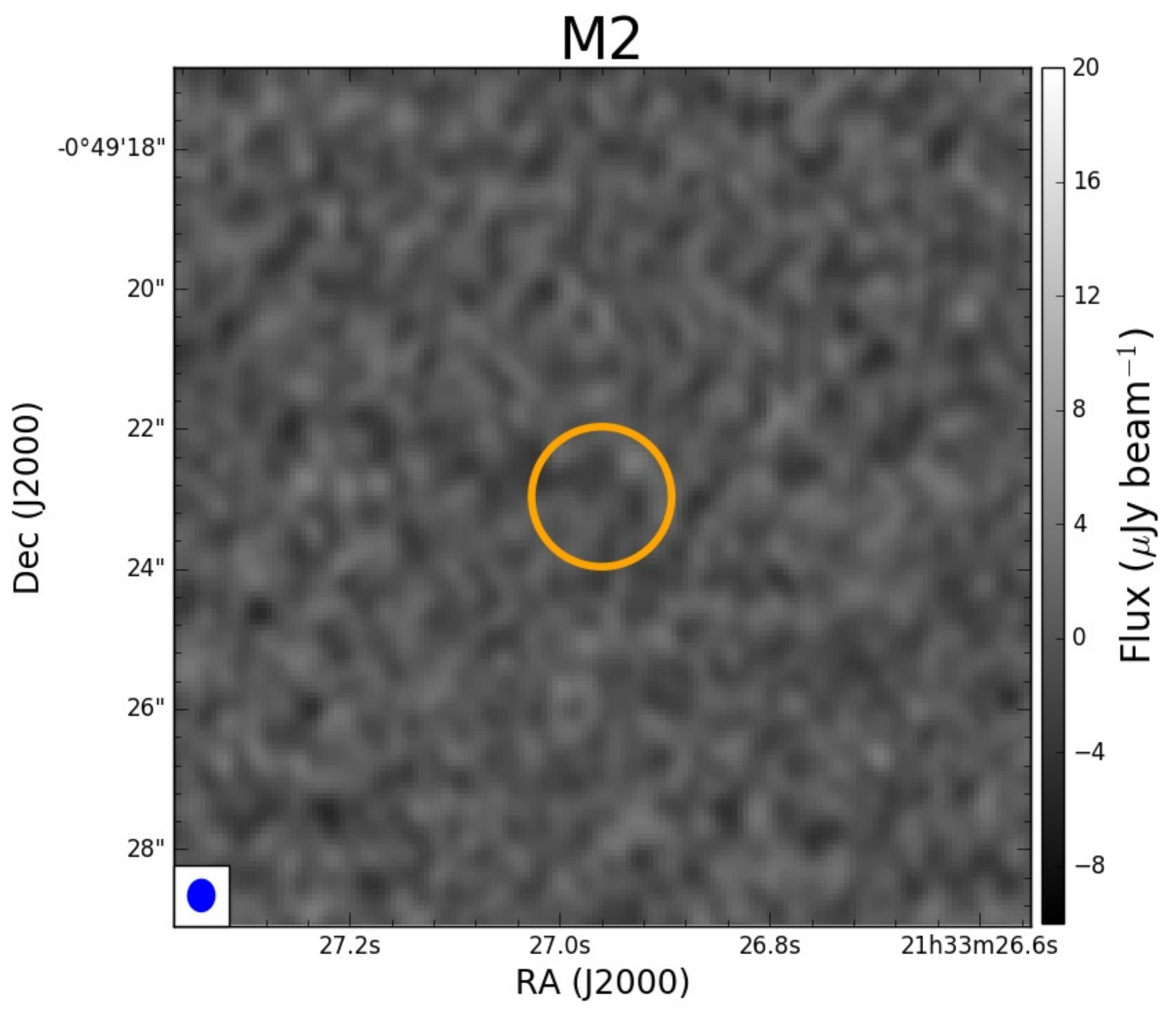} \\
 \includegraphics[width=.33\textwidth]{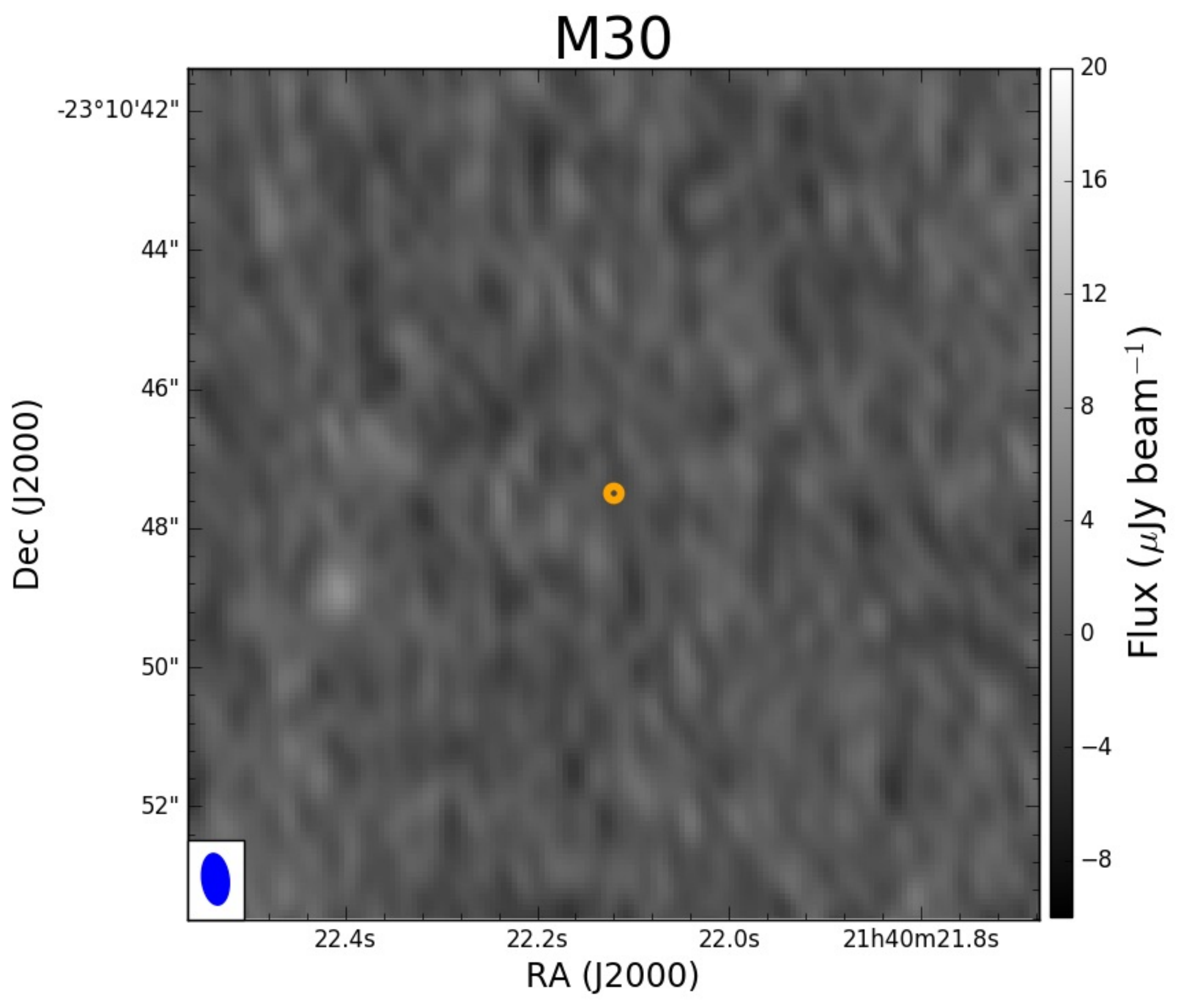}   
   \end{tabular}              
  \caption{VLA frequency-averaged images of the GCs listed in Table \ref{tab:vla}, showing the central $12.2^{\prime\prime} \times 12.2^{\prime\prime}$ for each cluster. The position of the cluster center is marked as an orange circle, and its size matched to the wander radius of a putative IMBH. Synthesized beams are shown in blue in the bottom-left corner of all images. \label{im:vla}}
\end{figure}

\begin{figure*}[htb]
\centering
  \begin{tabular}{@{}ccc@{}}
    \includegraphics[width=.33\textwidth]{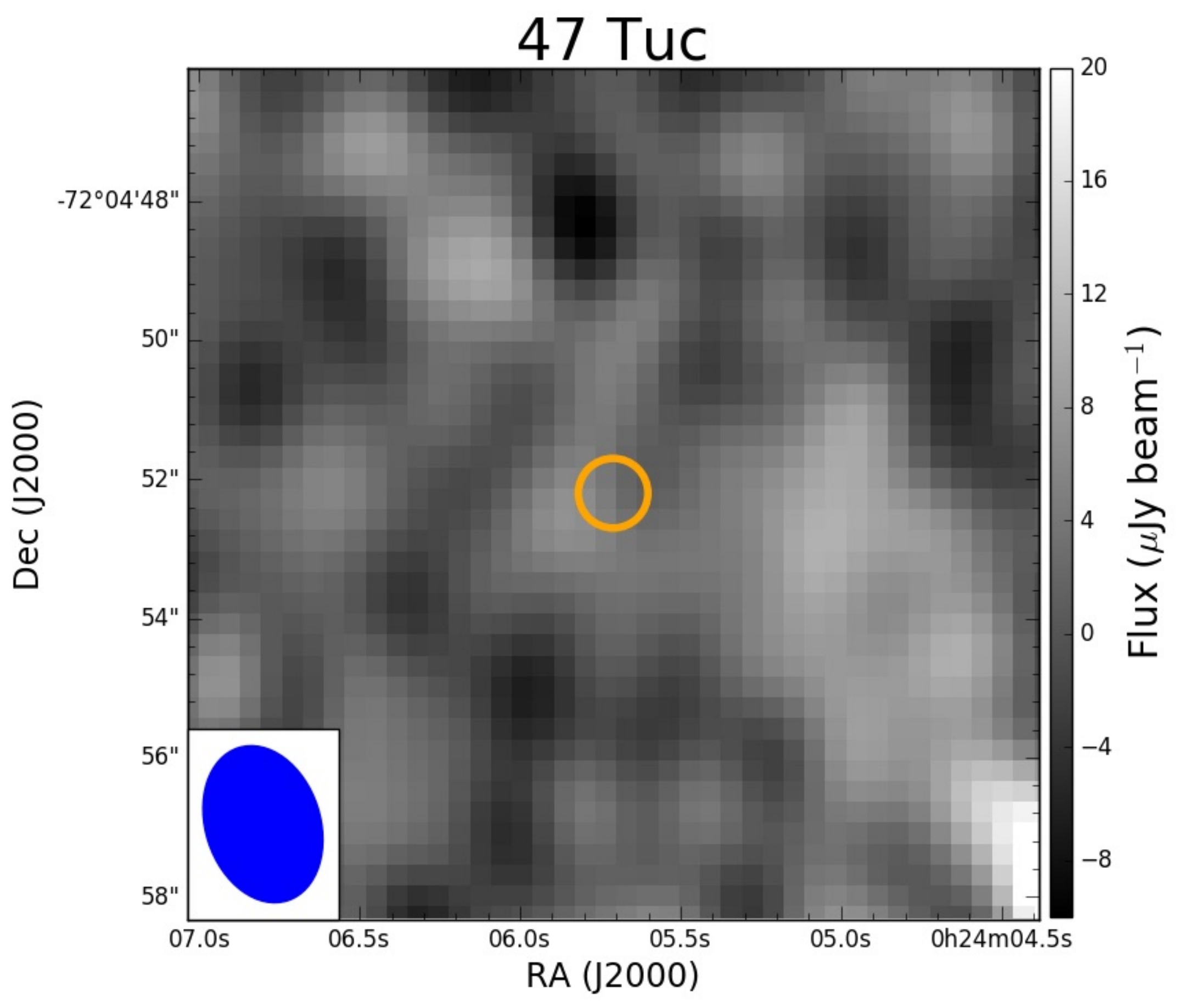} &
    \includegraphics[width=.33\textwidth]{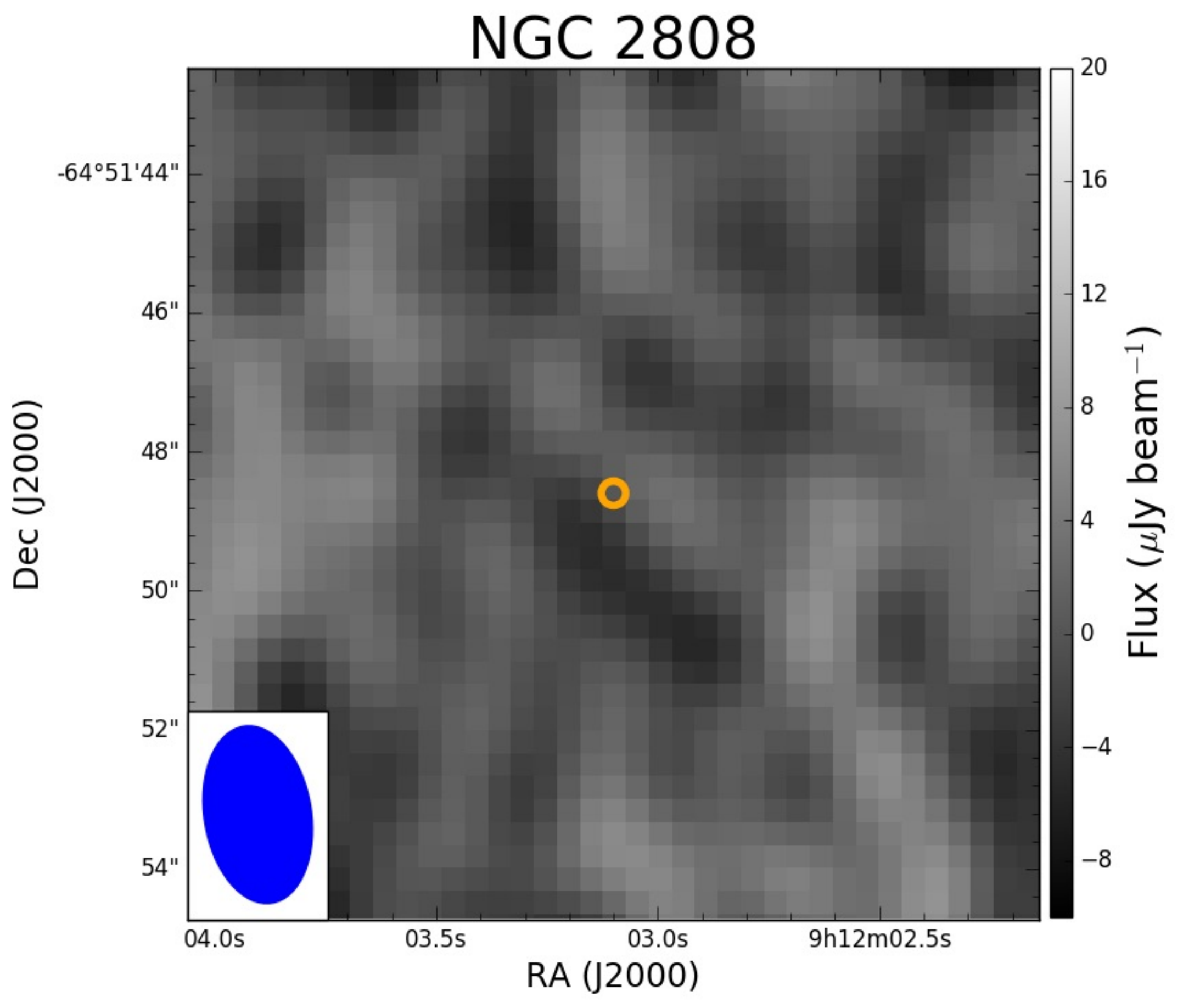} &
    \includegraphics[width=.33\textwidth]{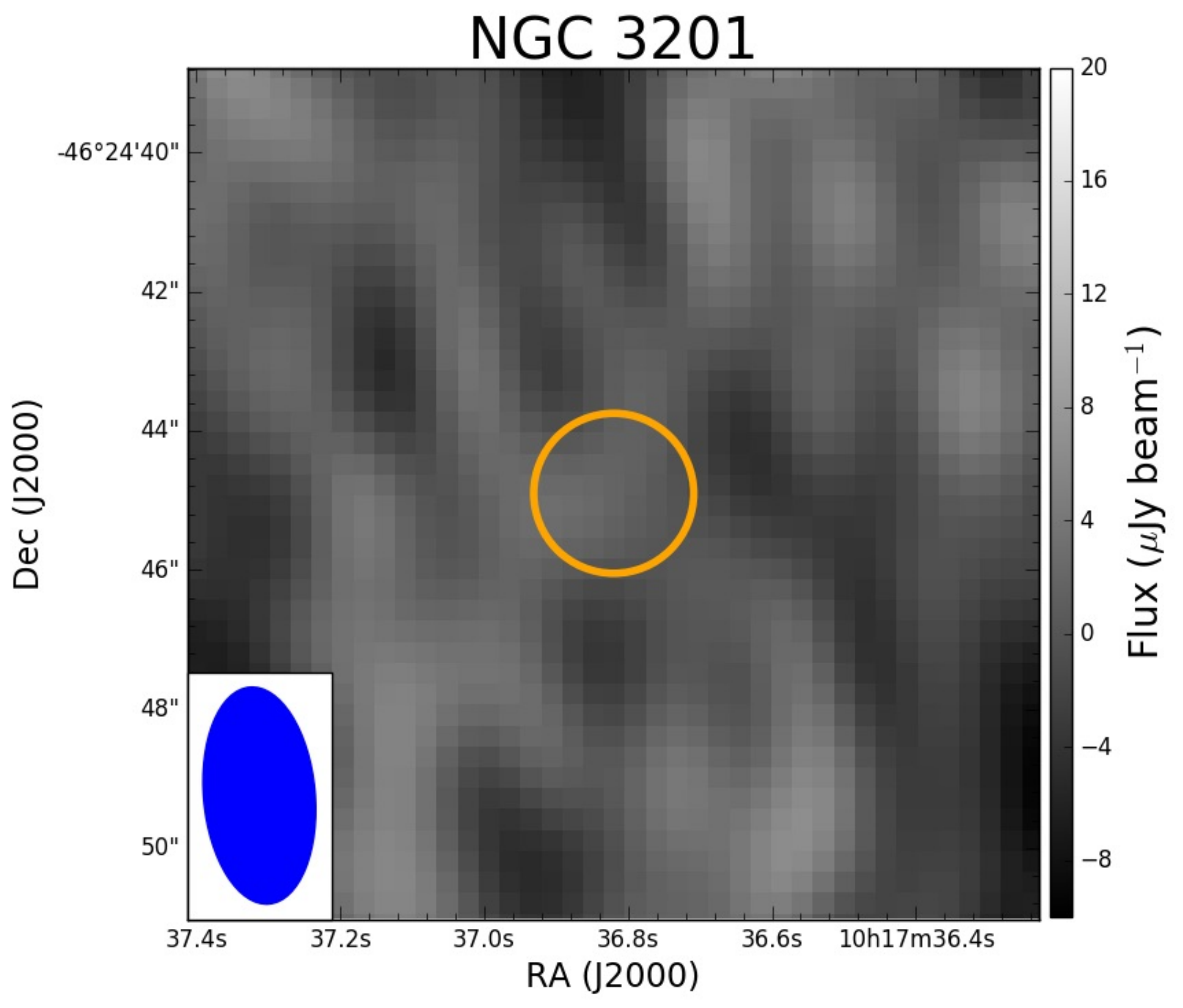} \\
    \includegraphics[width=.33\textwidth]{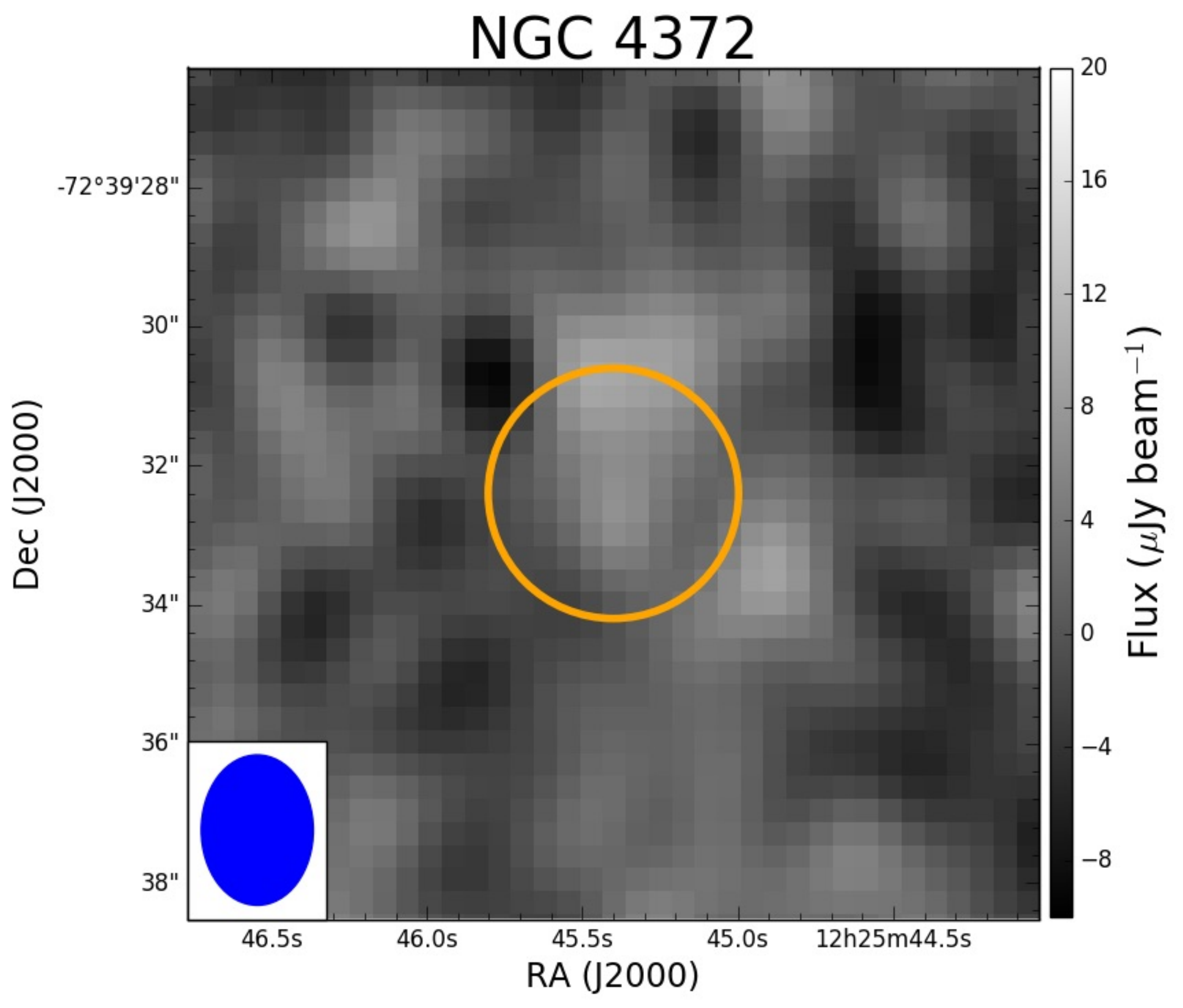}   &
    \includegraphics[width=.33\textwidth]{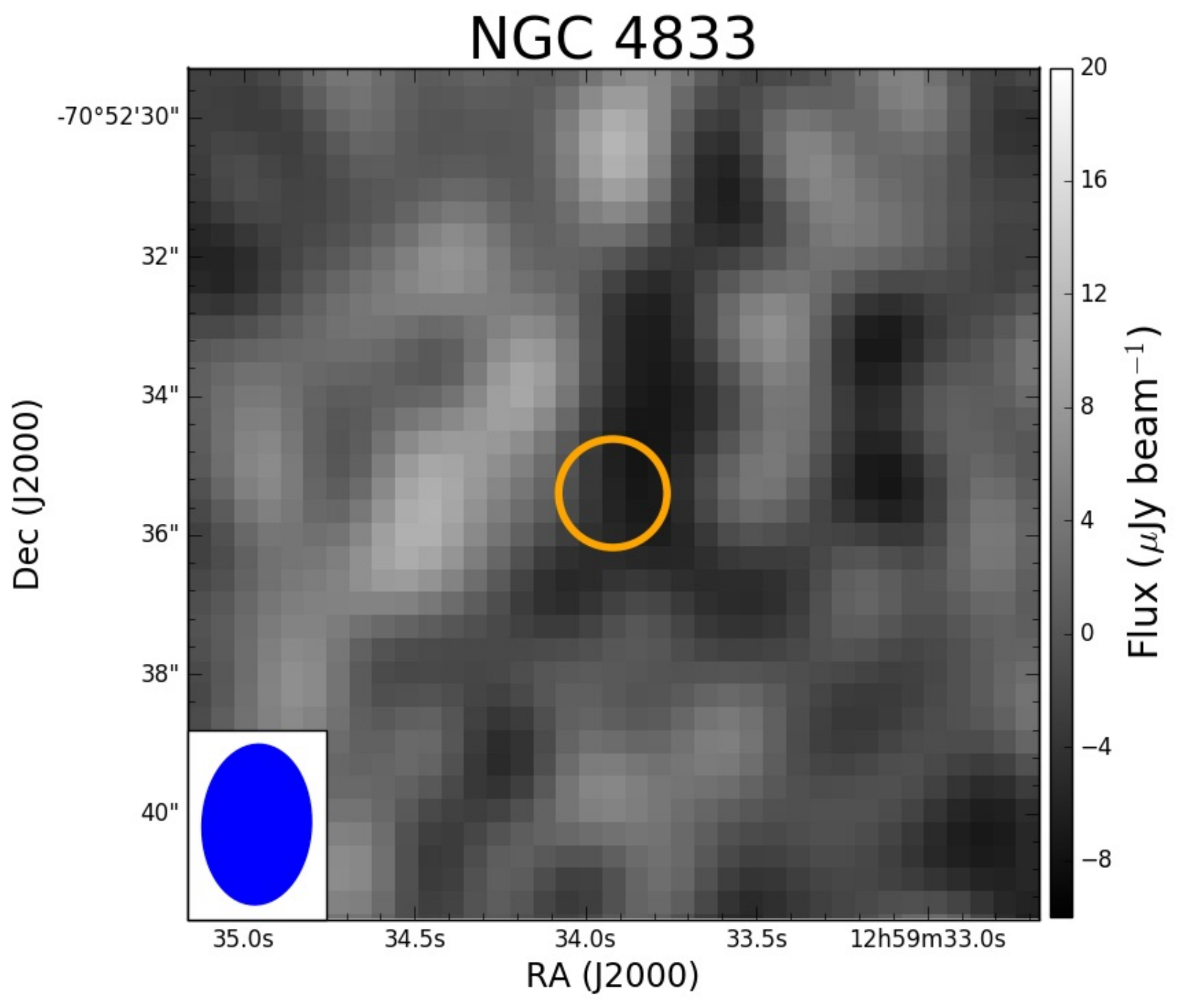} &
    \includegraphics[width=.33\textwidth]{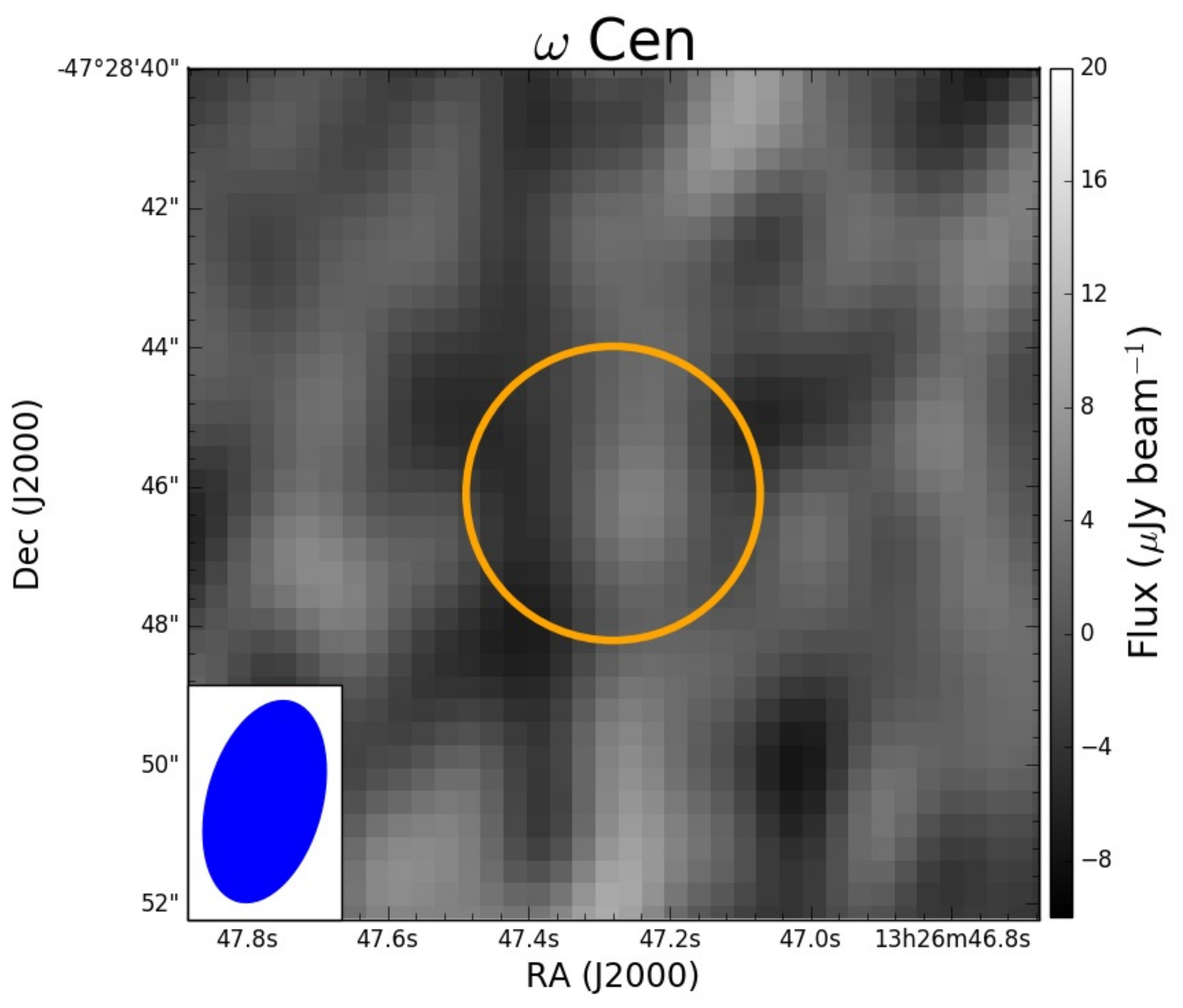} \\
    \includegraphics[width=.33\textwidth]{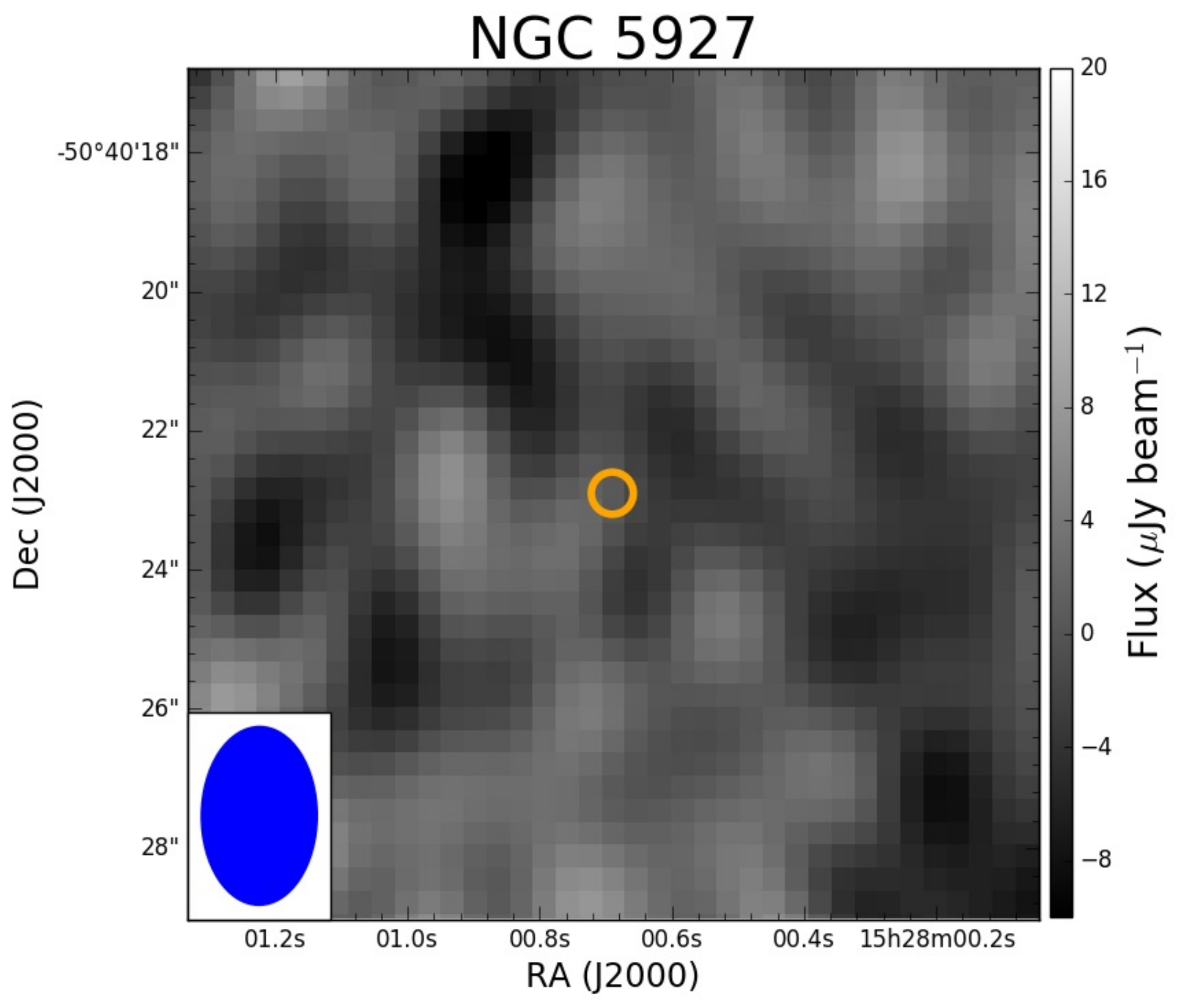} &
    \includegraphics[width=.33\textwidth]{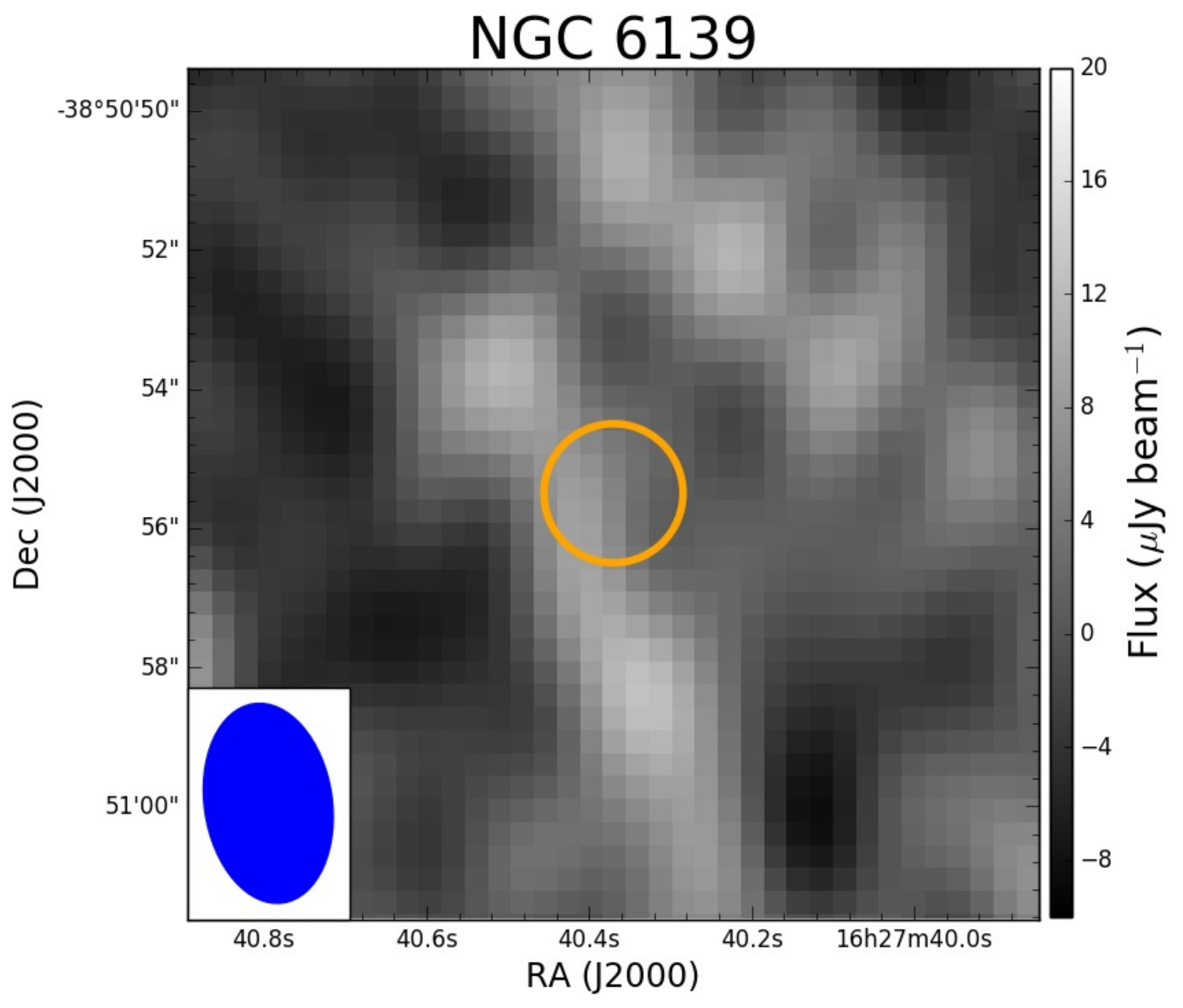}   &
    \includegraphics[width=.33\textwidth]{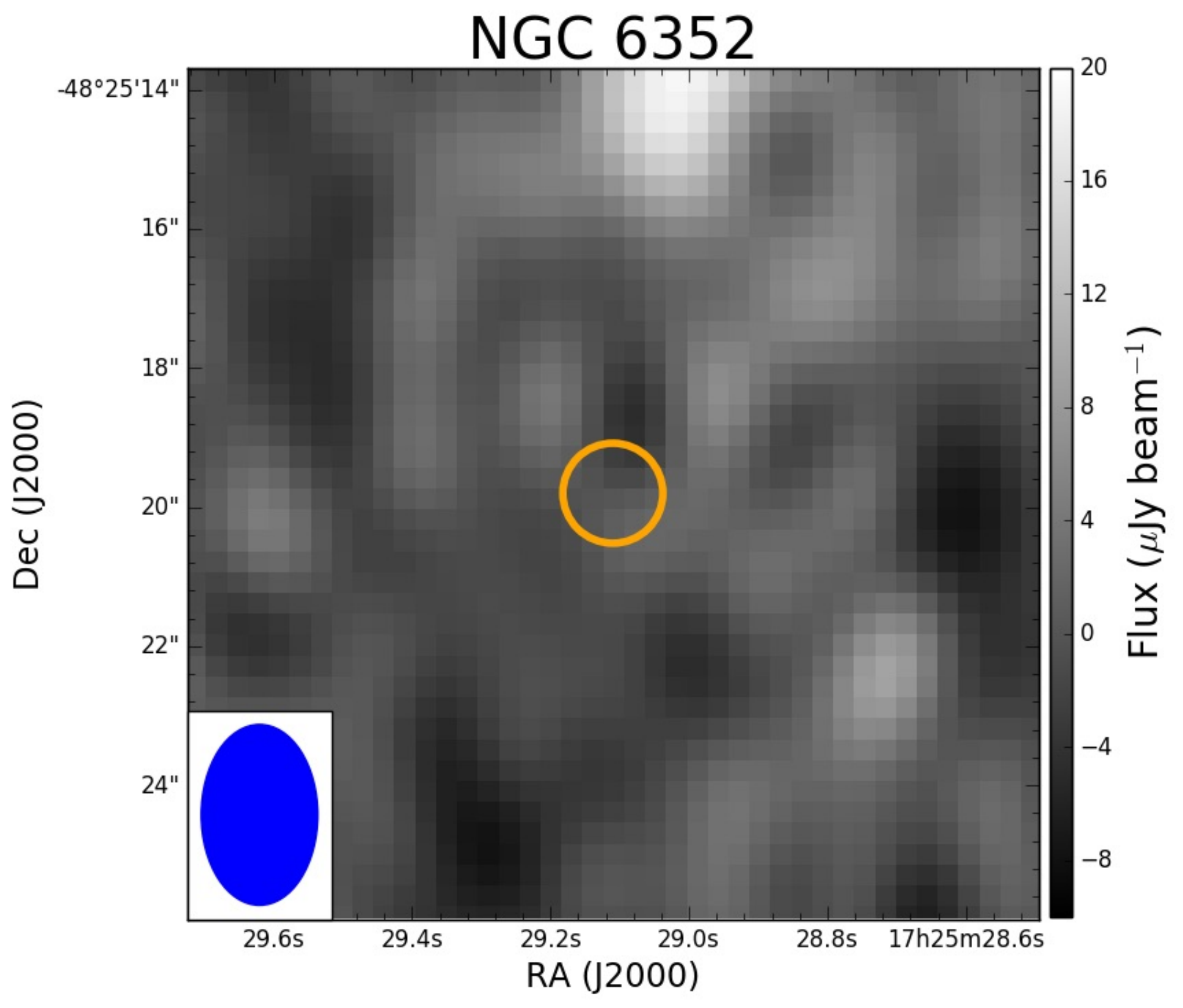} \\
    \includegraphics[width=.33\textwidth]{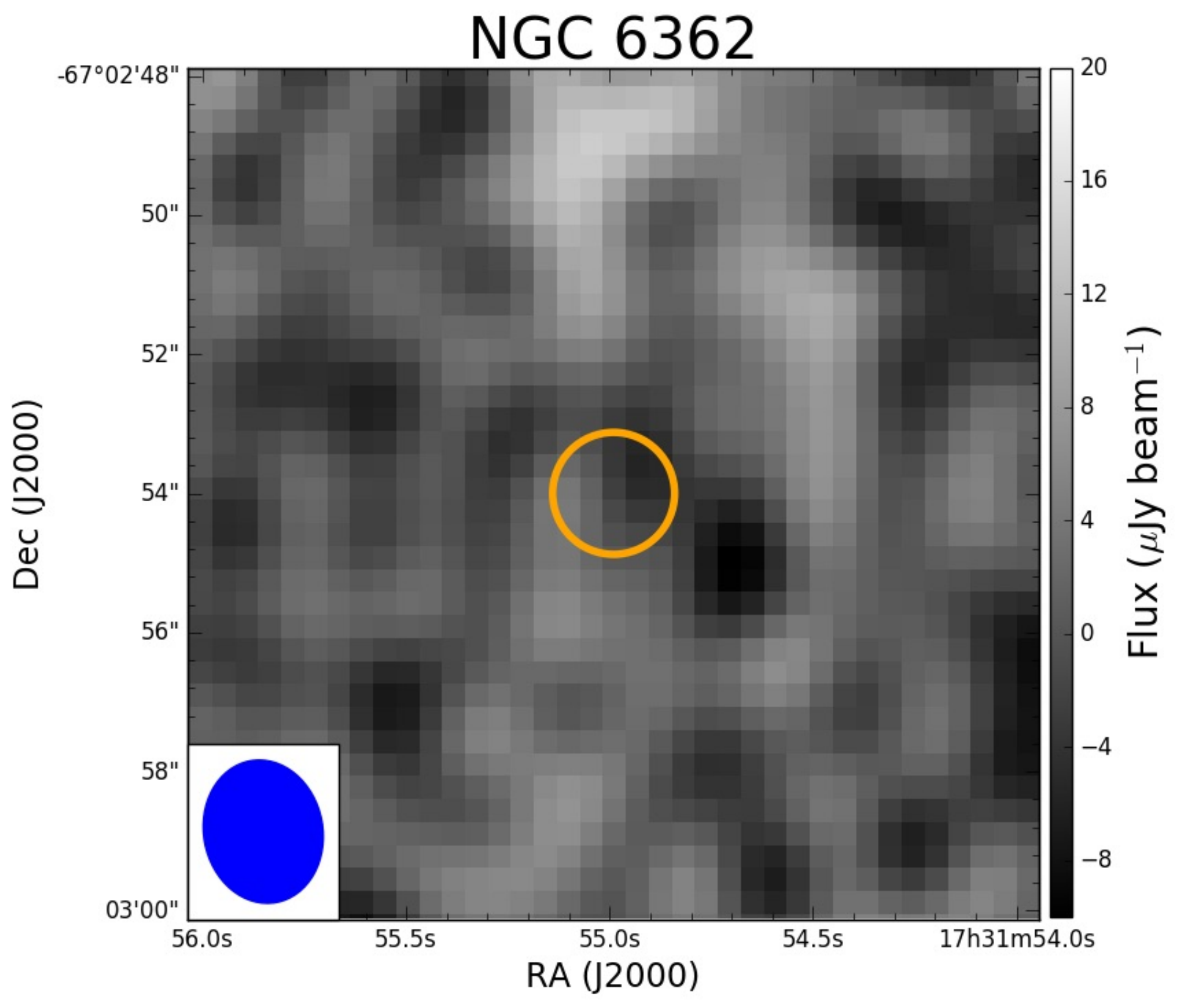} &
    \includegraphics[width=.33\textwidth]{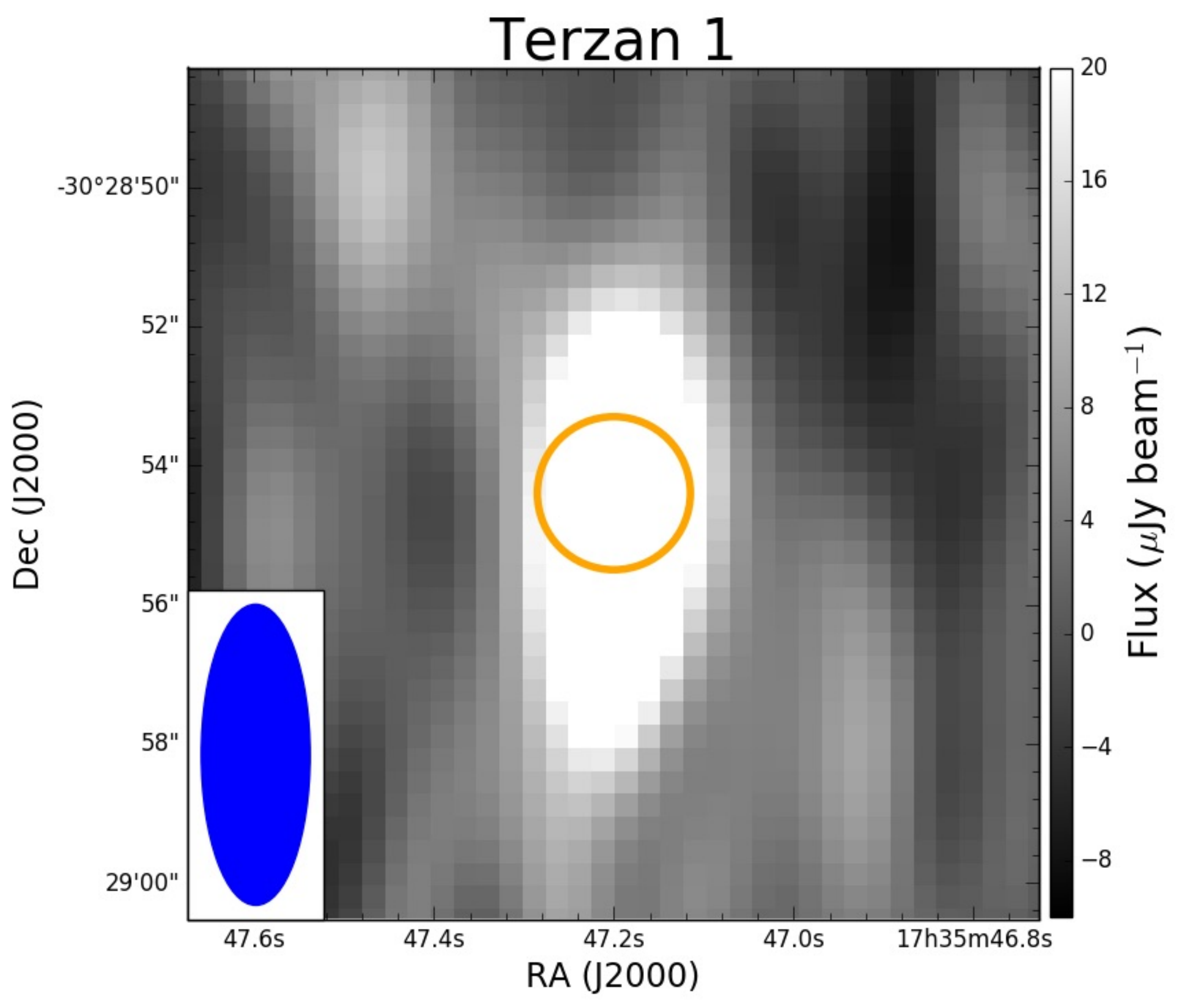} &
    \includegraphics[width=.33\textwidth]{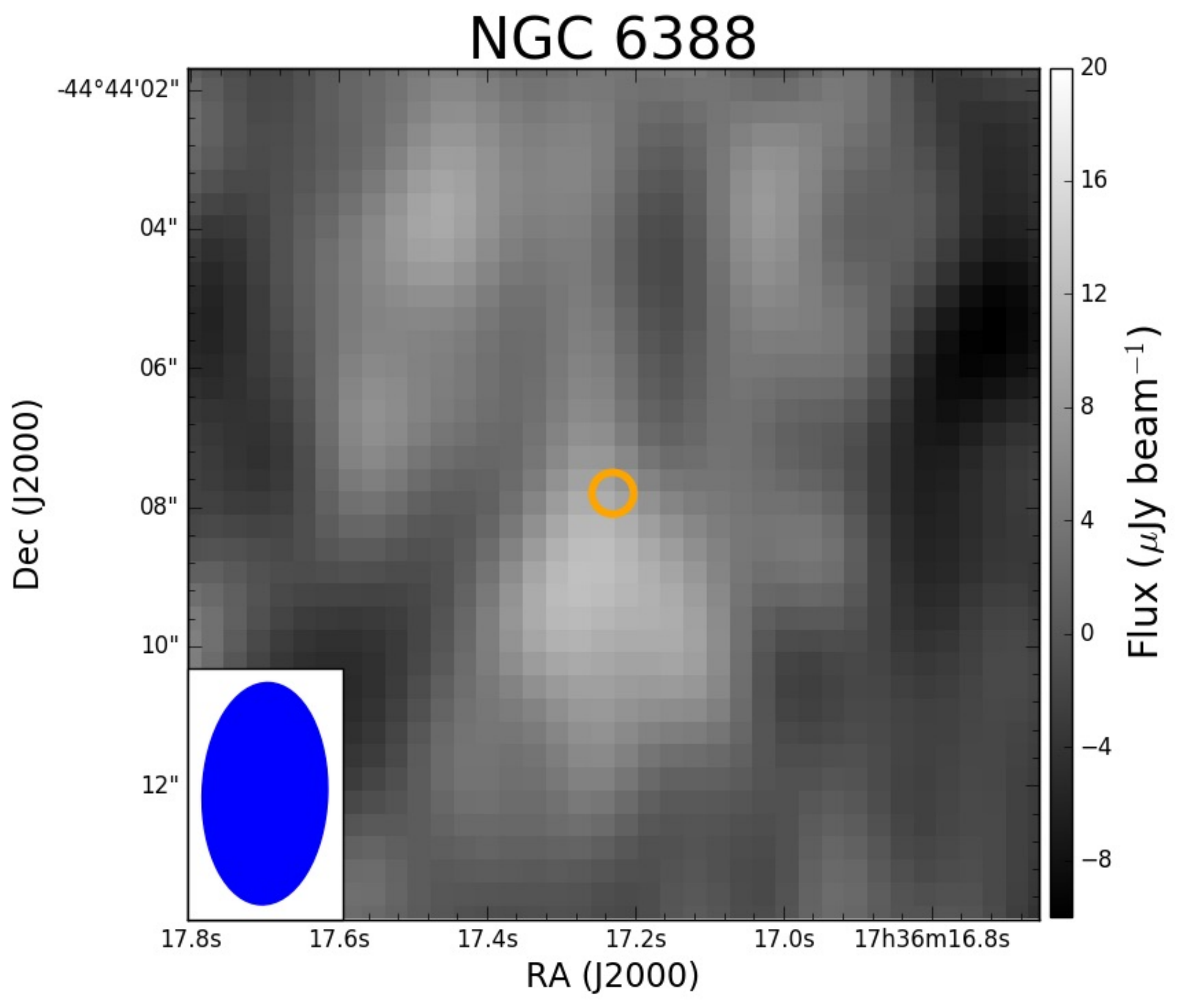} 
 
 \end{tabular}
 \end{figure*} 

 \begin{figure}[htb]
  \begin{tabular}{@{}ccc@{}}
     \includegraphics[width=.33\textwidth]{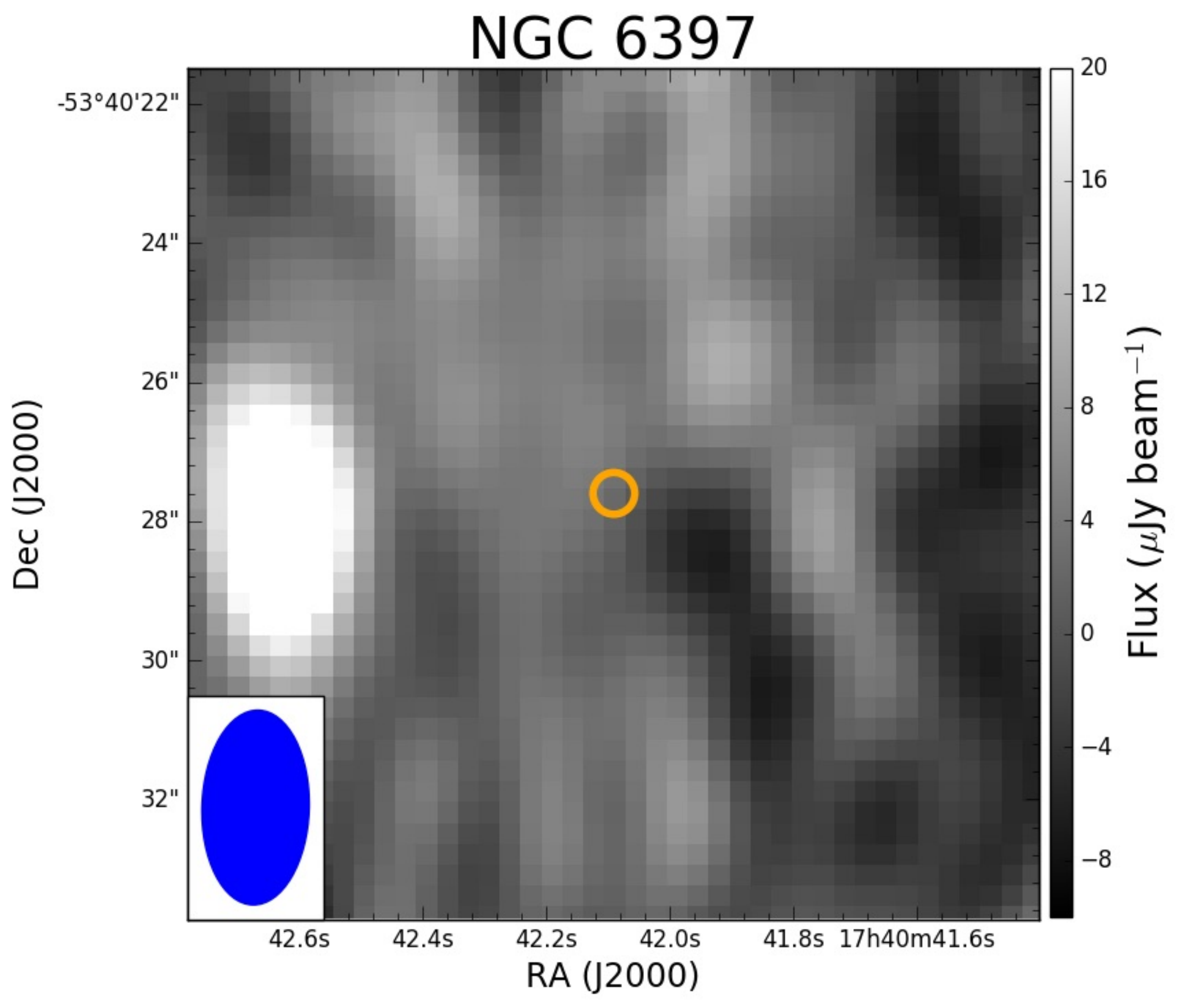} &
    \includegraphics[width=.33\textwidth]{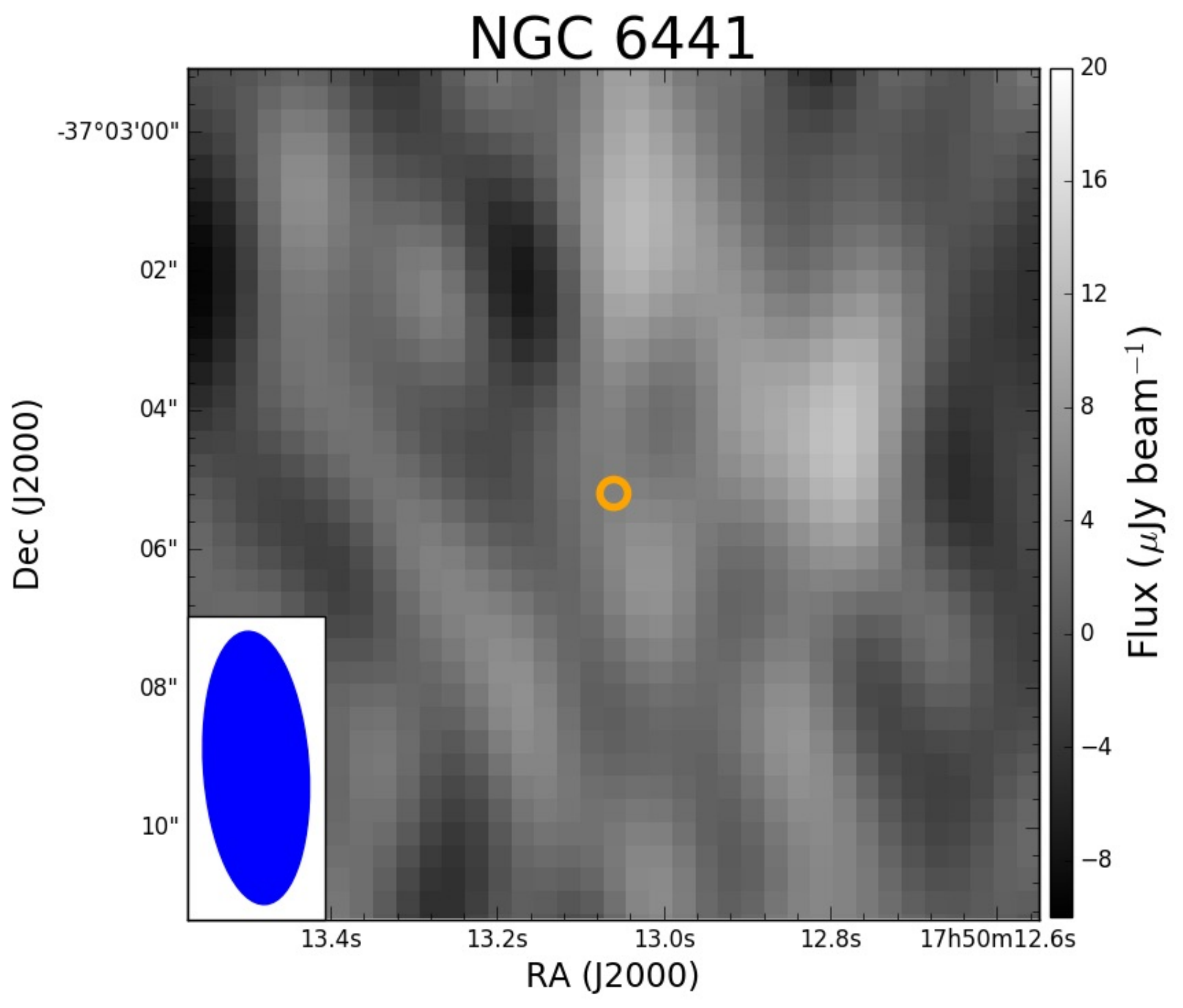} &
    \includegraphics[width=.33\textwidth]{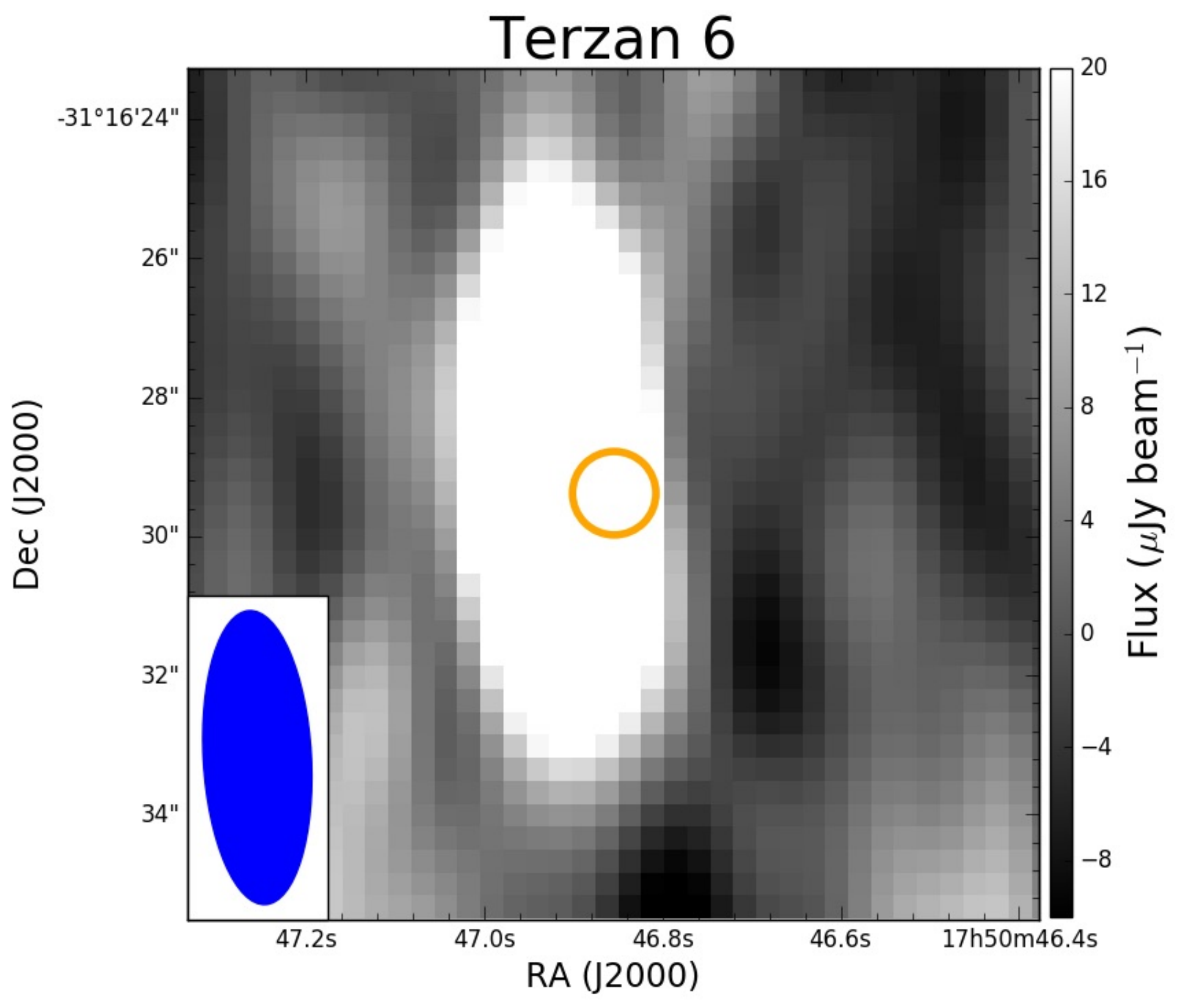} \\  
   \includegraphics[width=.33\textwidth]{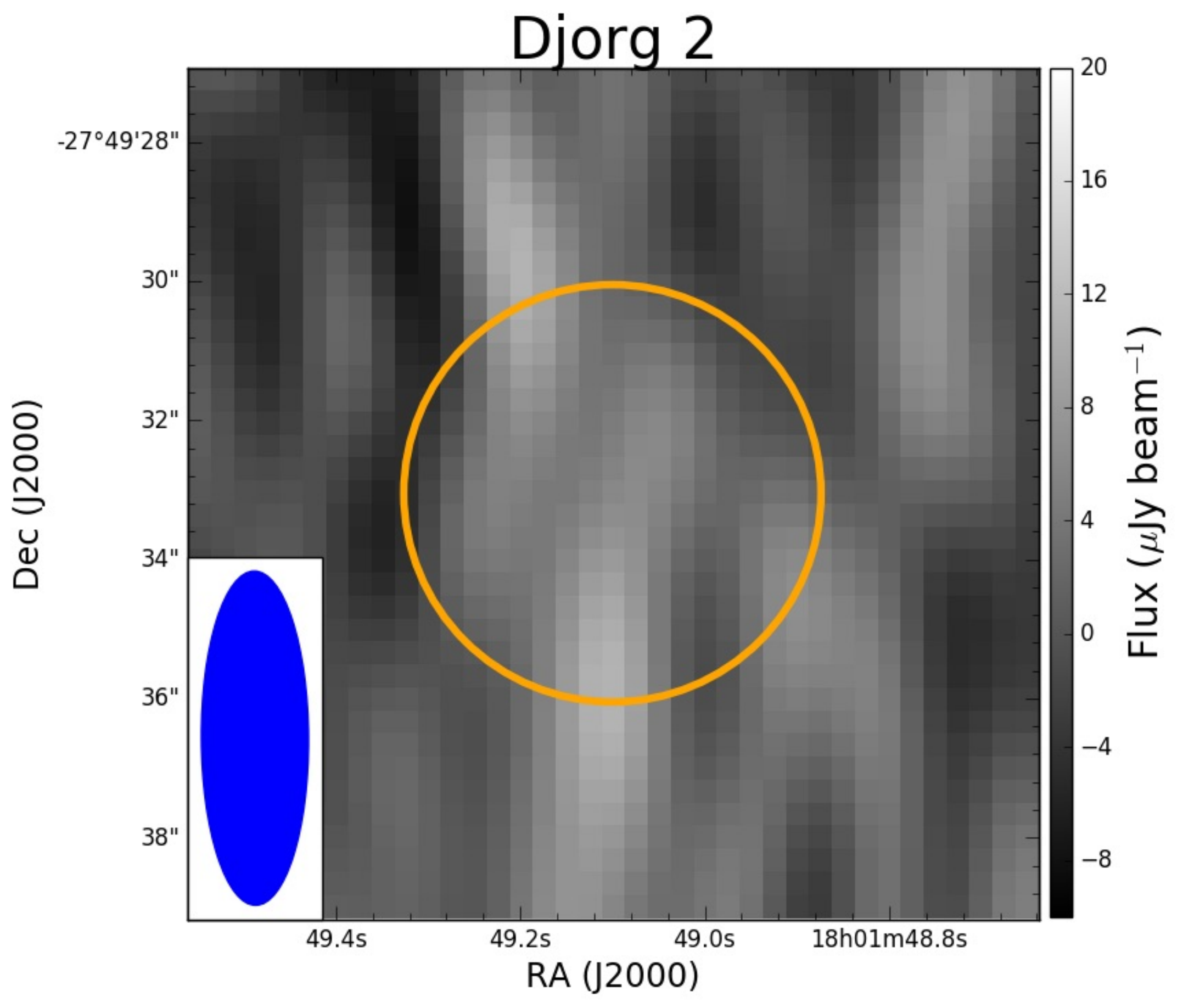} &
    \includegraphics[width=.33\textwidth]{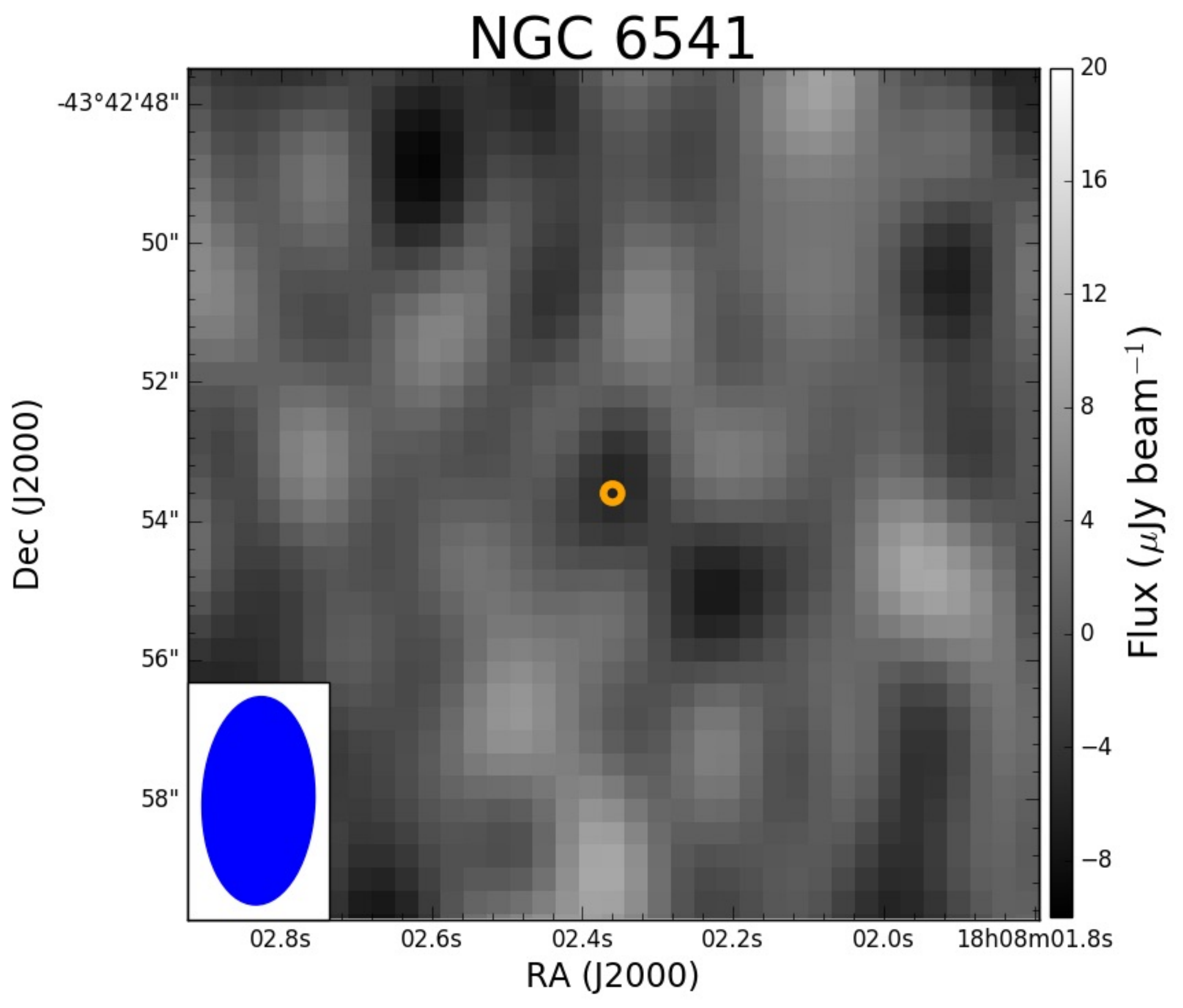} &
    \includegraphics[width=.33\textwidth]{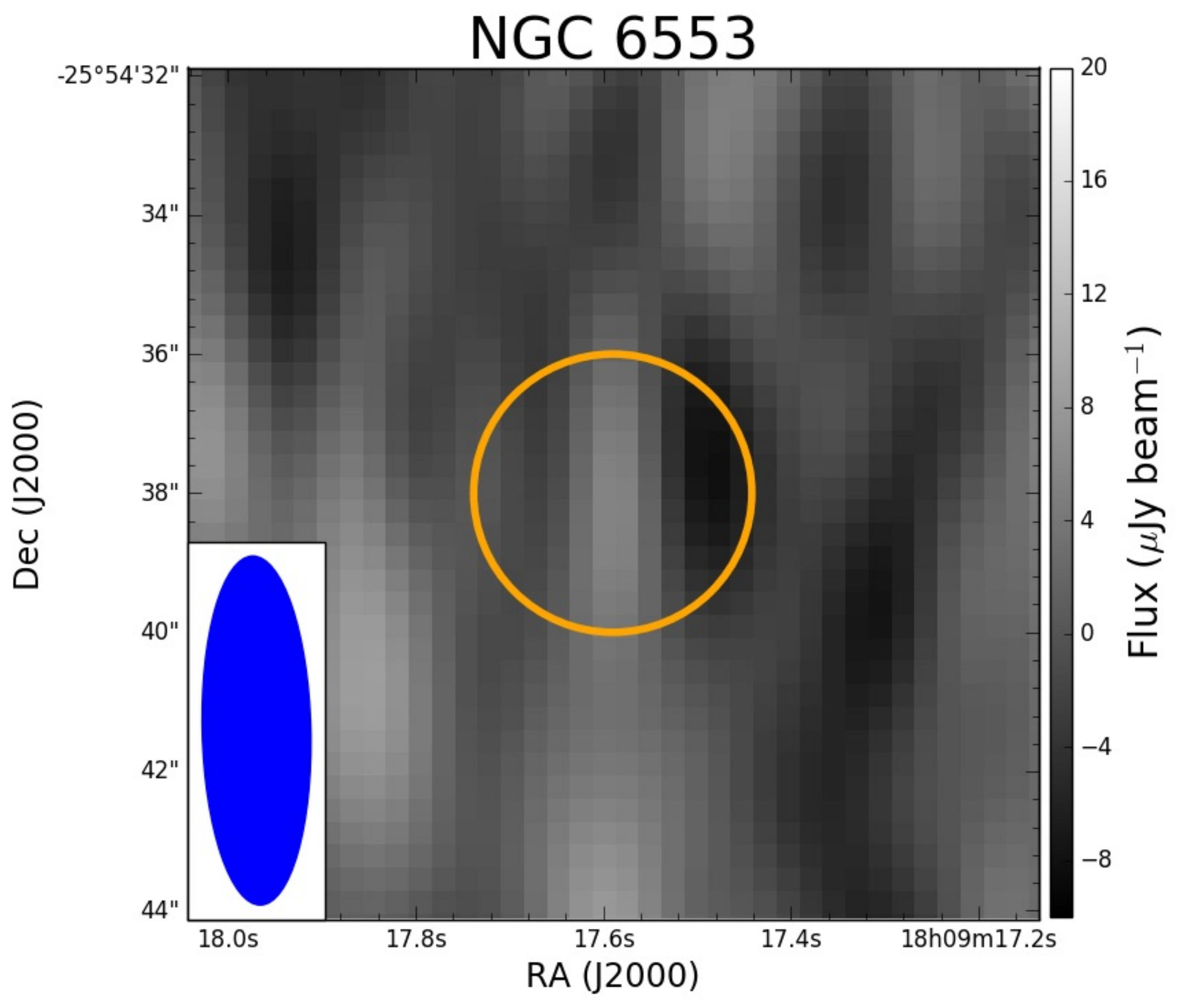}\\
 \includegraphics[width=.33\textwidth]{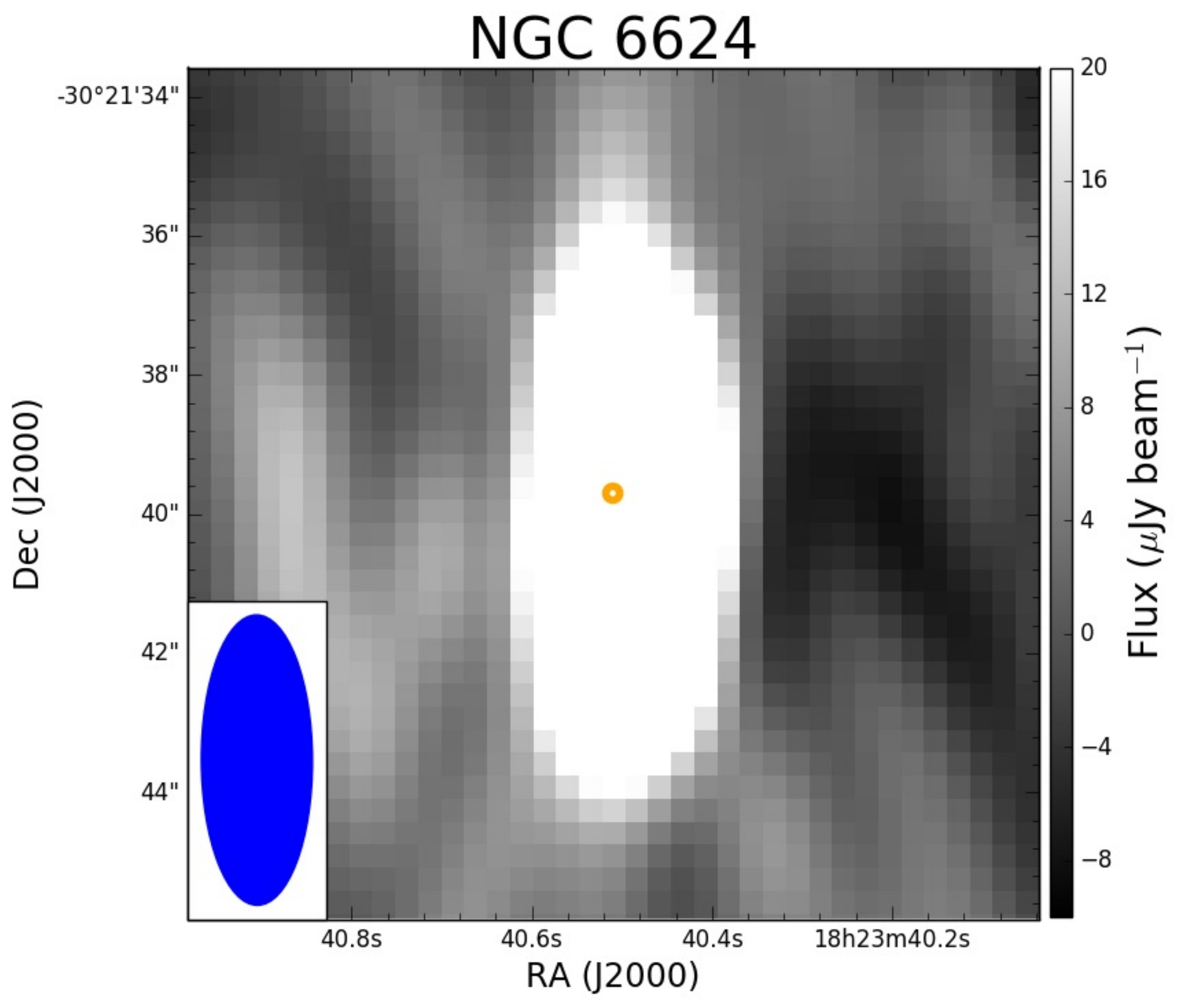} &
    \includegraphics[width=.33\textwidth]{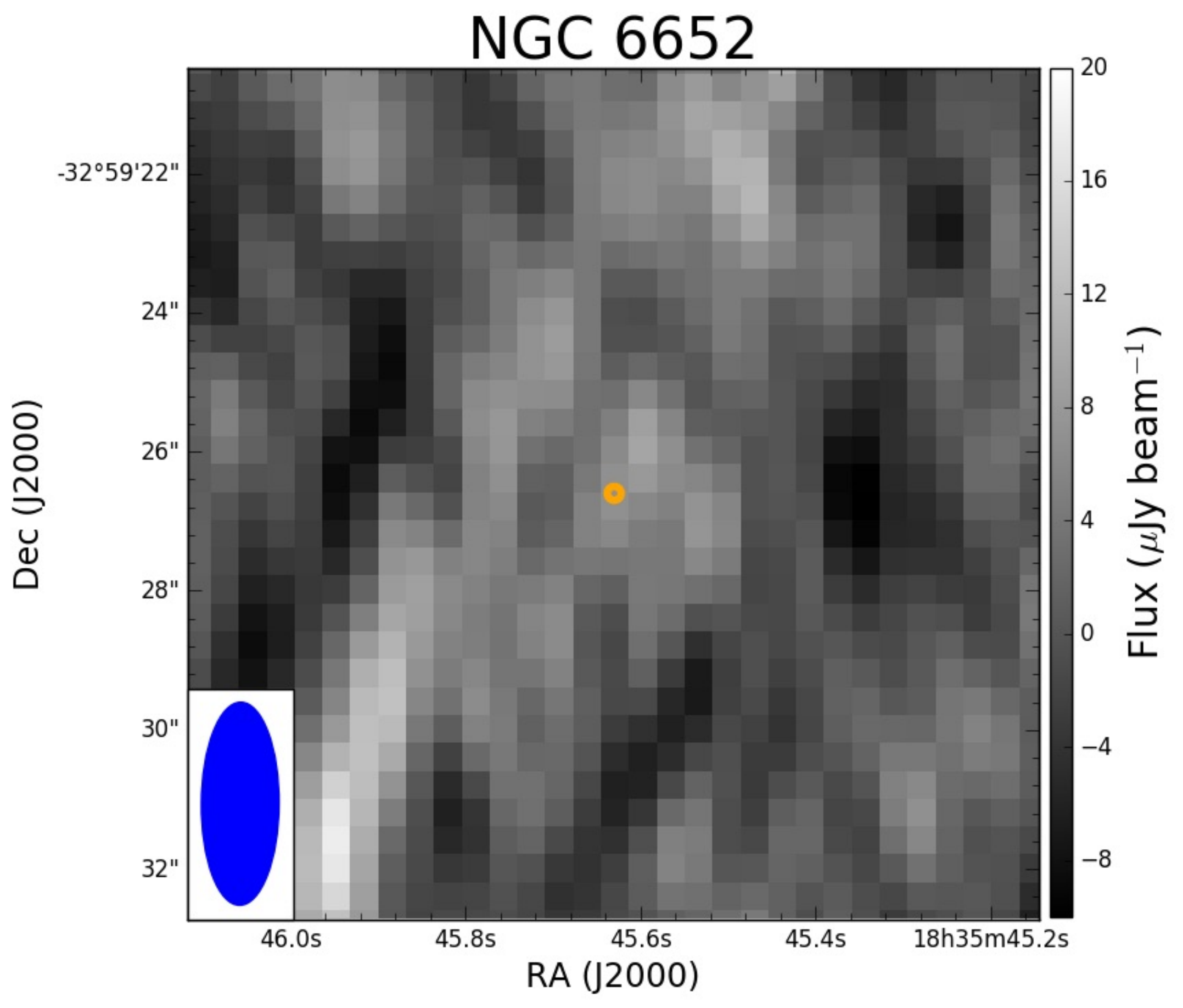} &
    \includegraphics[width=.33\textwidth]{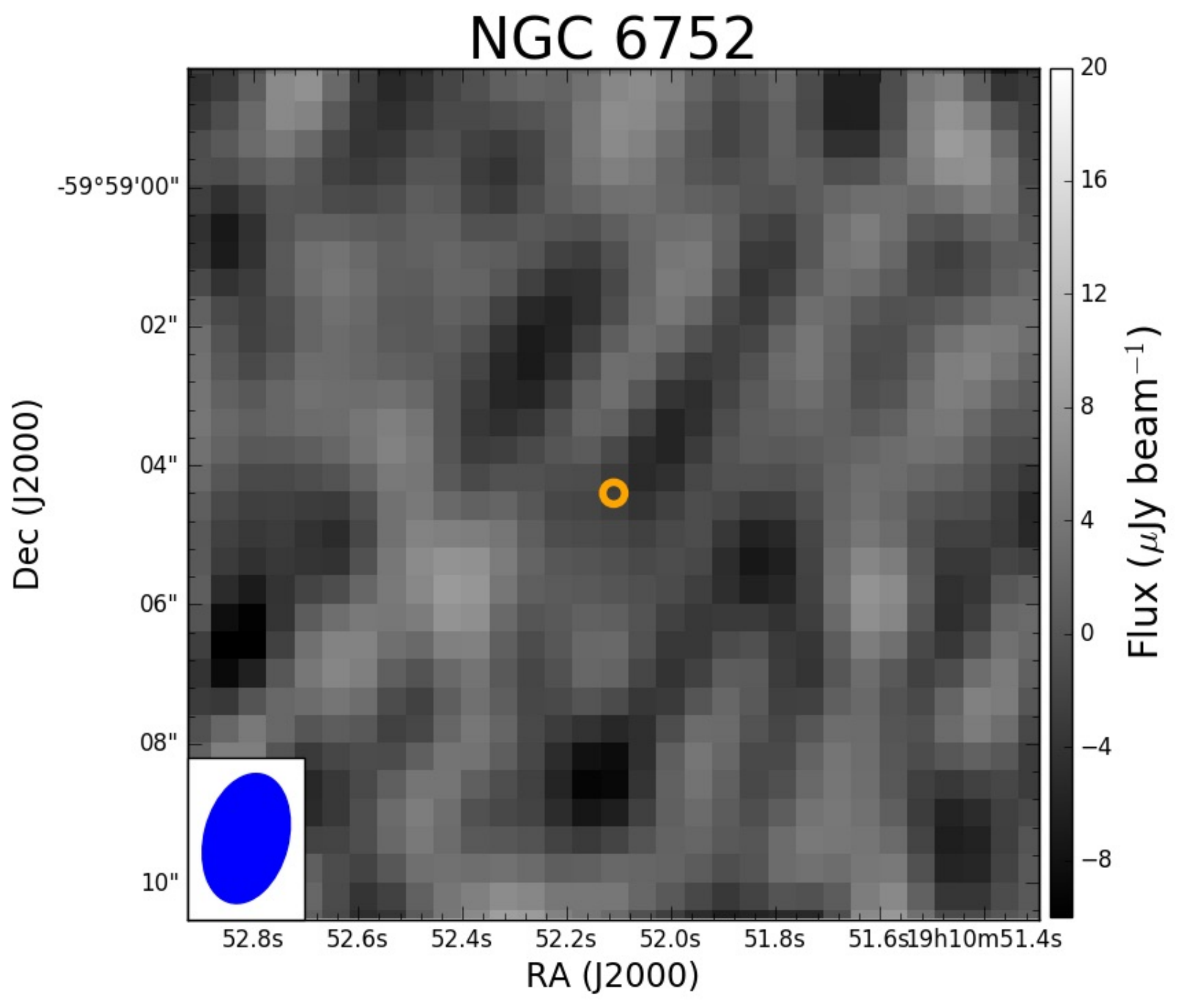} 
    \end{tabular}              
  \caption{ATCA frequency-averaged images of the GCs listed in Table \ref{tab:atca}. See Figure \ref{im:vla} for more details. \label{im:atca}}
\end{figure}

\begin{longrotatetable}
\begin{deluxetable}{lccccccccccccc}
\tabletypesize{\scriptsize}
\tablecolumns{13} 
\tablewidth{0pt} 
\rotate
\tablecaption{Results from VLA Data. We note that we assume 5 GHz as central frequency in order to calculate the radio luminosities and the mass upper limits. The sources used for the photometric centers and the distances are shown at the bottom of the Table.  The core radius of Liller 1 is adopted from \citet{2015saracino} while we use \cite{2010harris} for the rest of the clusters.    \label{tab:vla}}
\tablehead{
\colhead{ID}           & 
\colhead{RA (J2000)}          &
\colhead{DEC (J2000)}          & 
\colhead{Uncertainty}  &
\colhead{Position} &
\colhead{Distance} &
\colhead{Distance} &
\colhead{3$\sigma$ Flux  } & 
\colhead{ 3$\sigma$ $L_{R}$ } & 
\colhead{IMBH} &
\colhead{ 3$\sigma$ $L_{X}$ } &
\colhead{IMBH mass} &
\colhead{Core} &
\colhead{Brownian} 
 \\
\colhead{} &
\colhead{(h:m:s)} &
\colhead{(\phn{\arcdeg}~\phn{\arcmin}~\phn{\arcsec})} &
\colhead{(\phn{\arcsec})} & 
\colhead{reference}&
\colhead{(kpc)}& 
\colhead{reference} &
\colhead{density ($\mu$Jy)}&
\colhead{(erg/s)}&
\colhead{mass (M$_{\odot}$)}& 
\colhead{(erg/s)}& 
\colhead{fraction}& 
\colhead{radius (\phn{\arcsec})}&
\colhead{radius (\phn{\arcsec})}
}
\startdata                                                                                                                                                                                                                    
\textbf{M3}         				& 13 42 11.38   &28 22 39.1   & 1.0     &2 	&  10.1         & 4    	&$<$5.8           					&$<$3.5 $\times10^{27}$     		&$<$1460		&$<$7.8 $\times10^{30}$		&$<$0.37		& 22.2          &	0.28  		\\
\textbf{M5}         				& 15 18 33.21  & 02 04 51.8 & 0.2     	&2 	&  7.7      & 1    	&$<$4.2            						&$<$1.5 $\times10^{27}$				&$<$1060		&$<$3.0 $\times10^{30}$		&$<$0.28		& 26.4          &	0.38  		\\
\textbf{M4 }        				& 16 23 35.03   &-26 31 33.8 & 1.0      &2 	&  1.8          & 30    &$<$5.3    							&$<$1.0 $\times10^{26}$				&$<$390 		&$<$1.5 $\times10^{29}$		&$<$0.40		& 69.6          &	1.65  			\\           
\textbf{M107}       				& 16 32 31.86   &-13 03 13.6  & 0.1     &1 	&  6.1        & 4    	&$<$5.6            					&$<$1.3 $\times10^{27}$				&$<$990 		&$<$2.4 $\times10^{30}$		&$<$1.14		& 33.6          &	0.50 			\\
\textbf{M13}        				& 16 41 41.21   &36 27 35.6   & 0.4     &2 	&  7.6         & 4    	&$<$4.9            					&$<$1.7 $\times10^{27}$				&$<$1110		&$<$3.4 $\times10^{30}$		&$<$0.24		& 37.2          &	0.52 			\\
\textbf{M12}        				& 16 47 14.18   &-01 56 54.7  & 0.8     &1 	&  5.2        & 4    	&$<$4.3            					&$<$7.0 $\times10^{26}$         	&$<$800 		&$<$1.3 $\times10^{30}$		&$<$0.92		& 47.4          &   0.78 	                \\
\textbf{M10}      					& 16 57 8.92  	&-04 05 58.0  &1.0    	&13	&  4.4    & 6    	&$<$4.9   								&$<$5.7 $\times10^{26}$       		&$<$740 		&$<$1.0 $\times10^{30}$		&$<$0.40		& 46.2          &  	0.80       	  	\\ 
M62      							& 17 01 12.98   &-30 06 49.0  &0.2  	&2 	&  6.7    & 29    	&$<$6.6   								&$<$1.8 $\times10^{27}$ 			&$<$1130		&$<$3.6 $\times10^{30}$		&$<$0.16		& 13.2          &  	0.18                \\ 
\textbf{M19}      					& 17 02 37.80   &-26 16 04.7  &1.1  	&5 	&  8.2    & 5    	&$<$4.8   								&$<$1.9 $\times10^{27}$       		&$<$1170		&$<$4.0 $\times10^{30}$		&$<$0.17		& 25.8          &  	0.35                \\
\textbf{N6304}    					& 17 14 32.25   &-29 27 43.3  &0.2  	&1 	&  5.9    & 7    	&$<$5.8   								&$<$1.2 $\times10^{27}$         	&$<$980 		&$<$2.3 $\times10^{30}$		&$<$0.35		& 12.6          &  	0.18 	        	\\
\textbf{M92}      					& 17 17 07.43   &43 08 09.26  &0.1  	&2 	&  8.9    & 1  		&$<$3.6									&$<$1.7 $\times10^{27}$         	&$<$1110		&$<$3.4 $\times10^{30}$		&$<$0.41		& 15.6          &  	0.22 	        	\\
\textbf{N6325}\tablenotemark{c}    	& 17 17 59.21   &-23 45 57.6  &2.0    	&6 	&  6.5     & 8    	&$<$4.5   								&$<$1.1 $\times10^{27}$	    		&$<$960 		&$<$2.2 $\times10^{30}$		&$<$0.92		& 1.8           &  	0.03	        	\\
\textbf{M9}       					& 17 19 11.78   &-18 30 58.5  &2.0    	&10	&  7.8    & 10  	&$<$3.8   								&$<$1.4 $\times10^{27}$         	&$<$1030		&$<$2.7 $\times10^{30}$		&$<$0.40		& 27.0          &  	0.40 	        	\\
Liller 1 							& 17 33 24.56   &-33 23 22.4  &0.3  	&3 	&  8.1     & 12   	&$<$6.8   								&$<$2.7 $\times10^{27}$         	&$<$1320		&$<$5.8 $\times10^{30}$		&$<$0.20		& 5.4           &	0.07		        \\
\textbf{M14}      					& 17 37 36.10 	&-03 14 45.3  &0.5 		&6 	&  9.3      &6  	&$<$3.9									&$<$2.0 $\times10^{27}$				&$<$1190		&$<$4.2 $\times10^{30}$		&$<$0.15		& 47.4          &  	0.64 	                \\ 
Terzan 5 							& 17 48 04.85 	&-24 46 44.6  &1.0    	&11	&  5.9      &19 		&$<$9.8   							&$<$2.0 $\times10^{27}$				&$<$1190		&$<$4.2 $\times10^{30}$		&$<$0.21		& 9.6           &  	0.13 	        	\\  
\textbf{N6440}    					& 17 48 52.70  	&-20 21 36.9  &1.1 		&5 	&  8.5      &20    	&$<$6.5   								&$<$2.8 $\times10^{27}$    			&$<$1340		&$<$6.1 $\times10^{30}$		&$<$0.30		& 8.4           & 	0.11 	        	\\          
\textbf{N6522}\tablenotemark{c}    	& 18 03 34.89 	&-30 02 03.2  &     $\sim 2$\tablenotemark{a}	&  	&  7.7     & 6    	&$<$9.5   		&$<$3.4 $\times10^{27}$    			&$<$1430		&$<$7.4 $\times10^{30}$		&$<$0.36		& 3.0           &	0.04			\\ 
\textbf{N6539}    					& 18 04 49.68 	&-07 35 09.1  &0.32 	&6 	&  7.8      &20  	&$<$4.4   								&$<$1.6 $\times10^{27}$				&$<$1090		&$<$3.2 $\times10^{30}$		&$<$1.22		& 22.8          &  	0.32 	        	\\ 
\textbf{N6544}    					& 18 07 20.12 	&-24 59 53.6  &0.95 	&8 	&  3.0      &20  	&$<$5.3   								&$<$2.9 $\times10^{26}$				&$<$570 		&$<$4.7 $\times10^{29}$		&$<$0.90		& 3.0           &  	0.06	        	\\  
\textbf{M28}      					& 18 24 32.73 	&-24 52 13.0  &0.7  	&2 	&  5.5      &20    	&$<$5.1   								&$<$9.2 $\times10^{26}$				&$<$890 		&$<$1.8 $\times10^{30}$		&$<$0.24		& 14.4          &   	0.23 			\\ 
\textbf{M22 }     					& 18 36 23.94 	&-23 54 17.1  &0.8  	&1 	&  3.1      &28    	&$<$5.0   								&$<$2.9 $\times10^{26}$				&$<$570 		&$<$4.7 $\times10^{29}$		&$<$0.14		& 79.8          &  	1.56        	\\         
\textbf{N6712}    					& 18 53 04.30 	&-08 42 22.0  &0.5  	&9 	&  8.0        &9    	&$<$4.8   							&$<$1.8 $\times10^{27}$				&$<$1150		&$<$3.8 $\times10^{30}$		&$<$0.90		& 45.6          & 	0.63        	\\       
M54									& 18 55 03.33 	&-30 28 47.5  &0.1  	&1 	&  23.9     &15		&$<$7.1   								&$<$2.4 $\times10^{28}$				&$<$2990		&$<$6.1 $\times10^{31}$		&$<$0.21		& 5.4           &  	0.04        	\\ 
\textbf{N6760}    					& 19 11 12.01 	&01 01 49.7   &0.5 		&6 	&  7.4      &20    	&$<$4.5   								&$<$1.5 $\times10^{27}$				&$<$1060 		&$<$3.0 $\times10^{30}$		&$<$0.42		& 20.4          &  	0.29        	\\         
\textbf{M55 }     					& 19 39 59.71 	&-30 57 53.1  &0.8  	&1 	&  5.7     &4   	&$<$5.1   								&$<$1.0 $\times10^{27}$  			&$<$910 		&$<$1.9 $\times10^{30}$		&$<$0.48		& 108.0         &  	1.67        	\\ 
M15\tablenotemark{c}      			& 21 29 58.33 	&12 10 01.2   &0.2      &1  &  10.3    & 1  	&$<$6.3         						&$<$4.0 $\times10^{27}$    			&$<$ 1530   	&$<$9.0 $\times10^{30}$		&$<$0.34		& 8.4             	&  	 0.10           	\\       
\textbf{M2}       					& 21 33 26.96 	&-00 49 22.9  &1.0    	&12	&  11.5     &6  	&$<$3.9   								&$<$3.1 $\times10^{27}$				&$<$1390		&$<$6.8 $\times10^{30}$		&$<$0.24		& 19.2          &  	0.24        	\\         
\textbf{M30}\tablenotemark{c}      	& 21 40 22.12 	&-23 10 47.5  &0.1  	&1 	&  8.6     &4  		&$<$3.9									&$<$1.7 $\times10^{27}$				&$<$1120		&$<$3.5 $\times10^{30}$		&$<$0.84		& 3.6           &  	0.05        	\\         
\enddata                                                                                                                                                                                                                                                                                                                                                                                                    
\tablerefs{
(1) \cite{2010goldsbury}; (2) \cite{2013miocchi}; (3) \cite{2015saracino}; (4) \cite{1987djorg}; (5) \cite{1995picard}; (6) \cite{1986shawl};(7) \cite{2003zand}; (8) \cite{2014cohen}; (9) \cite{2006noyola}; (10) \cite{2015vanderbeke} ; (11) \cite{2010lanzoni} ; (12) \cite{2009dalessandro} ; (13) \cite{2011dalessandro}; (14) \cite{2007cohen}; (15) \cite{2015watkins};  (16) \cite{2016watkins} ; (17) \cite{2017baumgardt}; (18) \cite{1999ferraro}; (19)\cite{2007valenti} ; (20) \cite{2010harris}; (21) \cite{2005valenti}; (22) \cite{2003ortolani}; (23) \cite{2001paltrinieri}; (24) \cite{2013ferro} ; (25) \cite{2007ortolani}; (26) \cite{2015saracino}; (27) \cite{1999ortolani}; (28)\cite{2013kunder} ; (29) \cite{2010contreras}; (30) \cite{2013kaluzny}}
\tablenotetext{a}{The center listed in the \cite{2010harris} catalog from \cite{1986shawl} is inconsistent with the apparent center of the cluster in 2MASS. We redetermine the center
using this 2MASS images, and this value is the one listed in Table \ref{tab:vla}. The uncertainty in the value is not well-determined but we estimate $\sim 2\arcsec$.}
\tablenotetext{b}{GCs used for the stacking analysis are indicated in boldface}
{\tablenotetext{c}{Core collapsed GCs as adapted by \cite{1995trager} catalog}
}\end{deluxetable}
\end{longrotatetable}

\begin{longrotatetable}
\begin{deluxetable}{lccccccccccccc}
\tabletypesize{\scriptsize}
\tablecolumns{11} 
\tablewidth{0pt} 
\rotate
\tablecaption{Results from ATCA Data. We note that we assume 5 GHz as central frequency in order to calculate the radio luminosities and the mass upper limits. The sources used for the photometric centers and the distances are shown at the bottom of the Table.  The core radii are adopted from  \cite{2010harris}. \label{tab:atca}}
\tablehead{
\colhead{ID}           & 
\colhead{RA (J2000)}          &
\colhead{DEC (J2000)}          & 
\colhead{Uncertainty}  &
\colhead{Position} &
\colhead{Distance} &
\colhead{Distance} &
\colhead{3$\sigma$ Flux  } & 
\colhead{ 3$\sigma$ $L_{R}$ } & 
\colhead{IMBH} &
\colhead{ 3$\sigma$ $L_{X}$ } &
\colhead{IMBH mass} &
\colhead{Core} &
\colhead{Brownian} 
 \\
\colhead{} &
\colhead{(h:m:s)} &
\colhead{(\phn{\arcdeg}~\phn{\arcmin}~\phn{\arcsec})} &
\colhead{(\phn{\arcsec})} & 
\colhead{reference}&
\colhead{(kpc)}& 
\colhead{reference} &                    
\colhead{density ($\mu$Jy)}&
\colhead{(erg/s)}&
\colhead{mass (M$_{\odot}$)}& 
\colhead{(erg/s)}& 
\colhead{fraction}& 
\colhead{radius (\phn{\arcsec})}&
\colhead{radius (\phn{\arcsec})}
}    
\startdata
\textbf{47 Tuc}          				& 00 24 05.71   &-72 04 52.2  & 0.5  &2 	        &       4.6    &15	         &$<$11.1 		&$<$1.4$\times10^{27}$      	&$<$1040       	&$<$2.8$\times10^{30}$  &$<$ 0.13   		& 21.6       &  	0.32			 \\   
\textbf{N2808}           				& 09 12 03.10   &-64 51 48.6  & 0.1  &1 	        &       9.4    &15	         &$<$8.8    	&$<$4.7$\times10^{27}$          &$<$1620       	&$<$1.0$\times10^{31}$  &$<$ 0.22  		& 15.0       &  	0.17	\\
\textbf{N3201}           				& 10 17 36.82   &-46 24 44.9  & 1.0  &1 	        &       4.9    &18	         &$<$9.3  		&$<$1.3$\times10^{27}$          &$<$1020       	&$<$2.7$\times10^{30}$  &$<$ 0.68  		& 78.0       &  	1.15	\\
\textbf{N4372}           				& 12 25 45.40   &-72 39 32.4  & 1.8  &6 	        &       6.3    &18	         &$<$9.9  		&$<$2.4$\times10^{27}$          &$<$1250       	&$<$4.9$\times10^{30}$  &$<$ 0.50  		& 105.0      &  	1.39	\\
\textbf{N4833}           				& 12 59 33.92   &-70 52 35.4  & 0.3  &1 	        &       6.7    &18	         &$<$9.7    	&$<$2.6$\times10^{27}$          &$<$1300       	&$<$5.5$\times10^{30}$  &$<$ 0.53  		& 60.0       &  	0.77	\\
\textbf{$\omega$ Cen}    				& 13 26 47.28   &-47 28 46.1  & 0.1  &1 	        &       4.9    &15	         &$<$8.8		&$<$1.3$\times10^{27}$          &$<$1000       	&$<$2.5$\times10^{30}$  &$<$ 0.03 		& 142.2      &  	2.11	\\
\textbf{N5927 }          				& 15 28 00.69   &-50 40 22.9  & 0.2  &1 	        &       7.9    &15	         &$<$10.7 		&$<$4.0$\times10^{27}$          &$<$1530       	&$<$9.0$\times10^{30}$  &$<$ 0.43  		& 25.2	     &		0.30	\\
N6139   								& 16 27 40.37 	&-38 50 55.5  & 1.0  &6 			&       10.4   &21	   	 	 &$<$10.8 		&$<$7.0$\times10^{27}$			&$<$1880     	&$<$1.6$\times10^{31}$  &$<$ 0.52 		& 9.0        &  	0.09		\\ 
\textbf{N6352}   						& 17 25 29.11 	&-48 25 19.8  & 0.6  &1 			&       5.6    &20	   	 	 &$<$8.9  		&$<$1.7$\times10^{27}$			&$<$1110     	&$<$3.4$\times10^{30}$  &$<$ 1.18 		& 49.8       &  	0.70		\\ 
\textbf{N6362 }  						& 17 31 54.99 	&-67 02 54.0  & 0.5  &1 			&       7.6    &20	   	 	 &$<$7.8    	&$<$2.7$\times10^{27}$			&$<$1320  		&$<$5.8$\times10^{30}$  &$<$ 0.90 		& 67.8       &  	0.87		    \\  
Terzan 1\tablenotemark{b}				& 17 35 47.20 	&-30 28 54.4  & 1.1  &5 			&      	5.2    &13	         &$<$11.7    	&$<$1.9$\times10^{27}$    		&$<$1160     	&$<$3.9$\times10^{30}$  &$<$ 0.52 		& 2.4	     &		0.03	\\ 
N6388   								& 17 36 17.23 	&-44 44 07.8  & 0.3  &1 			&      	10.9   &15	       	 &$<$8.4 		&$<$6.0$\times10^{27}$    		&$<$1770     	&$<$1.4$\times10^{31}$  &$<$ 0.17 		& 7.2        &  	0.07		\\ 
\textbf{N6397}\tablenotemark{b}   		& 17 40 42.09 	&-53 40 27.6  & 0.3  &1 			&      	2.3    &15	      	 &$<$10.5 		&$<$3.3$\times10^{26}$			&$<$610       	&$<$5.7$\times10^{29}$  &$<$ 0.69 		& 3.0        &  	0.05		    \\  
\textbf{N6441}   						& 17 50 13.06 	&-37 03 05.2  & 0.2  &1 			&      	13.4   &19	         &$<$10.8 		&$<$1.2$\times10^{28}$			&$<$2270     	&$<$2.9$\times10^{31}$	&$<$ 0.18 	& 7.8        & 		0.07		\\ 
Terzan 6\tablenotemark{b}    			& 17 50 46.85	&-31 16 29.3  & 0.6  &7 	        &       6.8    &20	         &$<$12.6    	&$<$3.5$\times10^{27}$          &$<$1450      	&$<$7.7$\times10^{30}$  &$<$ 1.24 		& 3.0        &  	0.03		\\ 
Djorg 2 								& 18 01 49.1  	&-27 49 33.0  & 3.0  &10 			&      	7.0    &23	         &$<$10.6 		&$<$3.1$\times10^{27}$			&$<$1390     	&$<$6.8$\times10^{30}$	&$<$ 1.29 	& 19.8       &  	0.24		\\         
\textbf{N6541 }  						& 18 08 02.36 	&-43 42 53.6  & 0.1  &1 			&      	7.5    &20	         &$<$11.7 		&$<$3.9$\times10^{27}$			&$<$1520     	&$<$8.8$\times10^{30}$  &$<$ 0.55 		& 10.8       &  	0.13		\\ 
\textbf{N6553 }  						& 18 09 17.59 	&-25 54 38.0  & 2.0  &14			&      	6.0    &20	         &$<$10.3 		&$<$2.2$\times10^{27}$			&$<$1230     	&$<$4.7$\times10^{30}$  &$<$ 0.52 		& 31.8       &  	0.42		\\         
N6624\tablenotemark{b}   				& 18 23 40.51 	&-30 21 39.7  & 0.1  &1 			&       8.4    &19	         &$<$9.9     	&$<$4.2$\times10^{27}$			&$<$1550     	&$<$9.4$\times10^{30}$  &$<$ 2.12 		& 3.6        &  	0.04		\\ 
N6652   								& 18 35 45.63 	&-32 59 26.6  &	0.1  &1         	&		10.0   &20           &$<$10.7 		&$<$6.4$\times10^{27}$			&$<$1820      	&$<$1.5$\times10^{31}$	&$<$ 2.30 	& 6.0        &  	0.06		    \\       
\textbf{N6752}\tablenotemark{b}   		& 19 10 52.11 	&-59 59 04.4  & 0.1  &1 			&      	4.0    &15	      	 &$<$10.5 		&$<$1.0$\times10^{27}$			&$<$920      	&$<$1.9$\times10^{30}$  &$<$ 0.38 		& 10.2       &  	0.15		\\     
\enddata                                                                                                                                                                                                                                                                                                                                                                                     
\tablerefs{
(1) \cite{2010goldsbury}; (2) \cite{2013miocchi}; (3) \cite{2015saracino}; (4) \cite{1987djorg}; (5) \cite{1995picard}; (6) \cite{1986shawl};(7) \cite{2003zand}; (8) \cite{2014cohen}; (9) \cite{2006noyola}; (10) \cite{2015vanderbeke} ; (11) \cite{2010lanzoni} ; (12) \cite{2009dalessandro} ; (13) \cite{2011dalessandro}; (14) \cite{2007cohen}; (15) \cite{2015watkins};  (16) \cite{2016watkins} ; (17) \cite{2017baumgardt}; (18) \cite{1999ferraro}; (19) \cite{2007valenti} ; (20) \cite{2010harris}; (21) \cite{1998zinn}; (22) \cite{2016bogdanov}; (23) \cite{2010valenti}}
\tablenotetext{a}{GCs used for the stacking analysis are indicated in boldface.}
{\tablenotetext{b}{Core collapsed GCs as adapted by \cite{1995trager} catalog} 
}\end{deluxetable}
\end{longrotatetable}

\end{document}